\documentclass[twocolumn,twocolappendix]{aastex702}

\usepackage{amsmath}
\usepackage{graphicx}
\usepackage{threeparttable}
\usepackage{txfonts}
\usepackage{url}
\usepackage{hyperref}
\bibpunct[; ]{(}{)}{;}{a}{}{,}
\hypersetup{allcolors=blue}

\begin{document}
\title{Chandra Lensing-cluster Ultradeep Extragalactic Survey (CLUES) I: A 2 Ms Point-Source Catalog of the Abell 2744 Field 
}

\author[0009-0003-1260-4143]{Shouyi Wang}
\email[show]{sywang0302@gmail.com}
\affiliation{Department of Astronomy, School of Physics, Peking University, Beijing 100871, People's Republic of China}
\affiliation{Kavli Institute for Astronomy and Astrophysics, Peking University, Beijing 100871, People's Republic of China}

\author[0000-0002-4436-6923]{Fan Zou}
\email[show]{fanzou01@gmail.com}
\affiliation{Department of Astronomy, University of Michigan, 1085 S University, Ann Arbor, MI 48109, USA}

\author[0000-0001-5802-6041]{Elena Gallo}  \email{egallo@umich.edu}
\affiliation{Department of Astronomy, University of Michigan, 1085 S University, Ann Arbor, MI 48109, USA}

\author[0000-0002-9036-0063]{Bin Luo} \email{bluo@nju.edu.cn}
\affiliation{School of Astronomy and Space Science, Nanjing University, Nanjing, Jiangsu 210093, People’s Republic of China}
\affiliation{Key Laboratory of Modern Astronomy and Astrophysics (Nanjing University), Ministry of Education, Nanjing 210093, People’s Republic of China}

\author[0000-0002-0167-2453]{W. N. Brandt}
\email{wnbrandt@gmail.com}
\affiliation{Department of Astronomy and Astrophysics, 525 Davey Lab, The Pennsylvania State University, University Park, PA 16802, USA}
\affiliation{Institute for Gravitation and the Cosmos, The Pennsylvania State University, University Park, PA 16802, USA}
\affiliation{Department of Physics, 104 Davey Laboratory, The Pennsylvania State University, University Park, PA 16802, USA}

\author[0009-0005-3823-9302]{Yuxuan Pang}
\email{pangyuxuan@ucas.ac.cn}
\affiliation{School of Astronomy and Space Science, University of Chinese Academy of Sciences (UCAS), Beijing 100049, China}

\author[0000-0002-8460-0390]{Tommaso Treu}
\email{tt@astro.ucla.edu}
\affiliation{Physics and Astronomy Department, University of California, Los Angeles, CA 90095, USA}

\author[0000-0002-7350-6913]{Xue-Bing Wu}
\email{wuxb@pku.edu.cn}
\affiliation{Department of Astronomy, School of Physics, Peking University, Beijing 100871, People's Republic of China}
\affiliation{Kavli Institute for Astronomy and Astrophysics, Peking University, Beijing 100871, People's Republic of China}

\author[0000-0002-5678-1008]{Dieu D. Nguyen}
\email{dieun@umich.edu}
\affiliation{Department of Astronomy, University of Michigan, 1085 S University, Ann Arbor, MI 48109, USA}

\author[0000-0002-4140-1367]{Guido Roberts-Borsani}
\email{g.robertsborsani@ucl.ac.uk}
\affiliation{Department of Physics \& Astronomy, University College London, London, WC1E 6BT, UK}

\author[0009-0007-6655-366X]{Shengzhe Wang}
\email{wangsz@bao.ac.cn}
\affiliation{National Astronomical Observatories, Chinese Academy of Sciences, Beijing 100101, China}
\affiliation{School of Astronomy and Space Science, University of Chinese Academy of Sciences (UCAS), Beijing 100049, China}

\author[0000-0002-9587-6683]{Weiwei Xu} 
\email{}
\affiliation{National Astronomical Observatories, Chinese Academy of Sciences, Beijing 100101, China}
\affiliation{School of Astronomy and Space Science, University of Chinese Academy of Sciences (UCAS), Beijing 100049, China}
  
\author[0009-0000-8524-8344]{Zihao Zuo}
\email{zuo@umich.edu}
\affiliation{Department of Astronomy, University of Michigan, 1085 S University, Ann Arbor, MI 48109, USA}

\begin{abstract}
In the first paper of the Chandra Lensing-cluster Ultradeep Extragalactic Survey (CLUES), we present an ultradeep 2~Ms X-ray survey of the Abell~2744 field constructed from 101 archival Chandra ACIS-I observations. With an exposure comparable to the Chandra Deep Fields and further boosted by strong-lensing magnification from the Abell~2744 cluster at $z=0.3$, this field represents the third deepest extragalactic X-ray survey of the sky and also has rich synergy with extensive coverage by the Hubble Space Telescope and James Webb Space Telescope. We present the Chandra data reduction and detect sources in the soft (0.5--2~keV), hard (2--7~keV), and full (0.5--7~keV) bands over a total area of $369~\mathrm{arcmin^2}$. We perform dedicated image fitting with Chandra point spread functions to optimize point-source detections and improve \mbox{X-ray} positions and further screen the detections to address the impact of the central bright, structured intracluster medium. A total of 327~\mbox{X-ray} point sources are detected and cataloged, including their X-ray photometry and basic spectral properties. Detailed simulations are also conducted, based on which the expected 50\% flux completeness reaches $6.3\times10^{-16}$, $2.8\times10^{-16}$, and $4.9\times10^{-16}~\mathrm{erg~cm^{-2}~s^{-1}}$ in the full, soft, and hard bands, respectively. The source number density as a function of flux is consistent with those in blank-field surveys within a factor of $\approx2$, with a slight excess above $\approx10^{-14}~\mathrm{erg~cm^{-2}~s^{-1}}$. All \mbox{X-ray} data products are publicly released, including the catalog, \mbox{X-ray} images, exposure maps, background maps, and sensitivity maps.
\end{abstract}

\section{Introduction}
\label{sec: intro}
Over the past two to three decades, deep cosmic \mbox{X-ray} surveys have significantly advanced our understanding of the \mbox{X-ray} sky and the high-energy phenomena out to the distant universe (e.g., \citealt{Brandt2015, Brandt2024} and references therein), especially thanks to powerful modern \mbox{X-ray} facilities, such as the Chandra \mbox{X-ray} Observatory (Chandra), \mbox{X-ray} Multi-Mirror Mission (XMM-Newton), and extended ROentgen Survey with an Imaging Telescope Array (eROSITA). These surveys have resolved the majority of the cosmic \mbox{X-ray} background (e.g., \citealt{Luo2017}), revealed the \mbox{X-ray} binary populations out to cosmological distances (e.g., \citealt{Lehmer2016}), constrained galaxy clusters and groups (e.g., \citealt{Allen2011}; S. Wang et al. submitted), and, notably, provided rich information on the population of active galactic nuclei (AGNs). AGNs generally dominate the number of detected \mbox{X-ray} sources in current surveys, except for the very faint end below \mbox{X-ray} fluxes of $\approx10^{-17}~\mathrm{erg~cm^{-2}~s^{-1}}$ \citep{Luo2017}.\par
These \mbox{X-ray} surveys have been designed following a wedding-cake pattern, where different fields span different layers from deep pencil-beam to shallower wide-area. The deepest pencil-beam layer has been anchored primarily by Chandra due to its extremely low background and a sharp spatial resolution. Especially, the 7~Ms Chandra Deep Field-South (CDF-S; \citealt{Luo2017}) and 2~Ms Chandra Deep Field-North (CDF-N; \citealt{Xue2016}) represent the deepest \mbox{X-ray} views ever to the sky. The next deepest field is the central region of the Extended Groth Strip with 800~ks Chandra observations (AEGIS-XD; \citealt{Nandra2015}), which is shallower than the CDFs by a factor of several.\par
In this work, we present the Chandra Lensing-cluster Ultradeep Extragalactic Survey (CLUES), which covers a new ultradeep pencil-beam field, Abell~2744 (also known as the Pandora's Cluster), with 2~Ms archival Chandra observations in total. Unlike previous \mbox{X-ray} surveys that primarily target blank fields, Abell~2744 is a giant galaxy cluster at $z=0.3$. \citet{Chadayammuri2024} have conducted detailed analyses of Abell~2744 itself with these same Chandra observations and revealed a complex merging picture and diffuse \mbox{X-ray} emission from the hot intracluster medium (ICM). This work instead focuses on other compact \mbox{X-ray} sources in this field that usually are not physically associated with Abell~2744, although some may also be from the cluster members.\par
Compared to blank fields, this field has a unique feature that Abell~2744 exerts strong lensing around its center and hence magnifies background sources. Faint sources that would be undetectable without lensing can thus be magnified and reach the detection limits of current instruments (e.g., \citealt{Treu2010} and references therein). Abell~2744 is notably the best characterized lensing cluster \citep{Bergamini2023} with a larger lensing area at a fixed magnification $\mu$ compared to other similar lensing clusters such as Abell S1063 \citep{Atek2025}. Abell~2744 has also been extensively targeted by several large programs of the Hubble Space Telescope (HST) and the James Webb Space Telescope (JWST), including but not limited to the Hubble Frontier Fields (HFF; \citealt{Lotz2017}), the HST Beyond Ultra-deep Frontier Fields and Legacy Observations (BUFFALO; \citealt{Steinhardt2020}), the JWST Ultradeep NIRSpec and NIRCam Observations before the Epoch of Reionization (UNCOVER; \citealt{Bezanson2024}), the GLASS-JWST survey \citep{Treu2022}, Medium-band Astrophysics with the Grism of NIRCam in Frontier Fields (MAGNIF; \citealt{Fu2025}), and the upcoming Deep Amplified JWST Experiment (DAJE) survey (GO-9645; PIs T. Treu, A. Fontana, and G. Roberts-Borsani). Gravitational lensing has also helped Chandra deliver a number of possible AGN candidates based on putative \mbox{X-ray} detections in the field, such as GHZ9 \citep{Kovacs2024} and UHZ1 \citep{Bogdan2024}, which were listed as one of the most groundbreaking discoveries by Chandra \citep{Slane2025}, although some of these claims are disputed \citep{Alvarez-Marquez2026, Zou2026}.\par
Therefore, the Chandra CLUES survey has significant legacy values due to the following two reasons. First, its 2~Ms exposure is comparable to the CDF-N, and the lensing also improves the effective, de-magnified sensitivity in the center. Therefore, this survey will also represent one of the deepest \mbox{X-ray} views of the sky. Second, it has good synergy with the existing/upcoming deep HST and JWST data in this field. The \mbox{X-rays} can help constrain likely AGN phenomena discovered by HST/JWST, and even \mbox{X-ray} non-detections are often also informative given the \mbox{X-ray} depth. For example, several early galaxies at $z>9$ have been identified by JWST in this field (e.g., \citealt{Castellano2022, RobertsBorsani2023, RobertsBorsani2026, Boyett2024, Fujimoto2024, Napolitano2025}), partly thanks to the lensing magnification. Given the challenges of securely identifying AGN activity with traditional rest-UV and optical diagnostics (e.g., from HST and JWST), additional \mbox{X-ray} constraints serve as an extremely valuable additional constraint and diagnostic (e.g., \citealt{Bogdan2024, Kovacs2024, Zou2026}; B. P\'erez-D\'iaz et al. submitted).\par
In this Chandra CLUES survey, we will systematically reduce all the Chandra data and analyze the corresponding sources. This work is the first of the Chandra CLUES series, where we focus on identifying and cataloging \mbox{X-ray} point sources and presenting flux sensitivity maps. Compared to blank fields, the diffuse ICM emission represents the main practical difficulty in our field, and we will carefully treat it to ensure reliable source detections. In the second of this series (S. Wang et al. in preparation), we will identify the optical/infrared counterparts of these \mbox{X-ray} point sources and classify their natures (stars, galaxies, or AGNs). For galaxies and AGNs, we will further analyze their multiwavelength spectral energy distributions to derive dedicated photometric redshifts and host-galaxy properties.\par
This work is structured as follows. Section~\ref{sec: chandra} presents the Chandra data reduction. Section~\ref{sec:psf_modeling} presents source detections based on local background modeling and point-spread function (PSF) fitting. Section~\ref{sec:simulations} conducts detailed, end-to-end simulations to quantify the expected completeness and reliability. Section~\ref{sec:catalog} presents the generation of the main point-source catalog. Section~\ref{sec:sensimap} quantifies the survey sensitivity and effective sky area. Section~\ref{sec:number_counts} measures the cumulative source number density as a function of \mbox{X-ray} flux. Section~\ref{sec: summary} summarizes this work. We adopt a flat $\Lambda$CDM cosmology with $H_0=67.7~\mathrm{km~s^{-1}~Mpc^{-1}}$ and $\Omega_M=0.31$ \citep{PlanckCollaboration2020}.

\section{Chandra Data Reduction and Preliminary Source Detection}
\label{sec: chandra}
The Abell~2744 field has been observed repeatedly by Chandra over two decades, spanning 2001 September 3 to 2024 May 29, although most observations were conducted between 2022 and 2024.\footnote{This paper employs a list of Chandra datasets, obtained by the Chandra \mbox{X-ray} Observatory, contained in the Chandra Data Collection \dataset[DOI:10.25574/cdc.633]{https://doi.org/10.25574/cdc.633}.} This complete archive contains 104 observations. We first
excluded the earliest 25~ks observation (ObsID~2212),
the only observation taken with ACIS-S, because
ACIS-S has a different detector geometry from ACIS-I. This choice maintains a
homogeneous ACIS-I data set.
The aim points of the 99 observations obtained during
2022--2024 are all within $0.31\arcmin$ of their mean
position. Among the four earlier ACIS-I observations,
ObsIDs~8477 and 8557 are offset from this position by
approximately $5.4\arcmin$, while ObsIDs~7915 and
7712 have smaller offsets of $1.3\arcmin$ and
$1.6\arcmin$, respectively.
Since the Chandra PSF size strongly evolves with the off-axis angle \citep[e.g.,][]{Kim2007}, the PSFs of the two highly off-axis observations (ObsIDs~8477 and 8557) at a given position would be significantly different from the remaining observations. To ensure an accurate characterization of the PSF map, we excluded ObsIDs~8477 and 8557 but retained ObsIDs~7915 and 7712. These two excluded observations have exposures of 45.91 and 27.81~ks,
respectively, totaling 73.72~ks. Such a procedure of retaining only observations with roughly similar aim points (within $\approx1'-2'$) was also employed in the CDF-S \citep{Luo2017} due to the same PSF stability reason. Our final data set therefore contains 101
ACIS-I observations, with a total exposure
of 2.102~Ms.

\subsection{Initial Processing and Image Generation}
\label{sec:xrayimgproc}
All 101 ACIS-I observations were taken in Very Faint
mode to minimize particle background and enhance the sensitivity for weak source detection.
The selected observations have closely aligned aim
points but span a range of roll angles, which broaden the final footprint relative to a single ACIS-I pointing and help average over CCD-gap effects in the merged mosaic.
The observation log is given in
Table~\ref{tab:chandra_obs}.
\par

\begingroup
\setlength{\tabcolsep}{3pt}
\begin{deluxetable*}{lccccccc@{\hspace{0.20cm}}|@{\hspace{0.20cm}}cccccccc}
\tabletypesize{\scriptsize}
\tablewidth{0pt}
\tablecaption{List of 2.1 Ms Chandra Abell 2744 field Observations \label{tab:chandra_obs}}
\tablehead{
\colhead{ObsID} & \colhead{Date} & \colhead{Raw} & \colhead{Clean} & \colhead{R.A.} & \colhead{Decl.} & \colhead{Roll} & \colhead{Offset} & \colhead{ObsID} & \colhead{Date} & \colhead{Raw} & \colhead{Clean} & \colhead{R.A.} & \colhead{Decl.} & \colhead{Roll} & \colhead{Offset} \\
\colhead{(1)} & \colhead{(2)} & \colhead{(3)} & \colhead{(4)} & \colhead{(5)} & \colhead{(6)} & \colhead{(7)} & \colhead{(8)}&\colhead{(1)} & \colhead{(2)} & \colhead{(3)} & \colhead{(4)} & \colhead{(5)} & \colhead{(6)} & \colhead{(7)} & \colhead{(8)} \\
 &  & \colhead{(ks)} & \colhead{(ks)} & \colhead{(deg)} & \colhead{(deg)} & \colhead{(deg)} & \colhead{(arcmin)} &  &  & \colhead{(ks)} & \colhead{(ks)} & \colhead{(deg)} & \colhead{(deg)} & \colhead{(deg)} & \colhead{(arcmin)}
}
\startdata
7712 & 2007-09-10 & 8.07 & 8.07 & 3.62116 & $-30.39757$ & 31.89 & 1.59 & 25953 & 2022-09-17 & 24.76 & 24.76 & 3.59222 & $-30.40204$ & 17.69 & 0.14 \\
7915 & 2006-11-08 & 18.62 & 18.62 & 3.61242 & $-30.39067$ & 315.47 & 1.27 & 25954 & 2022-04-24 & 13.40 & 13.40 & 3.58793 & $-30.39813$ & 138.70 & 0.19 \\
25277 & 2023-10-02 & 18.68 & 18.67 & 3.59140 & $-30.40379$ & 335.20 & 0.21 & 25955 & 2023-07-20 & 43.42 & 43.42 & 3.58920 & $-30.39785$ & 96.81 & 0.16 \\
25278 & 2022-12-02 & 9.78 & 9.78 & 3.58895 & $-30.40537$ & 302.20 & 0.31 & 25956 & 2022-09-02 & 13.90 & 13.80 & 3.59167 & $-30.40050$ & 48.57 & 0.06 \\
25279 & 2022-09-06 & 24.46 & 24.46 & 3.59210 & $-30.40116$ & 34.21 & 0.10 & 25957 & 2022-09-08 & 21.80 & 21.80 & 3.59252 & $-30.40101$ & 34.21 & 0.11 \\
25907 & 2022-11-08 & 36.80 & 36.80 & 3.59093 & $-30.40513$ & 314.71 & 0.29 & 25958 & 2022-05-04 & 12.32 & 12.32 & 3.58870 & $-30.39814$ & 131.20 & 0.16 \\
25908 & 2022-09-23 & 22.61 & 22.61 & 3.59174 & $-30.40263$ & 6.94 & 0.15 & 25959 & 2023-08-05 & 15.39 & 15.39 & 3.59049 & $-30.39804$ & 84.66 & 0.14 \\
25909 & 2023-05-24 & 19.33 & 19.33 & 3.58706 & $-30.39733$ & 125.49 & 0.25 & 25960 & 2023-07-08 & 24.76 & 24.76 & 3.58862 & $-30.39740$ & 104.02 & 0.20 \\
25910 & 2022-09-25 & 19.31 & 19.31 & 3.59229 & $-30.40231$ & 13.21 & 0.15 & 25961 & 2023-09-09 & 18.84 & 18.84 & 3.59203 & $-30.40091$ & 34.26 & 0.09 \\
25911 & 2022-04-19 & 16.86 & 16.86 & 3.58724 & $-30.39805$ & 148.00 & 0.22 & 25962 & 2023-09-11 & 21.81 & 21.81 & 3.59264 & $-30.40063$ & 31.03 & 0.11 \\
25912 & 2022-04-18 & 15.38 & 15.38 & 3.58737 & $-30.39798$ & 148.25 & 0.21 & 25963 & 2022-11-26 & 37.59 & 37.59 & 3.58991 & $-30.40541$ & 305.48 & 0.31 \\
25913 & 2022-09-03 & 19.64 & 19.53 & 3.59207 & $-30.40018$ & 46.11 & 0.08 & 25964 & 2023-09-05 & 20.32 & 20.32 & 3.59237 & $-30.40003$ & 42.49 & 0.10 \\
25914 & 2022-10-15 & 28.20 & 28.20 & 3.59269 & $-30.40293$ & 348.21 & 0.19 & 25965 & 2023-07-07 & 35.65 & 35.65 & 3.58833 & $-30.39729$ & 104.49 & 0.21 \\
25915 & 2022-09-03 & 21.08 & 21.08 & 3.59177 & $-30.40035$ & 44.98 & 0.07 & 25966 & 2023-08-13 & 18.84 & 18.84 & 3.59014 & $-30.39867$ & 90.20 & 0.10 \\
25916 & 2023-09-03 & 22.20 & 22.20 & 3.59245 & $-30.39963$ & 46.60 & 0.11 & 25967 & 2022-08-01 & 33.64 & 33.64 & 3.59080 & $-30.39911$ & 88.19 & 0.07 \\
25917 & 2023-06-22 & 35.62 & 34.63 & 3.58841 & $-30.39753$ & 112.08 & 0.20 & 25968 & 2022-07-12 & 27.46 & 26.47 & 3.58893 & $-30.39779$ & 110.21 & 0.17 \\
25918 & 2022-09-13 & 20.63 & 20.63 & 3.59163 & $-30.40203$ & 26.18 & 0.12 & 25969 & 2022-10-09 & 27.72 & 27.23 & 3.59275 & $-30.40331$ & 342.70 & 0.21 \\
25919 & 2022-06-13 & 25.29 & 25.29 & 3.58920 & $-30.39841$ & 118.21 & 0.13 & 25970 & 2022-06-12 & 24.75 & 24.75 & 3.58923 & $-30.39830$ & 116.59 & 0.14 \\
25920 & 2022-06-13 & 30.51 & 30.24 & 3.58861 & $-30.39789$ & 118.21 & 0.18 & 25971 & 2022-05-04 & 12.62 & 12.62 & 3.58858 & $-30.39809$ & 131.20 & 0.17 \\
25921 & 2023-08-04 & 16.87 & 16.84 & 3.59034 & $-30.39828$ & 85.46 & 0.12 & 25972 & 2022-05-18 & 31.75 & 30.40 & 3.58757 & $-30.39785$ & 134.20 & 0.21 \\
25922 & 2022-06-14 & 31.37 & 31.37 & 3.58886 & $-30.39824$ & 118.20 & 0.15 & 25973 & 2022-11-11 & 18.15 & 18.15 & 3.59094 & $-30.40511$ & 314.71 & 0.29 \\
25923 & 2022-09-04 & 10.94 & 10.60 & 3.59251 & $-30.40066$ & 40.21 & 0.11 & 26280 & 2022-01-18 & 11.75 & 11.70 & 3.58631 & $-30.40336$ & 275.20 & 0.28 \\
25924 & 2022-09-07 & 21.80 & 21.80 & 3.59249 & $-30.40109$ & 34.21 & 0.11 & 27347 & 2022-09-09 & 21.96 & 21.96 & 3.59256 & $-30.40090$ & 34.21 & 0.11 \\
25925 & 2022-09-02 & 23.60 & 22.97 & 3.59161 & $-30.40073$ & 47.49 & 0.06 & 27449 & 2022-09-24 & 9.78 & 9.78 & 3.59164 & $-30.40285$ & 6.01 & 0.16 \\
25926 & 2023-07-12 & 61.18 & 61.18 & 3.58875 & $-30.39719$ & 99.20 & 0.21 & 27450 & 2022-09-26 & 9.78 & 9.78 & 3.59122 & $-30.40292$ & 2.54 & 0.16 \\
25927 & 2023-09-16 & 20.53 & 20.53 & 3.59245 & $-30.40200$ & 13.20 & 0.14 & 27556 & 2022-11-15 & 25.15 & 25.15 & 3.59054 & $-30.40553$ & 311.08 & 0.31 \\
25928 & 2022-05-03 & 15.87 & 15.87 & 3.58898 & $-30.39814$ & 131.20 & 0.15 & 27563 & 2023-06-08 & 11.70 & 11.70 & 3.58818 & $-30.39758$ & 116.20 & 0.20 \\
25929 & 2022-08-26 & 27.72 & 27.72 & 3.59209 & $-30.39941$ & 54.21 & 0.10 & 27575 & 2022-12-02 & 19.65 & 19.65 & 3.58926 & $-30.40516$ & 302.21 & 0.30 \\
25930 & 2022-11-15 & 19.92 & 19.92 & 3.59058 & $-30.40554$ & 311.44 & 0.31 & 27678 & 2023-01-27 & 12.42 & 12.42 & 3.58590 & $-30.40361$ & 265.20 & 0.31 \\
25931 & 2022-04-23 & 14.59 & 14.59 & 3.58794 & $-30.39802$ & 138.49 & 0.19 & 27679 & 2023-01-28 & 11.93 & 11.93 & 3.58573 & $-30.40294$ & 266.70 & 0.29 \\
25932 & 2022-05-05 & 14.08 & 14.08 & 3.58871 & $-30.39813$ & 131.20 & 0.16 & 27680 & 2023-01-28 & 13.21 & 13.21 & 3.58583 & $-30.40284$ & 265.21 & 0.29 \\
25933 & 2023-08-15 & 23.88 & 23.39 & 3.59075 & $-30.39856$ & 80.19 & 0.11 & 27681 & 2023-01-29 & 9.78 & 9.78 & 3.58600 & $-30.40295$ & 265.21 & 0.28 \\
25934 & 2022-04-21 & 19.30 & 19.01 & 3.58762 & $-30.39801$ & 146.64 & 0.20 & 27739 & 2023-10-01 & 21.31 & 21.31 & 3.59103 & $-30.40353$ & 339.20 & 0.19 \\
25935 & 2023-08-20 & 24.08 & 24.08 & 3.59127 & $-30.39905$ & 70.19 & 0.09 & 27780 & 2023-08-21 & 14.90 & 14.90 & 3.59194 & $-30.39847$ & 67.33 & 0.13 \\
25936 & 2023-01-26 & 12.92 & 12.92 & 3.58594 & $-30.40327$ & 264.71 & 0.30 & 27856 & 2023-05-25 & 15.88 & 15.88 & 3.58714 & $-30.39737$ & 125.27 & 0.25 \\
25937 & 2022-11-27 & 30.78 & 30.77 & 3.59012 & $-30.40542$ & 304.81 & 0.31 & 27857 & 2023-05-26 & 12.92 & 12.92 & 3.58754 & $-30.39758$ & 124.54 & 0.22 \\
25938 & 2022-11-26 & 18.66 & 18.66 & 3.59005 & $-30.40580$ & 305.03 & 0.33 & 27896 & 2023-06-10 & 13.73 & 13.70 & 3.58797 & $-30.39769$ & 117.62 & 0.20 \\
25939 & 2023-01-28 & 14.32 & 14.32 & 3.58574 & $-30.40298$ & 264.71 & 0.29 & 27974 & 2023-08-05 & 28.71 & 28.71 & 3.59049 & $-30.39831$ & 85.11 & 0.12 \\
25940 & 2023-08-10 & 27.72 & 27.72 & 3.59085 & $-30.39842$ & 82.19 & 0.11 & 28370 & 2023-08-13 & 20.73 & 20.73 & 3.59061 & $-30.39831$ & 82.19 & 0.12 \\
25941 & 2023-06-09 & 32.65 & 32.65 & 3.58835 & $-30.39756$ & 114.20 & 0.20 & 28483 & 2023-08-19 & 20.22 & 20.22 & 3.59133 & $-30.39930$ & 70.20 & 0.07 \\
25942 & 2022-05-04 & 15.18 & 15.18 & 3.58835 & $-30.39786$ & 131.20 & 0.18 & 28872 & 2023-09-01 & 13.11 & 12.62 & 3.59222 & $-30.39933$ & 50.28 & 0.11 \\
25943 & 2023-08-31 & 16.69 & 16.69 & 3.59275 & $-30.39882$ & 50.92 & 0.15 & 28886 & 2023-09-10 & 9.96 & 9.96 & 3.59250 & $-30.40062$ & 32.25 & 0.10 \\
25944 & 2022-09-08 & 21.62 & 21.62 & 3.59195 & $-30.40133$ & 34.21 & 0.10 & 28887 & 2023-09-10 & 19.85 & 19.85 & 3.59229 & $-30.40067$ & 33.19 & 0.09 \\
25945 & 2022-09-27 & 17.03 & 17.03 & 3.59134 & $-30.40297$ & 358.21 & 0.17 & 28910 & 2023-10-25 & 25.75 & 25.75 & 3.59062 & $-30.40367$ & 326.01 & 0.20 \\
25946 & 2023-07-01 & 29.69 & 29.69 & 3.58859 & $-30.39766$ & 107.19 & 0.19 & 28920 & 2023-09-25 & 15.28 & 15.13 & 3.59259 & $-30.40219$ & 4.66 & 0.16 \\
25947 & 2023-09-24 & 14.90 & 14.90 & 3.59235 & $-30.40239$ & 6.17 & 0.16 & 28934 & 2023-09-29 & 19.83 & 19.83 & 3.59200 & $-30.40330$ & 345.20 & 0.20 \\
25948 & 2022-09-30 & 27.87 & 27.87 & 3.59074 & $-30.40309$ & 358.21 & 0.17 & 28951 & 2023-10-05 & 12.90 & 12.90 & 3.59098 & $-30.40315$ & 347.85 & 0.17 \\
25949 & 2023-10-27 & 20.82 & 20.82 & 3.59024 & $-30.40387$ & 324.39 & 0.21 & 28952 & 2023-10-08 & 13.79 & 13.79 & 3.59178 & $-30.40364$ & 339.20 & 0.21 \\
25950 & 2023-06-30 & 29.69 & 28.62 & 3.58843 & $-30.39753$ & 108.08 & 0.20 & 29207 & 2024-05-29 & 19.65 & 19.65 & 3.58833 & $-30.39753$ & 122.95 & 0.20 \\
25951 & 2022-11-18 & 28.71 & 28.71 & 3.58985 & $-30.40475$ & 317.21 & 0.27 & 29427 & 2024-05-29 & 18.66 & 18.66 & 3.58798 & $-30.39738$ & 122.74 & 0.22 \\
25952 & 2023-09-27 & 10.84 & 10.84 & 3.59215 & $-30.40265$ & 359.92 & 0.16 &  &  &  &  &  &  &  &  \\
\enddata
\tablecomments{Columns are: (1) Chandra observation ID; 
(2) observation start date; (3) total exposure time in ks; (4) cleaned exposure time in ks; (5)--(6) Right Ascension and Declination (ICRS) of the aim point; (7) roll angle in degrees; (8) angular separation of each observation's aim point from the exposure-weighted mean aim point of the 101 retained observations, (R.A., Decl.) = (3.59051$^\circ$, $-30.40031^\circ$), in arcminutes. }
\end{deluxetable*}
\endgroup

Data reduction was performed with \texttt{CIAO}~4.18 and \texttt{CALDB}~4.12.4 \citep{CIAO2006}. We first reprocessed each observation with \texttt{chandra\_repro}, enabling \texttt{check\_vf\_pha=yes} for the Very Faint mode data. For each reprocessed event file, we extracted a background light curve and applied \texttt{deflare} with a $3\sigma$ clipping criterion to identify intervals affected by background flares. The resulting good-time intervals were applied to generate cleaned event files. We then retained only events on the ACIS-I CCDs 0--3 with standard ASCA grades 0, 2, 3, 4, and 6, producing the final filtered event files used for imaging and source detection. The cumulative exposure before flare filtering is 2.102~Ms, while the final cleaned exposure is 2.094~Ms; thus only 7.9~ks, or approximately 0.4\% of the exposure, is removed by flare filtering. Individual cleaned exposures range from 8.1 to 61.2~ks. 
\par
To correct the astrometry of the Chandra observations, we performed the following steps:
\begin{enumerate}
    \item We first produced a provisional full-band (0.5--7~keV) mosaic by merging all observations with \texttt{merge\_obs}, without applying any astrometric correction. We then ran \texttt{wavdetect} on this first-pass merged image using a $\sqrt{2}$ sequence of wavelet scales (1, 1.414, 2, 2.828, 4, 5.656, and 8 pixels) and a significance threshold of $10^{-6}$. However, given the structured ICM emission, \texttt{wavdetect} may return more spurious sources superimposed on this ICM background (e.g., see Figure~10 in \citealt{Evans2024} for an extreme example). To ensure reliable detections, we further calculated the binomial no-source probability $P_B$ (\citealt{Weisskopf2007}) and only kept those above $5\sigma$ for this astrometric correction purpose.

    \item We used the resulting first-pass X-ray source list to construct the astrometric reference sample. We selected compact, bright X-ray sources with net counts $\geq 100$ and 90\% PSF sizes $\leq 5\arcsec$, and cross-matched them with the Legacy Survey DR10 optical catalog \citep{Dey2019} within $1\arcsec$. The optical positions of the matched counterparts were adopted as a reference list for \texttt{fine\_astro}, tying the X-ray astrometry to the Legacy Survey DR10 reference frame.

    \item We then ran \texttt{fine\_astro} on the clean event files for each observation. Source detection within \texttt{fine\_astro} used the same $\sqrt{2}$ sequence of wavelet scales and retained compact X-ray sources with net counts $>5$ and 50\% PSF sizes $<3\arcsec$ to match to the optical reference list. This step produced the astrometrically corrected event files used in the final merge.

    \item To verify the final astrometric accuracy, we created a post-correction merged full-band image and ran \texttt{wavdetect} again. We matched the resulting X-ray source positions to Gaia DR3 sources \citep{GaiaCollaboration2023} within $1\arcsec$ and found 17 matches. The mean and median separations are $0.22\arcsec$ and $0.11\arcsec$, respectively.
\end{enumerate}

\begin{figure}
  \centering
  \includegraphics[width= 3.3in]{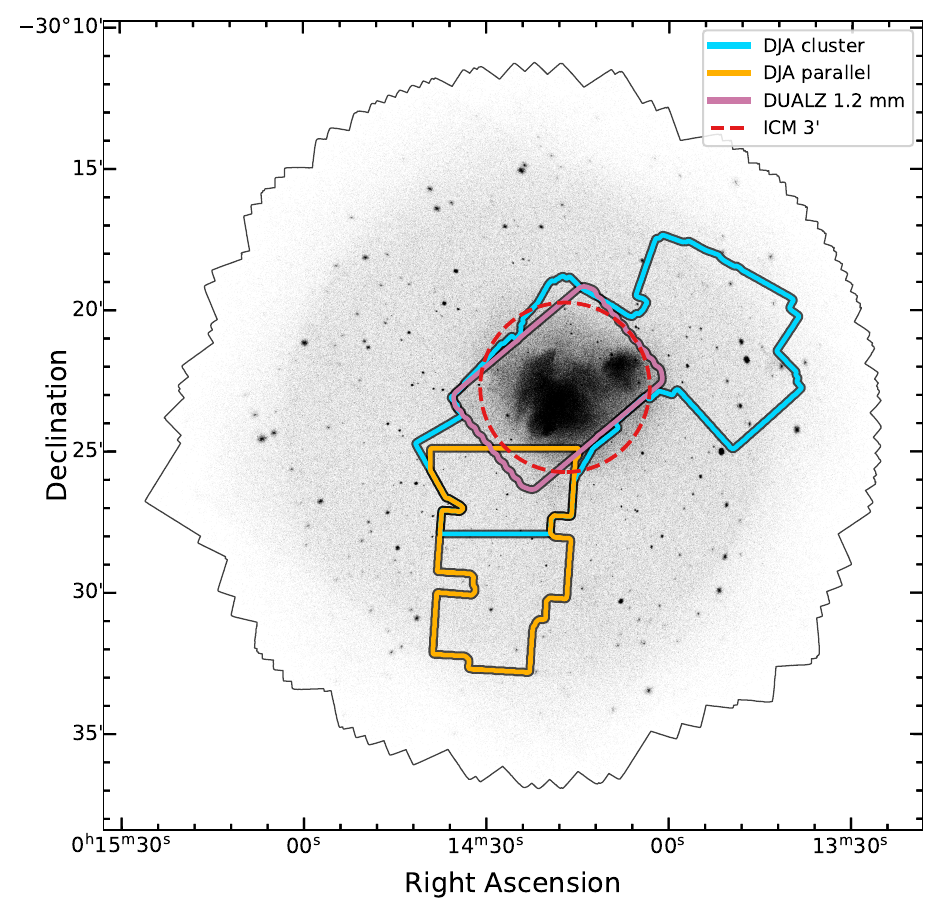}
   \caption{Merged Chandra full-band (0.5--7~keV) image of Abell~2744. The black contour marks the boundary of the full Chandra coverage with nonzero exposure, while the cyan and orange contours show the footprints of the DJA cluster and parallel fields, respectively. These footprints are derived from the weight maps of the combined JWST detection images and smoothed for display. The cluster detection image combines the F277W, F300M, F335M, F356W, F360M, F410M, F430M, F444W, F460M, and F480M bands, while the parallel detection image combines the F277W, F356W, F444W, and F480M bands. The magenta contour shows the DUALZ 1.2~mm footprint with a relative primary-beam response of at least 20\%. The red dashed circle indicates the central $3\arcmin$ ICM region.} \label{fig:ximage}
 \end{figure}

\begin{figure}
  \centering
  \includegraphics[width= 3.3in]{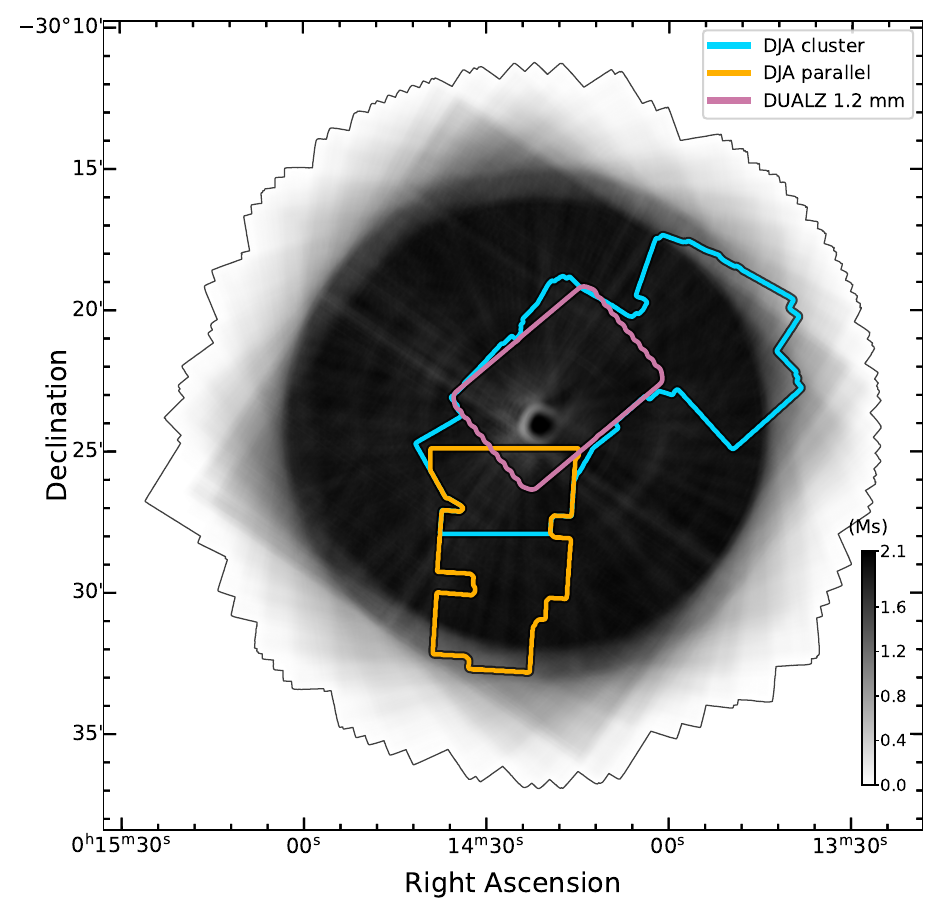}
    \caption{Full-band (0.5--7~keV) exposure map of the Abell~2744 field, shown with a linear grayscale in units of Ms. The black contour marks the boundary of the full Chandra coverage with nonzero exposure, while the cyan and orange contours show the DJA cluster and parallel JWST footprints, respectively. These footprints are derived from the same combined detection-image weight maps, using the filters listed in Figure~\ref{fig:ximage}. The magenta contour shows the DUALZ 1.2~mm footprint defined in Figure~\ref{fig:ximage}. The radial features and the central ring of reduced exposure arise primarily from the ACIS-I CCD gaps projected across observations with different aim points and roll angles.} \label{fig:xexpmap}
 \end{figure}

Using astrometrically corrected event files, we created merged event files, images, exposure maps, and PSF maps in the full (0.5--7 keV), soft (0.5--2 keV), and hard (2--7 keV) bands, using \texttt{merge\_obs}. The corresponding PSF maps were computed for a 90\% encircled energy fraction. 
We define the survey footprint as the union of the regions with exposure greater than 10\% of the maximum value in each band.
This union covers approximately $369~\mathrm{arcmin^2}$.\footnote{The areas satisfying this exposure criterion are approximately $364$, $341$, and $369~\mathrm{arcmin^2}$ in the full, soft, and hard bands, respectively. This threshold corresponds to the exposure cut used by \texttt{wavdetect} for source detection.}
The exposure-weighted mean aim point of the merged data set is (R.A., Decl.) = (3.59051$^\circ$, $-30.40031^\circ$).
The full-band raw image and the corresponding exposure map are shown in Figures~\ref{fig:ximage} and \ref{fig:xexpmap}. For comparison, we also show the footprints of the combined JWST detection images and the 1.2~mm mosaic from the Deep UNCOVER--ALMA Legacy High-$z$ (DUALZ) Survey \citep{Fujimoto2025DUALZ}. The JWST footprints are derived from products in the DAWN JWST Archive\footnote{\url{https://dawn-cph.github.io/dja/}} (DJA; \citealt{Valentino2023DJA}), which has systematically compiled and reduced the archival HST and JWST data. The DJA divides Abell~2744 into the ``cluster'' and ``parallel'' subfields, as labeled in the figures. The DUALZ footprint is defined by a relative primary-beam response of at least 20\% and covers approximately $4\arcmin\times6\arcmin$ around the cluster. We also mark the region where the ICM emission is strongest using a circle with a radius of $3\arcmin$, centered at (R.A., Decl.) = (3.57114$^\circ$, $-30.37911^\circ$). \par

\begin{figure*}
  \centering
  \includegraphics[width= \textwidth]{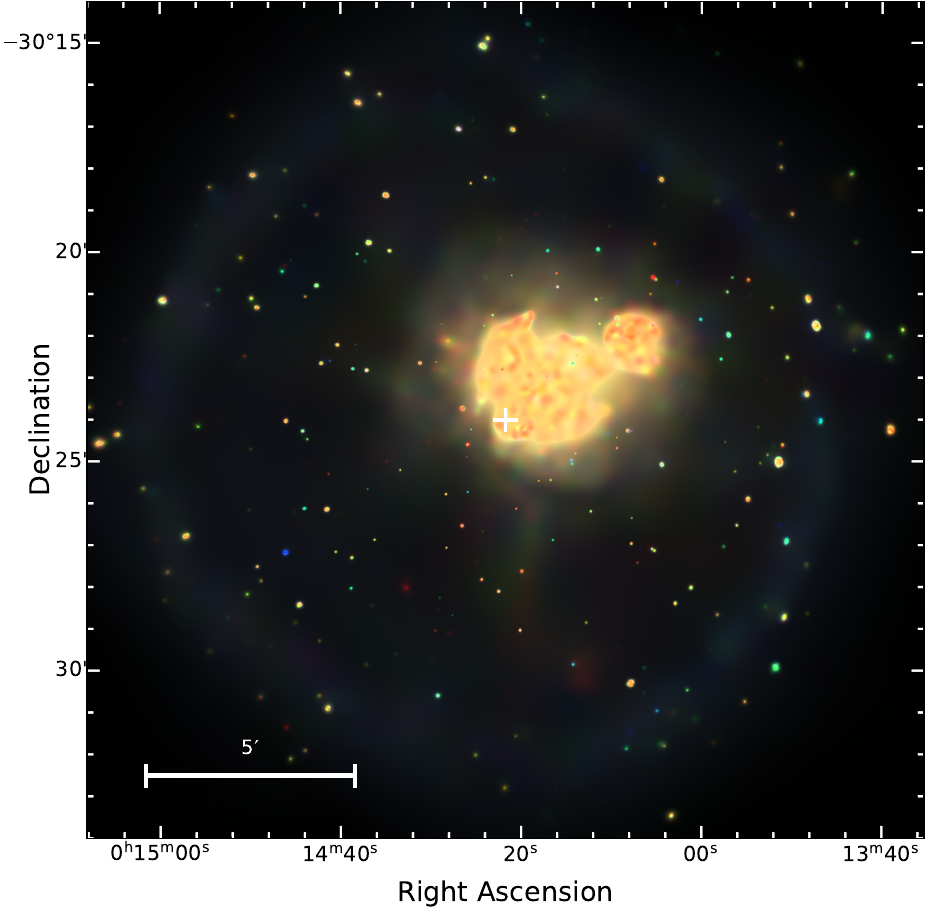}
\caption{Adaptively smoothed false-color X-ray image of the central $20\arcmin \times 20\arcmin$ region of Abell~2744. Red, green, and blue represent the 0.5--2, 2--4, and 4--7~keV bands, respectively. The extended emission near the field center is dominated by the intracluster medium, while the white cross marks the exposure-weighted mean aim point of the 101 Chandra observations.} \label{fig:x3color}
 \end{figure*}

We also generated smoothed images in the 0.5--2 keV, 2--4 keV, and 4--7 keV bands following the method in Section 3.1 of \citet{Luo2017} using the CIAO tool \texttt{csmooth}. These smoothed images were then combined to produce a color composite using the \texttt{Astropy} \citep{Astropy2022} \texttt{make\_lupton\_rgb} function. We show the central $20 \arcmin \times 20 \arcmin$ region in Figure~\ref{fig:x3color}.  Although many of the X-ray sources are clearly visible in the adaptively smoothed images, our source searching was performed on the raw images. Figure~\ref{fig:x3color} shows bright ICM emission near the field center, extending toward the upper right, reflecting the complex merging picture of Abell~2744 itself, as discussed in \citet{Chadayammuri2024}. Note that the strong lensing peak (e.g., \citealt{Bergamini2023, Furtak2023}) is offset from the ICM emission peak, leaving valuable regions that are highly magnified but relatively less affected by the ICM background.

\subsection{Candidate Detection and Band Merging}
\label{sec:candidate_detection}
We first constructed band-specific candidate-source lists by running
\texttt{wavdetect} independently on the merged full (0.5--7~keV), soft (0.5--2~keV), and hard (2--7~keV) images
\citep{Freeman2002ApJS..138..185F}. We used a $\sqrt{2}$ sequence of wavelet scales (1, 1.414, 2, 2.828, 4, 5.656, 8, 11.314, and 16 pixels) and a false-positive probability threshold of $10^{-5}$, following the source-detection procedure adopted for the 7 Ms CDF-S survey \citep{Luo2017}. 
The \texttt{wavdetect} run made
use of a merged PSF map, created by choosing the minimum
PSF size at each pixel among all the PSF maps of
individual observations; such a process is optimized for point-source
detection.\footnote{See \url{https://cxc.harvard.edu/ciao/threads/wavdetect_merged/index.html\#min}}
By default,
\texttt{wavdetect} excludes pixels with exposure values below 10\% of the
maximum exposure in the corresponding band.\footnote{See \url{https://cxc.cfa.harvard.edu/ciao/ahelp/wavdetect.html\#plist.expthresh}}
Consequently, some low-exposure regions near the field boundaries were omitted.
  The initial full-, soft-, and hard-band candidate
lists contain 360, 253, and 295 sources, respectively.
We then merged the three filtered band-specific catalogs. Candidates were
matched using a radius of $2.5\arcsec$ within $6\arcmin$ of the aim point
and $4\arcsec$ at larger off-axis angles. We additionally inspected all
candidates more than $8\arcmin$ from the mean aim point and removed three that
were judged to be duplicate detections of neighboring candidates separated by
approximately $4$--$7\arcsec$.  For a source detected in multiple bands, we
adopted the full-band position when available, followed by the soft- and hard-band positions.  The merged source list used in the subsequent analysis
contains 410 unique candidates.

\section{PSF Fitting and Background Modeling}

\label{sec:psf_modeling}

For the 410 unique candidates, we evaluated the significance of each candidate using Poisson statistics, based on the locally estimated background and the Chandra PSF. Conventional aperture-based methods estimate the detection significance from the observed counts relative to the expected background within a predefined source region. Our approach additionally incorporates the local Chandra PSF into the source model, so that the detection significance depends not only on the excess counts but also on whether their spatial distribution is consistent with that expected for a point source.
The three bands were fitted independently. We evaluated the performance of this approach and calibrated the adopted selection threshold using the simulations presented in Section~\ref{sec:simulations}.

\subsection{Stacked PSF Construction}
\label{sec:stacked_psf}

For each candidate and energy band, we extracted a
$90\arcsec\times90\arcsec$ cutout from the merged event files and exposure maps.
We constructed maps of the 90\% encircled energy radius of the Chandra PSF (denoted as $R_{90}$)
separately for the full, soft, and hard bands. For each band, the PSF maps from
the individual observations were combined using the exposure times as
weights. These exposure-weighted PSF maps differ from those used for the
\texttt{wavdetect} candidate search in Section~\ref{sec:candidate_detection},
where the smallest PSF size among the contributing observations was adopted at each pixel
to optimize point source detection. Because the merged images
contain photons contributed by all observations, the exposure-weighted PSF maps
provide a more representative estimate of the effective PSF size in the merged
data. At each candidate position, we adopted a
circular source fitting aperture centered on the 
\texttt{wavdetect} position with a radius of $R_{90}$, which slightly varies across different bands.

The size, shape, and orientation of the Chandra PSF depend on the position of
a source relative to the aim point and the
roll angle in each observation. Because the 101 observations have different pointings,
the same sky source is sampled by different PSFs across the data set.  For each observation with nonzero exposure
at the candidate position, we generated a two-dimensional PSF model using the
\texttt{CIAO} \texttt{simulate\_psf} tool and \texttt{MARX}
\citep{Davis2012SPIE.8443E..1AD}, based on the astrometrically corrected event
file and the corresponding aspect solution. We adopted nominal monochromatic energies
of 2.3, 1.4, and 3.8 keV for the full, soft, and hard bands, respectively. The
individual PSFs were reprojected onto the common merged image grid, normalized
within an $85\times85$ pixel region, and combined using the local exposure
contribution of each observation as the weight. The resulting stacked PSF was
then renormalized and adopted as the PSF template for that candidate and energy
band.

\subsection{Local Background Modeling and Test of Polynomial Order}
\label{sec:local_psf_background}

The background in Abell~2744 is strongly affected by the bright and spatially
varying ICM emission, particularly in the central region. A single background
value estimated from a surrounding annulus may therefore not adequately
describe the local background within the source region. We instead fitted a
smooth model to the local count distribution in each cutout. This model provides
the expected background counts at each pixel across the source fitting region,
allowing the spatial variation of the ICM background to be incorporated directly
into the source analysis.

Similar strategies are used in other X-ray source detection pipelines, where
sources are masked and a smooth background is constructed before source
detection or PSF fitting \citep[e.g.,][]{Watson2009,Brunner2022,Evans2024}. Following
this approach, we fitted the local background separately for each of the 410
candidates in each of the three energy bands. Each fit used the corresponding
$90\arcsec\times90\arcsec$ count image and exposure map introduced in
Section~\ref{sec:stacked_psf}.

Within each cutout, we masked the target candidate and every other entry in the
\texttt{wavdetect} list whose mask overlapped the cutout. The mask radius
was evaluated separately for each candidate and energy band as
\begin{equation}
R_{\rm mask}=\max(2R_{90},3\arcsec).
\label{eq:background_mask}
\end{equation}
The region used to fit the background therefore consisted of all valid pixels
within the cutout that remained outside these masks. 

{For the remaining pixels, we modeled the logarithm of the local
background surface brightness with a two-dimensional polynomial.
For the adopted fourth-order model, the expected background counts
in pixel $i$ are
\begin{equation}
B_i = \widetilde{E}_i
\exp\left[
\sum_{m=0}^{4}\sum_{n=0}^{4-m}
a_{mn}u_i^m v_i^n
\right],
\label{eq:background_poly4}
\end{equation}
where $u_i$ and $v_i$ are scaled image coordinates centered on the
initial \texttt{wavdetect} position, $\widetilde{E}_i$ is the
normalized exposure map value, and $a_{mn}$ are the fitted
coefficients. The exponential ensures a positive background model,
while the exposure map accounts for variations in effective exposure.
We determined the coefficients by maximizing the Poisson likelihood
over the unmasked background pixels and kept the resulting background
model fixed during subsequent PSF fitting.}

We tested polynomial orders $p=2$--6 for each candidate in each of the three
energy bands, using the same count images, exposure maps, source masks, and
$R_{90}$ apertures. For each order, we compared the expected background counts
within $R_{90}$ with those from the fourth-order model. We focused on the 104 candidates within $3\arcmin$ of the adopted ICM center,
where the background is more structured. Based on the model selection by the Bayesian information criterion, $p=4$ is generally favoured, above which the statistical improvements saturate. Nevertheless, the expected background level within the central source aperture generally remains quite robust against the chosen polynomial order. The median absolute
fractional differences for orders two, three, five, and six were 3.4\%, 3.3\%,
0.6\%, and 2.7\%, respectively. The corresponding values at the 95th percentile
were 11.9\%, 11.9\%, 3.1\%, and 8.7\%. 

We also examined how the source selection changed with polynomial order by repeating the PSF fits in Section~\ref{sec:psf_fitting} for orders $p=2$--6, including the joint fits for overlapping candidates. Among the 104 candidates within $3\arcmin$ of the adopted ICM center, orders two and three added four and five sources, respectively, relative to order four under the same $\Delta C\geq25$ selection criterion (see Section~\ref{sec:psf_fitting}). Order five left the selected sample unchanged, while order six removed one source. Across all tested orders, six candidates (5.8\%) changed their selection status, and the remaining 98 candidates were consistently retained or rejected. 

The purpose of these tests was to evaluate how sensitive the inferred local background is to the choice of polynomial order, rather than to select an optimal order for each source. Since $p=4$ is generally statistically favoured in the central ICM region, we therefore adopted a fourth-order polynomial for all candidates and energy bands. This order is sufficiently flexible to capture the smooth spatial variation of the ICM background while remaining intermediate in complexity among the tested models. Using the same polynomial order also ensures a uniform background treatment throughout the catalog.

\subsection{PSF Fitting, Treatment of Overlapping Candidates, and Source Selection}
\label{sec:psf_fitting}

For each candidate and energy band, we defined the fitting region as a circular
aperture of radius $R_{90}$ centered on the \texttt{wavdetect} position. Within
the valid pixels of this region, we modeled the expected counts in pixel $i$ of
energy band $b$ as
\begin{equation}
\mu_{ib} =
B_{ib} +
A_b P_{ib}(\Delta x_b,\Delta y_b),
\label{eq:single_psf_model}
\end{equation}
where $\mu_{ib}$ is the model-predicted number of counts, $B_{ib}$ is the
expected local background, $P_{ib}$ is the stacked PSF shifted by
$(\Delta x_b,\Delta y_b)$, and $A_b\geq0$ is the normalization. The free
parameters were the two positional offsets and the PSF normalization. The PSF
centroid was allowed to shift from the \texttt{wavdetect} position but was
constrained to remain within the same $R_{90}$ aperture. The three bands were
fitted separately using the corresponding background model, PSF, and $R_{90}$
aperture for each band.

The number of counts in each image pixel was assumed to follow a Poisson
distribution. We therefore fitted the model by maximizing the joint Poisson
likelihood of the observed pixel counts. Equivalently, we minimized the
Poisson deviance, $D$, defined as
\begin{equation}
D =
2\sum_i
\left[
n_i\ln\left(\frac{n_i}{\mu_i}\right)-n_i+\mu_i
\right],
\label{eq:poisson_deviance}
\end{equation}
where $n_i$ and $\mu_i$ are the observed counts and the model-predicted counts
in pixel $i$, respectively, and the logarithmic term is defined to be zero
when $n_i=0$. For a fixed data set, the Poisson deviance differs from the Cash
statistic only by a data-dependent constant; therefore, differences in $D$ are
equivalent to differences in the Cash statistic
\citep{Cash1979ApJ...228..939C}.

For an isolated candidate, we defined the improvement over the background-only
model as
\begin{equation}
\Delta C_b =
D_{0,b}-D_{1,b}
=
2\left[
\ln\mathcal{L}_{1,b}
-
\ln\mathcal{L}_{0,b}
\right],
\label{eq:delta_c}
\end{equation}
where $D_{0,b}$ and $D_{1,b}$ are the deviances of the background-only and
background-plus-PSF models, respectively. A larger $\Delta C_b$ therefore
indicates a greater improvement after adding a PSF component.

Candidates with overlapping fitting apertures were modeled jointly because
the counts in the shared pixels can contain contributions from more than one
source. We linked two candidates when their $R_{90}$ apertures intersected in
at least one band, i.e., when
$d_{jk}<R_{90,jb}+R_{90,kb}$. Candidates connected through a chain of such
intersections were placed in the same overlap group. Our candidate list
contains 12 overlap groups involving 25 candidates: 11 pairs and one group of three.
The final catalog selection for these candidates was based on the joint fits
rather than on their individual fits.

For each overlap group and energy band, we fitted the PSF components
simultaneously over the union of the member $R_{90}$ apertures. A model with
$K$ components has the form
\begin{equation}
\mu_{ib} =
B_{ib} +
\sum_{k=1}^{K}
A_{kb} P_{ikb}(\Delta x_{kb},\Delta y_{kb}),
\qquad A_{kb}\geq0,
\label{eq:joint_psf_model}
\end{equation}
where each component has its own stacked PSF, position, and nonnegative
normalization. The centroid of each component was constrained to remain within
$R_{90}$ of the corresponding \texttt{wavdetect} position.

For a group of $N$ candidates, we considered models containing between zero
and $N$ source components. For each value of $K$, we fitted every possible
subset of $K$ candidate components and retained the model with the lowest
Poisson deviance, $D^{*}_{K,b}$. We defined the incremental improvement from
the best $(K-1)$-component model to the best $K$-component model as
\begin{equation}
\Delta C_{K,b} =
D_{K-1,b}^{*}-D_{K,b}^{*}.
\label{eq:multiplicity_delta_c}
\end{equation}
We increased the allowed source multiplicity sequentially and required
$\Delta C_K=\max_b(\Delta C_{K,b})\geq25$ at each step, where the maximum was
taken over the full, soft, and hard bands. For example, for an overlapping pair, a
one-component model was accepted if its improvement over the background-only
model satisfied this threshold. A two-component model was accepted only if
its improvement over the best one-component model also satisfied the
threshold. An overlapping pair could therefore contribute zero, one, or two
sources to the final catalog. The three-candidate group was treated
analogously.

For an isolated candidate, we used $\Delta C=\max_b(\Delta C_b)$ and retained
the candidate as a source when $\Delta C\geq25$. For each retained isolated
source, we adopted the fitted position from the band with the largest
$\Delta C_b$. In both the simulations and the observed counterpart sample,
these positions were, on average, closer to the reference positions than the
corresponding \texttt{wavdetect} positions (see
Section~\ref{sec:simulations} for details). The adopted threshold of
$\Delta C=25$ was calibrated using the simulations described in
Section~\ref{sec:simulations}.

\begin{figure}[t]
\centering
\includegraphics[width=\columnwidth]
{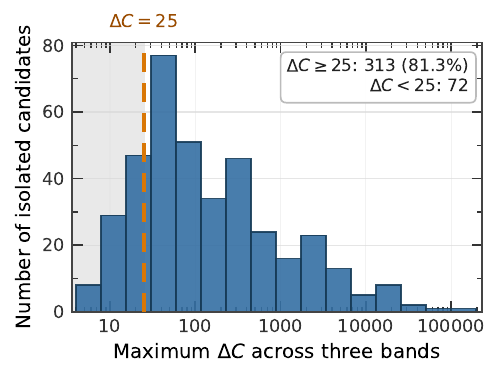}
\caption{
Distribution of $\Delta C=\max_b(\Delta C_b)$ for the 385 isolated preliminary
candidates in the observed Abell~2744 data. For each candidate, the plotted
value is the maximum among the independently fitted full, soft, and hard
bands. The 25 candidates belonging to overlap groups are not plotted because
their final selection was based on joint fits. The horizontal axis and
histogram bins are logarithmic. The orange dashed line marks the adopted
selection threshold of $\Delta C=25$, and the gray shaded region indicates
candidates below this threshold. A total of 313 isolated candidates (81.3\%)
satisfy $\Delta C\geq25$, while 72 fall below the threshold.
}
\label{fig:observed_delta_c_distribution}
\end{figure}

Of the 385 isolated candidates, 313 satisfied $\Delta C\geq25$. Their distribution of $\Delta C=\max_b(\Delta C_b)$ is shown in Figure~\ref{fig:observed_delta_c_distribution}.
The 12 overlap groups contain 25 preliminary candidates; the
selected model multiplicities across these groups summed to 14, so the overlap
groups contributed 14 sources to the final catalog. Combining the 313 isolated
sources and the 14 jointly fitted sources yields a final catalog of 327
sources. The
adopted PSF-fit positions of these sources were subsequently used
for the catalog photometry in Section~\ref{sec:catalog}.
 
\begin{figure*}[p]
\centering
\includegraphics[
    width=1.5\textwidth,
    height=0.9\textheight,
    keepaspectratio
]{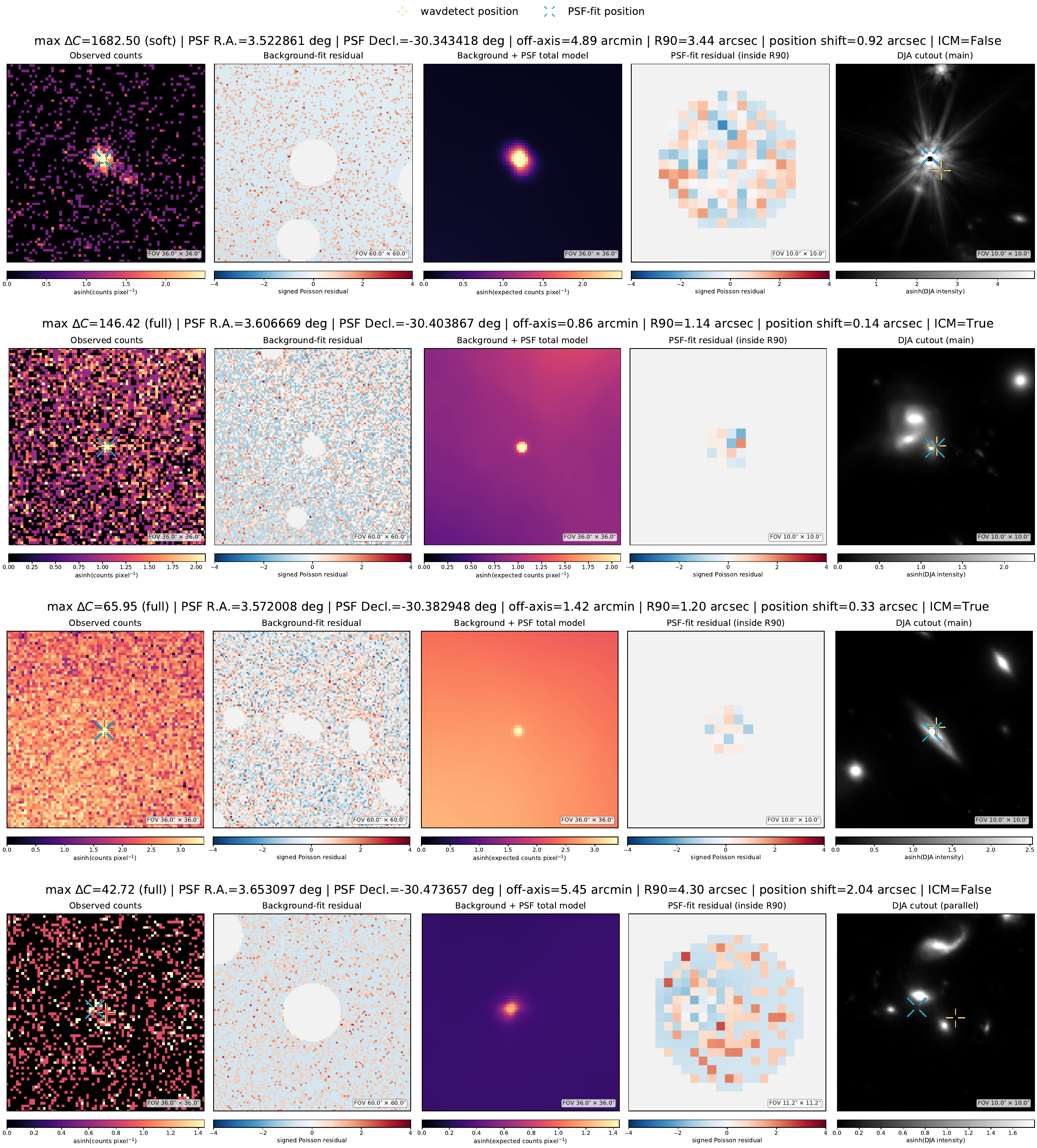}
\caption{
Diagnostics for the PSF fits of four isolated candidates, ordered from top to
bottom by decreasing $\Delta C$. For each candidate, we show the band with the
largest $\Delta C$ among the independently fitted full, soft, and hard bands.
From left to right, the panels show the observed counts, the signed Poisson
deviance residual from the fourth-order polynomial background fit, the
best-fit background-plus-PSF model, the signed Poisson deviance residual within
the band-specific $R_{90}$ fitting aperture, and the DJA detection-image
cutout. The DJA cutouts are taken from the combined JWST detection images
of the cluster and parallel fields, using the filters listed in
Figure~\ref{fig:ximage}. The panels have different angular sizes, as labeled.
White regions in the background residual were excluded from the
background fit. Blue and red residuals indicate fewer and more observed counts
than predicted, respectively. The yellow plus and cyan cross mark the initial
\texttt{wavdetect} and best-fit PSF positions, respectively. Here,
$\Delta C=D_{\rm bkg}-D_{\rm bkg+PSF}$, so that a larger value indicates a
stronger improvement after adding the PSF component. These images are shown
only as visual diagnostics. The sequence continues
in Figure~\ref{fig:psf_fit_diagnostics_2}.
}
\label{fig:psf_fit_diagnostics_1}
\end{figure*}

\begin{figure*}[p]
\centering
\includegraphics[
    width=1.2\textwidth,
    height=0.9\textheight,
    keepaspectratio
]{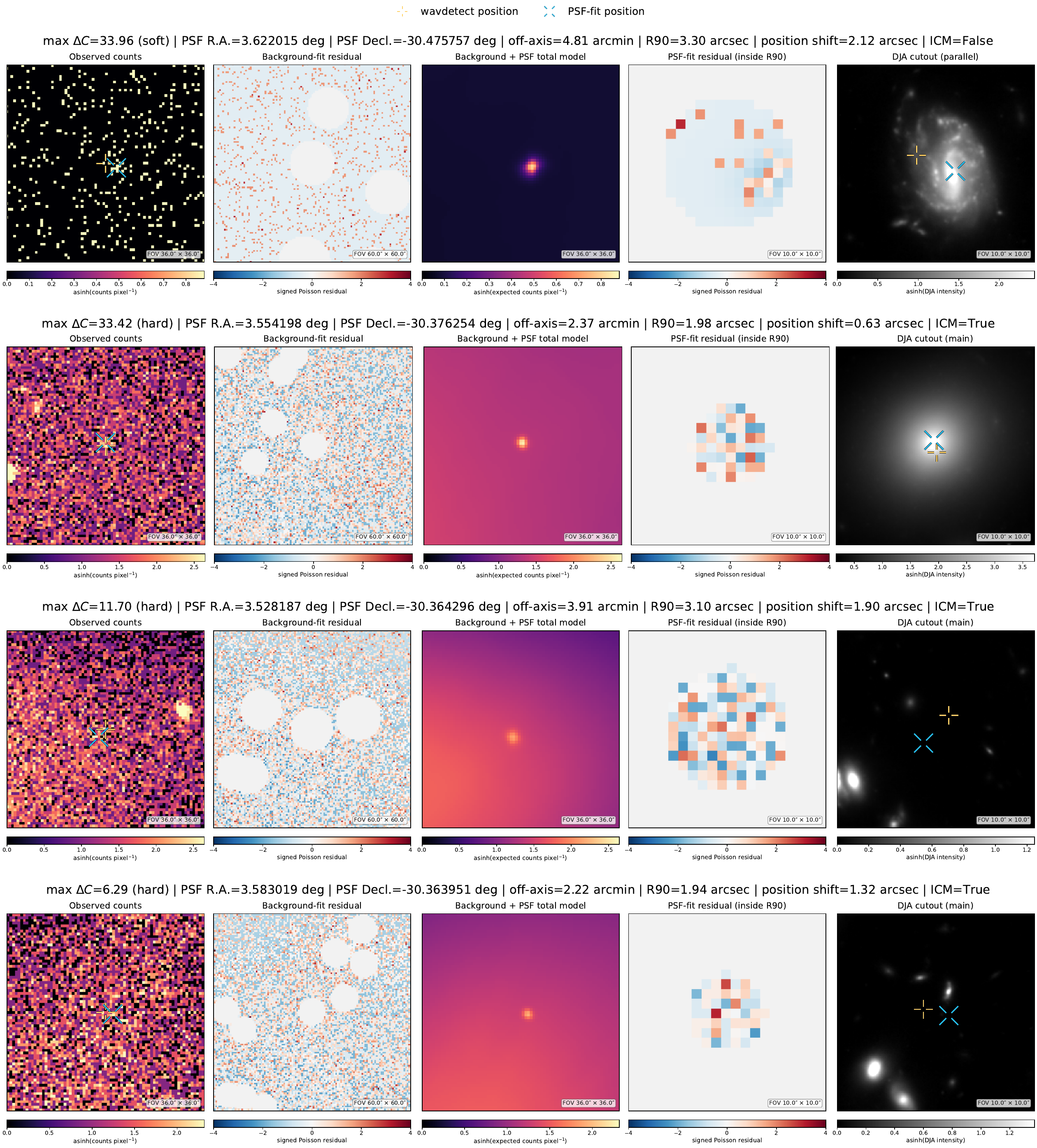}
\caption{
Same as Figure~\ref{fig:psf_fit_diagnostics_1}, but for four additional
candidates with smaller $\Delta C$, ordered from top to bottom by decreasing
$\Delta C$. The bottom two candidates have $\Delta C<25$ and are not
included in the main catalog.
}
\label{fig:psf_fit_diagnostics_2}
\end{figure*}

For illustration and diagnostic purposes, we calculated the signed Poisson deviance residual in
each fitted pixel,
\begin{equation}
r_i =
{\rm sgn}(n_i-\mu_i)
\left\{
2\left[
n_i\ln\left(\frac{n_i}{\mu_i}\right)-n_i+\mu_i
\right]
\right\}^{1/2}.
\label{eq:signed_deviance_residual}
\end{equation}
Figures~\ref{fig:psf_fit_diagnostics_1} and
\ref{fig:psf_fit_diagnostics_2} present PSF-fit diagnostics for eight isolated
candidates spanning a broad range of $\Delta C$. Each row presents a candidate
in the band with the largest $\Delta C$ among the independently fitted full,
soft, and hard bands. From left to right, the panels show the observed counts,
the signed Poisson deviance residual from the fourth-order polynomial
background fit, the best-fit background-plus-PSF model, the signed Poisson
deviance residual from this model within the $R_{90}$ fitting aperture, and the
corresponding DJA detection-image cutout. The candidates are ordered from top
to bottom by decreasing $\Delta C$, with the four higher-$\Delta C$ candidates
shown in Figure~\ref{fig:psf_fit_diagnostics_1} and the four lower-$\Delta C$
candidates shown in Figure~\ref{fig:psf_fit_diagnostics_2}. These images are
presented only as visual diagnostics and were not used as an additional
source-selection criterion.

\begin{figure}[t]
    \centering
    \includegraphics[width=\columnwidth]{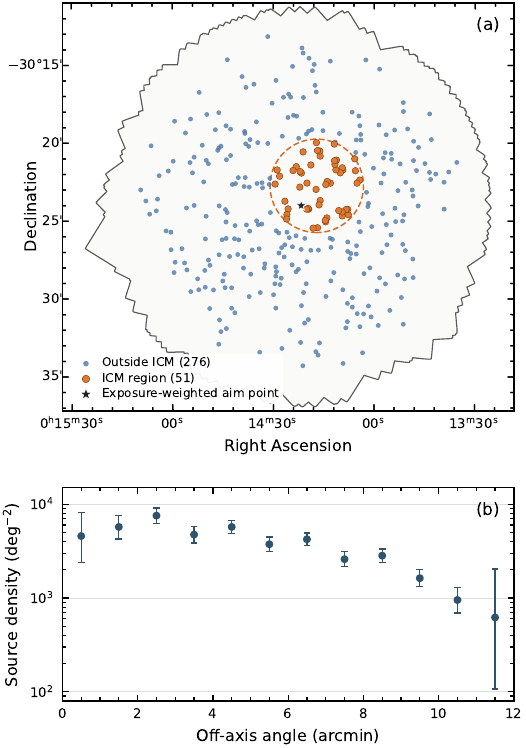}
    \caption{Spatial distribution and observed source density of the 327
    main-catalog sources. (a) Blue circles show sources outside the ICM region,
    while orange filled circles show sources within it. The orange
    dashed circle marks the adopted ICM region with a radius of $3\arcmin$, and
    the black star marks the exposure-weighted mean aim point. The gray area and
    its boundary show the full Chandra coverage with nonzero exposure.
    (b) Observed source density in $1\arcmin$ off-axis bins. Error bars show
    $1\sigma$ uncertainties. No correction for detection incompleteness or
    Eddington bias has been applied.}
    \label{fig:source_spatial_density}
\end{figure}

Figure~\ref{fig:source_spatial_density}(a) shows the spatial distribution of the retained 327 sources. Figure~\ref{fig:source_spatial_density}(b) displays the observed source sky density as a function of the off-axis angle. These apparent source densities have not been corrected for detection incompleteness or Eddington bias. The density reaches $7.6\times10^{3}~\mathrm{deg}^{-2}$ in the
$2$--$3\arcmin$ bin and decreases to below
$1.7\times10^{3}~\mathrm{deg}^{-2}$ beyond $9\arcmin$. The overall decline toward large off-axis angles mainly reflects the decreasing
sensitivity and broadening of the Chandra PSF.

\section{Simulations and Catalog Calibration}
\label{sec:simulations}

Following the approach commonly adopted in previous X-ray surveys \citep[e.g.,][]{Cappelluti2007,Cappelluti2009,Puccetti2009ApJS..185..586P,Xue2011,Xue2016,Luo2017}, we performed simulations of our two-stage source detection procedure, which consists of initial candidate detection with \texttt{wavdetect} followed by PSF fitting.  We used these simulations for two related purposes: to determine the $\Delta C$ threshold adopted for the main catalog based on the resulting completeness and reliability, and to test whether PSF fitting improves the source positions relative to the initial \texttt{wavdetect} centroids.

\subsection{Generation of Background Maps and Simulated Data}
\label{sec:simulation_construction}

First, we constructed background maps from the corresponding observed images in the full, soft, and hard bands. All 410 preliminary candidates initially from \texttt{wavdetect} were masked. Inside each source mask, we replaced the observed counts with the
expectation from the fourth-order polynomial model fitted in a
$90\arcsec\times90\arcsec$ cutout, as described in
Section~\ref{sec:local_psf_background}. As a consistency check, we also constructed a set of background maps with \texttt{dmfilth}, following the approach adopted in previous Chandra Deep-Field surveys \citep[e.g.,][]{Xue2011,Luo2017}. We used \texttt{dmfilth} to fill the masked regions with random counts drawn from the local background. The polynomial-filled maps and maps constructed with \texttt{dmfilth} are similar. Their total filled counts differ by
only 0.3\%, 0.8\%, and 0.5\% in the full, soft, and hard bands, respectively. Therefore, we adopted the polynomial-filled maps for internal
consistency with the local background model.

Second, we produced a mock catalog covering the survey footprint. The mock catalog extended well below the detection limit of the Abell~2744 field, down to a soft-band flux of $8\times10^{-18}$ erg~cm$^{-2}$~s$^{-1}$. Each simulated AGN and galaxy was assigned a soft-band flux randomly drawn from the soft-band logN--logS relations of the AGN and galaxy population models given in Equation 3 of \citet{Luo2017}. The soft-band fluxes of the simulated AGNs and galaxies were converted to full-band fluxes assuming power-law spectra with $\Gamma=1.4$ and 2.0, respectively.
The true $\log N$--$\log S$ relation in the Abell~2744 field may differ from the blank-field model of \citet{Luo2017} because of cluster members and gravitational lensing, but such differences are not expected to significantly affect the catalog calibration.

Third, for each mock catalog, we simulated event files for the 101 ACIS-I observations. Each simulated observation adopted the same aim point, roll angle, exposure time, and aspect solution as the corresponding real observation after astrometric correction (see Table~\ref{tab:chandra_obs}). We used \texttt{MARX} \citep{Davis2012SPIE.8443E..1AD} to convert the input source fluxes into Poisson realizations of dithered photons and to simulate their detection by ACIS-I. These event files therefore contain only events associated with the simulated point sources. The 101 simulated event files were then reprojected and merged using the \texttt{CIAO} tool \texttt{merge\_obs}. We added the resulting simulated source images to the corresponding background maps in the full, soft, and hard bands, respectively. Following \citet{Xue2011} and \citet{Luo2017}, we removed 0.5\% of the background counts before source injection to avoid double counting the unresolved input source population.
Thus, the resulting simulation closely mirrors the real Abell 2744 observation.

Finally, we ran \texttt{wavdetect} on the simulated combined raw images with a false-positive probability threshold of $10^{-5}$ to detect sources following the procedure described in Section~\ref{sec:candidate_detection}. We then performed the background modeling and PSF fitting described in Section~\ref{sec:psf_modeling} for all detected sources, obtaining a fitted position and the corresponding $\Delta C$ value in each band.

The above simulations were repeated ten times using independent mock source catalogs to reduce statistical uncertainties of our $\Delta C$ calibration.

\subsection{Completeness and Reliability}
\label{sec:simulation_completeness_reliability}

By comparing the detected sources with the input catalog, we assessed the completeness and reliability of the simulations. We matched detected sources to input sources separately in each band using the positions measured from PSF fitting.
To estimate the probability of chance associations, we repeated the matching after shifting the fitted positions by $60\arcsec$--$90\arcsec$ in random directions, retaining shifted positions within the survey footprint. The resulting chance-association rates were 6.2\%, 3.1\%, and 5.3\% in the full, soft, and hard bands, respectively. These values are sufficiently small compared with our expected reliability (see below), indicating that the high reliabilities below are genuine rather than driven by chance associations. At a given $\Delta C$ threshold, a detected source was classified as spurious if it passed the threshold but had no input counterpart satisfying the above matching criteria.

\begin{figure*}[t]
\centering
\includegraphics[width=\textwidth]
{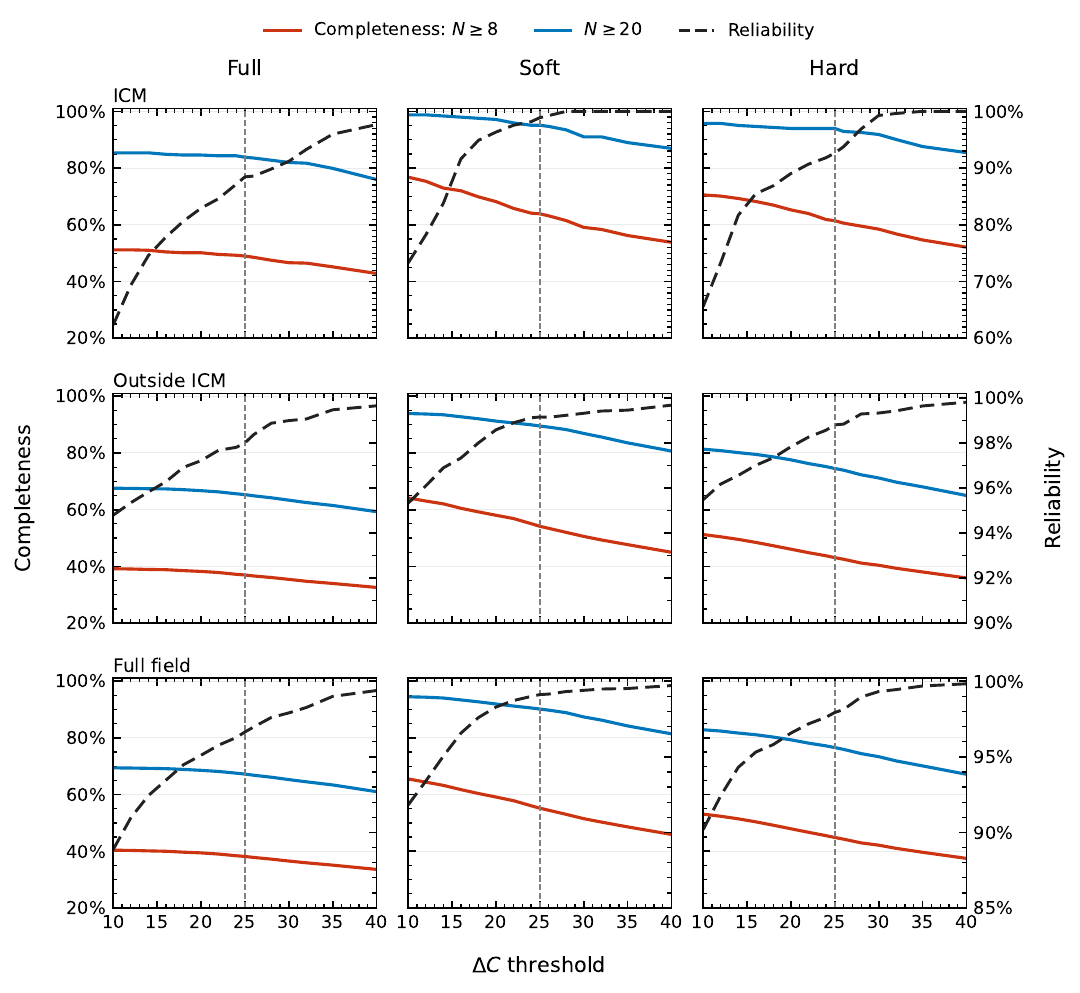}
\caption{
Completeness and reliability as functions of the $\Delta C$
threshold. Rows show the central $3\arcmin$ ICM
region, the area outside the ICM region, and the full $369~\mathrm{arcmin^2}$ survey footprint; columns show the
full, soft, and hard bands. Red and blue solid curves show completeness
for input source counts of $N_{\rm input}\geq8$ and 20, respectively,
on the left axis. Black dashed curves show reliability for all selected
detections, with no additional source-count cut, on the right axis. Note that, for visualization purposes, the plotted reliability axis ranges are different among the top, middle, and bottom panels.
The vertical gray dashed line marks
the adopted $\Delta C=25$ threshold.
Here, $N_{\rm input}$ denotes the input counts of each simulated source in
the corresponding band. Source numbers are pooled over ten simulations
before calculating completeness and reliability.
}
\label{fig:simulation_delta_c}
\end{figure*}

For a given input-count limit, completeness is defined as the fraction of input sources above this limit that are recovered.  A matched source is counted as recovered regardless of its fitted counts. Reliability is defined as $1-N_{\rm spurious}/N_{\rm selected}$, where $N_{\rm spurious}$ and $N_{\rm selected}$ are the numbers of spurious sources and all selected detections, respectively, at a given $\Delta C$ threshold, with no additional source-count cut. We sum the source numbers over the ten simulations before calculating completeness and reliability. Figure~\ref{fig:simulation_delta_c} presents the resulting curves as functions of $\Delta C$ within the ICM region, outside the ICM region, and over the entire field in the three bands. Completeness is evaluated for input sources with at least 20 and 8 counts, while reliability is evaluated for all selected detections.

As expected, Figure~\ref{fig:simulation_delta_c} shows that completeness decreases and reliability increases as the $\Delta C$ threshold increases. The completeness is higher for the 20-count sample.
At our adopted main-catalog threshold of $\Delta C\geq25$, the completeness levels within the central $3\arcmin$ ICM region are 83.8\% and 49.0\% in the full band, 95.1\% and 63.9\% in the soft band, and 94.0\% and 61.4\% in the hard band for sources with at least 20 and 8 input counts, respectively. Across the entire Abell~2744 field, the corresponding completeness levels are 67.2\% and 38.2\% in the full band, 90.2\% and 55.2\% in the soft band, and 76.6\% and 44.9\% in the hard band. The completeness is lower than in blank-field deep X-ray surveys because the deepest central region is affected by the bright ICM emission, whereas the outer region with little ICM emission has a lower effective exposure. Our main catalog contains 51 sources within the ICM region and 276 outside it. The simulated reliability ranges from 88.5\% to 98.9\% within the ICM region and from 98.0\% to 99.2\% outside it. Applying the simulated contamination fractions to the corresponding observed samples, we estimate approximately 5.8, 0.4, and 3.5 spurious detections within the ICM region in the full, soft, and hard bands, respectively. Outside the ICM region, the corresponding estimates are 5.4, 1.6, and 2.9. The total number of spurious
detections estimated from simulations is thus $\approx 14$ in our main catalog, including $\approx7$ within the ICM region and another $\approx7$ outside. Our reliability outside the ICM region is similar to the CDFs \citep{Xue2016, Luo2017} but is slightly worse within the ICM region, because compact fluctuations near steep surface-brightness gradients can be difficult to distinguish from point sources. 

\begin{figure*}[t]
\centering
\includegraphics[width=0.95\textwidth]
{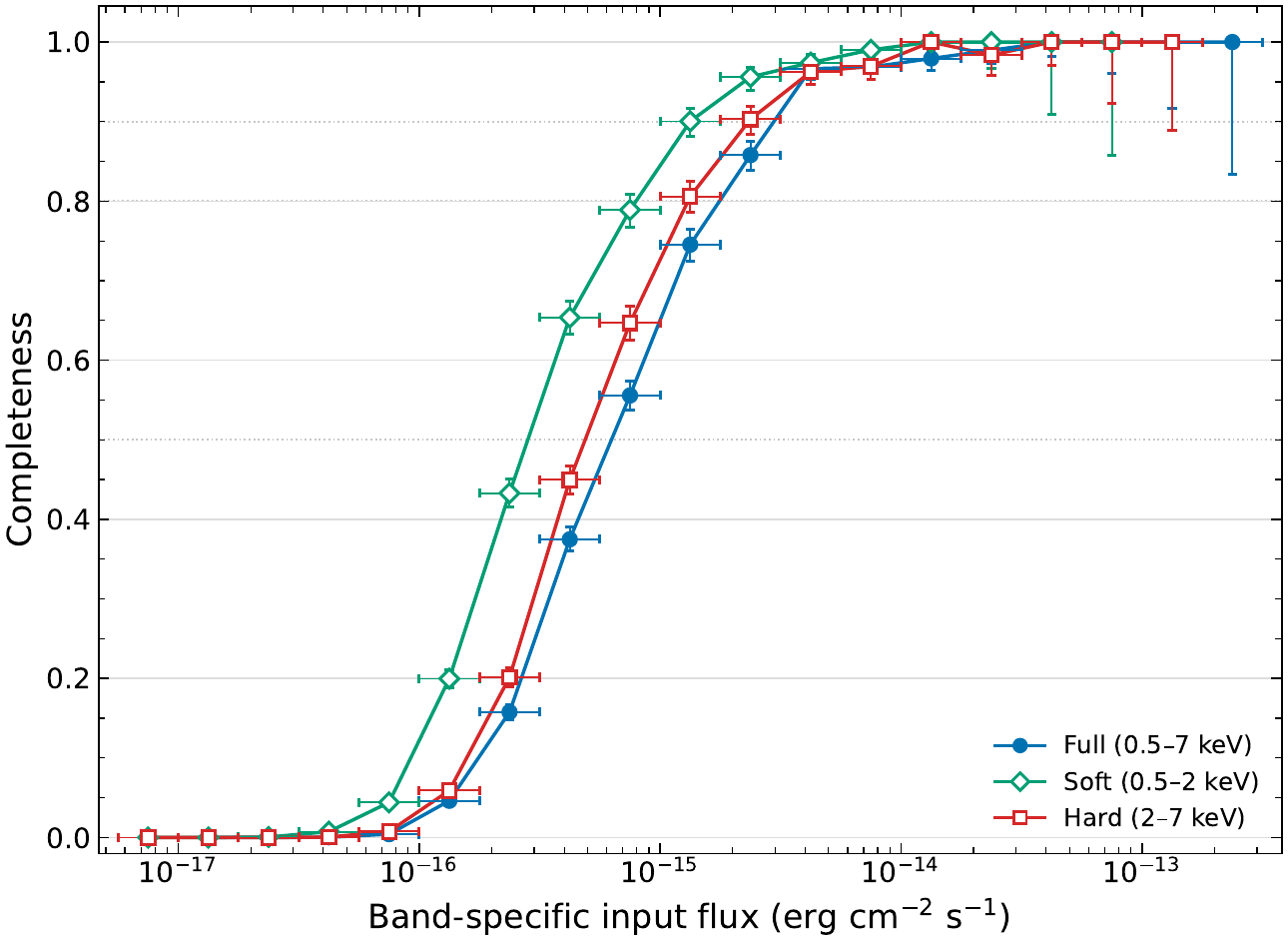}
\caption{
Completeness within the survey footprint at the adopted $\Delta C\geq25$ threshold as a
function of band-specific input flux.
 Solid lines connect adjacent measurements in
the same band. Blue circles, green diamonds, and red squares denote the full,
soft, and hard bands, respectively. Horizontal error bars show the flux-bin
widths, and vertical error bars show $1\sigma$ uncertainties.
No correction for gravitational magnification is applied.
}
\label{fig:simulation_flux_completeness}
\end{figure*}

At our adopted threshold of $\Delta C=25$, the full-field completeness as a function of band-specific input flux is shown in Figure~\ref{fig:simulation_flux_completeness}. These completeness measurements are obtained directly from the injection and recovery results of the ten simulations. We list the flux limits at four selected completeness levels in Table~\ref{tab:flux_completeness}. The same binned recovery fractions are used to derive the simulation-based sky coverage in Section~\ref{sec:simulation_sky_coverage}.

{We also evaluate the completeness and reliability of
overlapping sources across the entire field in our ten simulations.
For the completeness calculation, we require both the input source
and at least one neighbor with an intersecting $R_{90}$ aperture
to have at least 20 input counts in the corresponding band.
At $\Delta C\geq25$, the full-field completeness levels for these
sources are 48.2\%, 62.1\%, and 54.4\% in the full, soft, and hard
bands, respectively. For all detections selected from overlap
groups across the entire field, with no additional source-count
cut, the reliabilities at this threshold are 81.1\%, 93.4\%,
and 87.2\% in these bands, respectively.}

{The simulations may overestimate the reliability within
the ICM region because masking all 410 preliminary candidates and filling
their regions with the local polynomial background may remove
compact ICM structures that mimic point sources, including those
that could pass the $\Delta C\geq25$ threshold. Such structures
cannot be restored by random Poisson fluctuations once removed
from the background maps. The resulting underestimate of spurious
detections is difficult to quantify because compact ICM structures
cannot always be distinguished from genuine point sources using X-ray data alone.}

{We therefore empirically assess the adopted detection threshold using the
multiwavelength counterpart information from the 
DJA catalogs. We match the positions from PSF fitting to the DJA catalogs
within $1\arcsec$, requiring $\mathrm{mag\_auto}<24$, where
$\mathrm{mag\_auto}$ is the DJA stacked AB magnitude from the JWST NIRCAM F277W, F356W, and F444W bands. Of the 130 main-catalog sources within the DJA
footprint, 121 (93.1\%) have a counterpart satisfying these criteria.
We repeat the matching in 2000 realizations after shifting each X-ray
position by $10\arcsec$ in a random direction, retaining shifted
positions within the DJA footprint. The mean number of random matches
is 7.4, corresponding to a conservative contamination estimate of
6.1\% among the 121 matches.}

{The above matching criteria may miss counterparts that are fainter
than 24~mag or have larger positional offsets. An additional search
within $2\arcsec$ identifies six more potential counterparts down
to $\mathrm{mag\_auto}\approx25$~mag, while the remaining three
sources without sufficiently bright DJA counterparts all lie within
the ICM region. This is qualitatively consistent with the lower
reliability predicted by our simulations in this region. The high
counterpart fraction suggests that most of our cataloged X-ray
sources are real.
For comparison, among the 67 candidates within the DJA footprint
that do not pass the adopted X-ray selection, only nine (13.4\%)
have a counterpart with $\mathrm{mag\_auto}<24$ within $1\arcsec$.
Together, this empirical counterpart test and our simulations
support the adopted $\Delta C\geq25$ threshold. A detailed
multiwavelength counterpart analysis using the Bayesian matching
code \texttt{NWAY} \citep{Salvato2018} will be presented in the
second paper of this series.}

\begin{deluxetable}{cccc}
\tabletypesize{\scriptsize}
\tablewidth{3.35in}
\tablecaption{Flux Limits and Completeness Levels
\label{tab:flux_completeness}}
\tablehead{
\colhead{Completeness} &
\colhead{$f_{0.5-7~\mathrm{keV}}$} &
\colhead{$f_{0.5-2~\mathrm{keV}}$} &
\colhead{$f_{2-7~\mathrm{keV}}$} \\
\colhead{(\%)} &
\multicolumn{3}{c}{($\mathrm{erg~cm^{-2}~s^{-1}}$)}
}
\startdata
90 & $3.0\times10^{-15}$ & $1.3\times10^{-15}$ & $2.3\times10^{-15}$ \\
80 & $1.8\times10^{-15}$ & $7.9\times10^{-16}$ & $1.3\times10^{-15}$ \\
50 & $6.3\times10^{-16}$ & $2.8\times10^{-16}$ & $4.9\times10^{-16}$ \\
20 & $2.7\times10^{-16}$ & $1.3\times10^{-16}$ & $2.4\times10^{-16}$ \\
\enddata
\tablecomments{The flux limits correspond to the first crossing of each completeness level and are obtained by linear interpolation in $\log f$ between adjacent points of the binned simulation completeness curves at $\Delta C\geq25$.}
\end{deluxetable}

\subsection{Improvement of Source Positions}
\label{sec:simulation_positions}

\begin{figure*}[t]
\centering
\includegraphics[width=0.92\textwidth]
{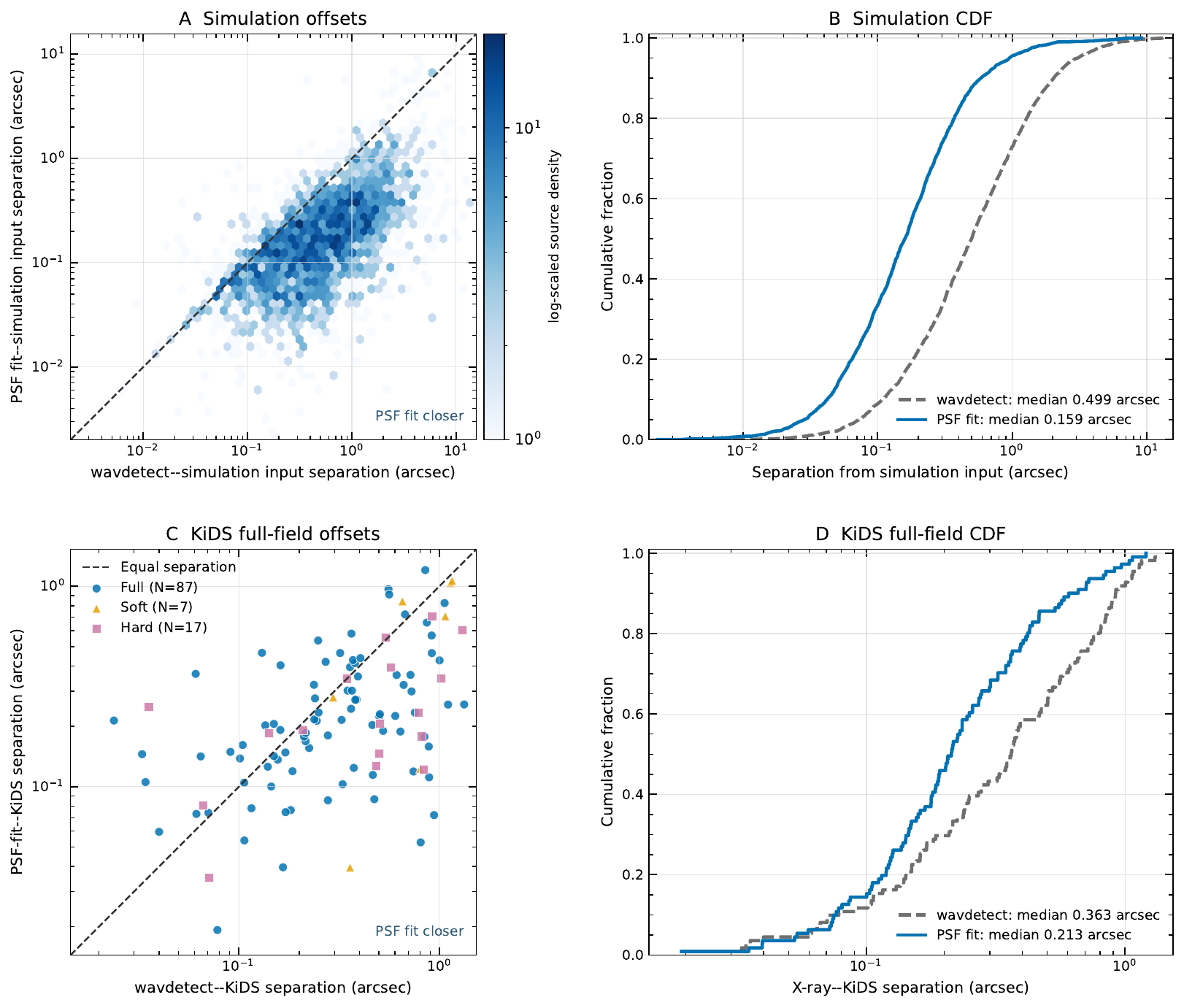}
\caption{
Comparison of the positional accuracy of \texttt{wavdetect} and PSF fitting.
Panels A and B show the 2644 simulation pairs within the survey footprint relative to the known input
positions, while panels C and D show the 111 observed pairs relative to the
Gaia-registered KiDS counterparts. Panels A and C compare the source-by-source
positional offsets, with points below the diagonal indicating smaller offsets
after PSF fitting. Panels B and D show the corresponding cumulative
distributions.
}
\label{fig:simulation_position_accuracy}
\end{figure*}

We next used the simulations to test whether the PSF fits improve the source
positions. For each recovered component with $\max(\Delta C_b)\geq25$, we identified the band with the largest $\Delta C_b$ and compared the \texttt{wavdetect} and PSF-fit positions in that band with the assigned input position. Requiring both the input and fitted positions to lie within the survey footprint gives 2644 matched components across the
ten simulations. 
The median separation from the input position decreases from
$0.499\arcsec$ for \texttt{wavdetect} to $0.159\arcsec$ for the PSF fit.
The PSF-fit
position is closer for 88.0\% of the matched components. Their comparisons are plotted in Figure~\ref{fig:simulation_position_accuracy}.

As an independent check using the observed data, we compared the two kinds of X-ray
positions with objects in the Kilo-Degree Survey
(KiDS; \citealt{Wright2024}), which includes near-infrared photometry from the
VISTA Kilo-degree INfrared Galaxy survey
(VIKING; \citealt{Edge2013}). We first registered the KiDS
coordinates to Gaia using 1480 matches and applied corrections of
$\Delta\mathrm{R.A.}=+0.097\arcsec$ and
$\Delta\mathrm{Decl.}=-0.131\arcsec$ to the KiDS coordinates. We
then selected KiDS objects with $K_s<21$ and
signal-to-noise ratios of at least 5. To avoid favoring either X-ray position, we selected the KiDS object nearest to the midpoint of the \texttt{wavdetect} and PSF-fit positions, requiring it to also be the nearest object to each position and within $1.5\arcsec$ of both. Final associations were restricted to be one-to-one.
 This
selection yields 111 X-ray--KiDS pairs. Their median separation decreases from
$0.363\arcsec$ for \texttt{wavdetect} to $0.213\arcsec$ for the PSF fit. The PSF-fit
position is closer for 67.6\% of the sample (see Figure~\ref{fig:simulation_position_accuracy}). A test using $10\arcsec$ shifts
in eight directions gives a small expected false-match fraction of 3.0\%.

The simulations and the observed counterpart comparison both show that the
PSF fits provide more accurate source positions. We therefore adopt the
PSF-fit position from the band with the largest $\Delta C$ for each source in
the main catalog.

\section{Main Chandra Source Catalog}
\label{sec:catalog}
For the 327 sources selected with $\Delta C\geq25$ in
Section~\ref{sec:psf_fitting}, we present the \mbox{X-ray} photometry and basic
spectral properties in this section. Section~\ref{sec:xray_photometry}
describes the net counts, hardness ratios, photon indices, and fluxes.
Section~\ref{sec:positional_uncertainties} derives empirical positional
uncertainties, and Section~\ref{sec: release}
describes the released catalog.

\subsection{X-Ray Photometry and Basic Spectral Analyses}
\label{sec:xray_photometry}

We used \texttt{srcflux} to extract the X-ray photometric properties of these sources in the full, soft, and hard bands at the PSF-fit positions, following the CIAO thread.\footnote{\url{https://cxc.cfa.harvard.edu/ciao/threads/fluxes\_multiobi/index.html}} Some previous deep Chandra surveys have also used ACIS Extract (AE; \citealt{Broos2010,Broos2012}) for source photometry and significance assessment (e.g., \citealt{Xue2016,Luo2017}). The core functions of \texttt{srcflux} and AE are similar, including source and background extraction, aperture photometry, response generation, and the combination of multiple observations, and their flux estimates are generally consistent when comparable settings are adopted.\footnote{\url{https://cxc.cfa.harvard.edu/ciao/guides/srcflux_for_ae_users.html}} Thus, our use of \texttt{srcflux} is broadly consistent with the methods used in previous deep-field catalogs based on AE.


To assess the dependence of our photometry on background estimation, we compared the background counts in source regions estimated by \texttt{srcflux} with those predicted by our polynomial models for the 51 sources within the central $3\arcmin$ ICM region. We refitted the fourth-order polynomial background models described in Section~\ref{sec:local_psf_background} at the positions used by \texttt{srcflux}, using the 1~keV merged PSF maps for source masking. The median absolute fractional differences, $|B_{\rm srcflux}/B_{\rm poly}-1|$, are 3.8\%, 6.0\%, and 3.8\% in the full, soft, and hard bands, respectively. The differences exceed 15\% for 6, 9, and 9 sources in these bands, with 12 independent sources exceeding this threshold in at least one band. These results indicate that background estimation can affect the photometry of a few sources, reflecting the sensitivity of the inferred ICM background to the choice of regions and models used to describe its spatial variation. Since spectral extraction requires a defined source aperture, we retain the \texttt{srcflux} extraction scheme to use consistent source and background regions for photometry and spectral analysis. However, fluxes of sources within the ICM region should be interpreted with caution, particularly for faint sources superimposed on bright ICM emission, whose inferred fluxes may depend strongly on the adopted background estimation.

For sources in crowded regions, \texttt{srcflux} automatically accounted for overlapping extraction regions using the \texttt{CIAO} tool \texttt{roi}: overlapping source regions were mutually excluded, and neighboring source regions were excluded from the corresponding background regions. The PSF fractions associated with the resulting extraction regions were taken into account through the aperture correction performed with \texttt{psfmethod = arfcorr}.

For each source in each of the three X-ray bands, if the source was detected in that band ($\Delta C_b \geq 25$), we reported the net counts and their associated $1\sigma$  uncertainties following \citet{1986ApJ...303..336G}; otherwise, we reported the 90\% confidence-level upper limit on the net counts using the Bayesian method of \citet{Kraft1991}. The resulting source-count distributions in the three bands are shown in the right panel of Figure~\ref{fig:photometry_distributions}.  The full-, soft-, and hard-band samples contain 320, 226, and 291 detections, with median net counts of 103.5, 58.0, and 79.2, respectively. Among the 327 cataloged sources, 200 are detected in all three bands, while 10, 6, and 1 are detected only in the full, soft, and hard bands, respectively. A total of 163 sources have more than 100 full-band net counts, including 26 sources with more than 1000 full-band net counts. Sources with only upper limits in a given band are not included in the corresponding count distribution.

We calculated the hardness ratio and its 68\% lower and upper bounds using
\texttt{FastHR}, which treats the source and background counts in a Bayesian
framework \citep[see Appendix~A of ][for details]{Zou2023}. The hardness ratio
is defined as
\begin{equation}
{\rm HR}=\frac{H-S}{H+S},
\end{equation}
where $S$ and $H$ are the background-corrected soft- and hard-band count
rates, respectively. 

\begin{figure}[t]
    \centering
    \includegraphics[width=\columnwidth]{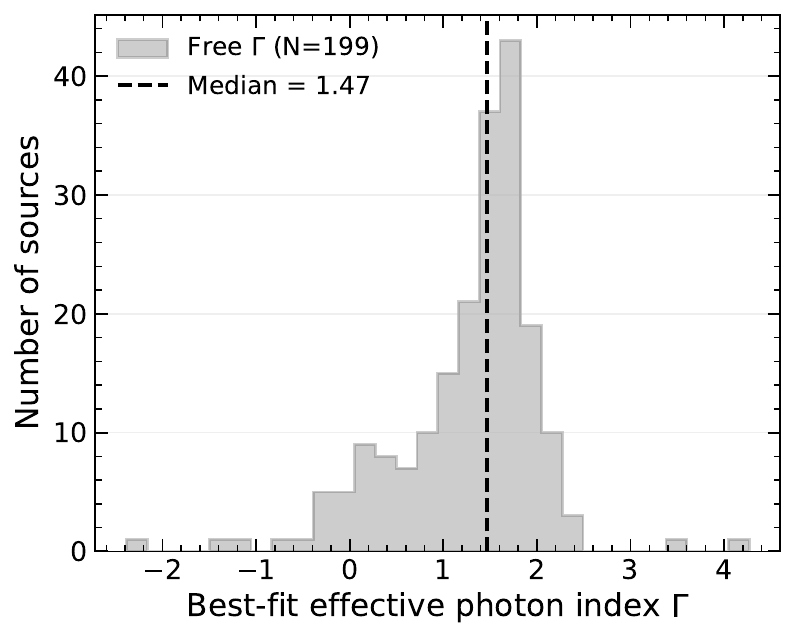}
    \caption{Distribution of the best-fit effective photon indices for the
    199 sources with at least 70 full-band net counts, for which
    $\Gamma$ was allowed to vary. The dashed line marks the median,
    $\Gamma=1.47$. The 128 sources fitted with
    $\Gamma=1.4$ fixed are not included.}
    \label{fig:gamma_distribution}
\end{figure}

\begin{figure*}[t]
    \centering
    \includegraphics[width=\textwidth]{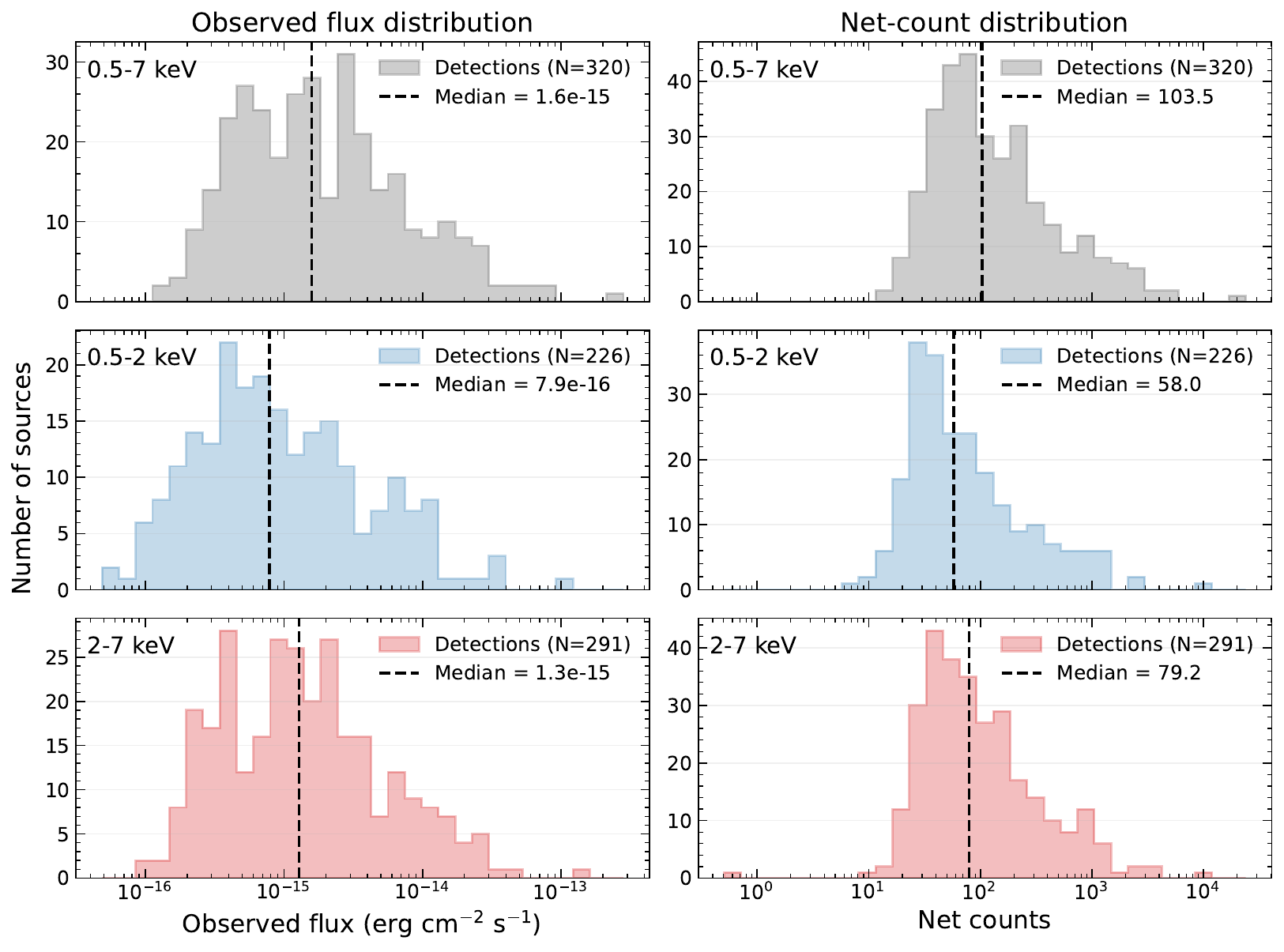}
    \caption{Distributions of the observed energy fluxes (left) and
   net counts (right) for the main catalog. Rows show
    the full, soft, and hard bands. Only sources detected at
    $\Delta C_b\geq25$ in the corresponding band are included; band upper limits
    are excluded. Dashed lines mark the medians, and each legend gives the
    number of detected sources in each band.}
    \label{fig:photometry_distributions}
\end{figure*}

We further fitted the merged source and background spectra and the
corresponding response files with \texttt{sherpa}
\citep{Freeman2001, Siemiginowska2024}. The spectra were grouped to contain at
least one count per bin and fitted over 0.5--7~keV using the W-statistic. We
adopted an absorbed power-law model, \texttt{phabs*cflux(powerlaw)}, with the
Galactic column density fixed at
$N_{\rm H}=1.34\times10^{20}~\mathrm{cm^{-2}}$
\citep{HI4PICollaboration2016}. For the 199 sources with at least 70 
full-band net counts, we allowed the effective photon index
$\Gamma$ to vary. For the remaining 128 sources, we fixed
$\Gamma=1.4$. The free-$\Gamma$ sample has a median best-fit photon index of $\Gamma=1.47$.
Figure~\ref{fig:gamma_distribution} shows the distribution of the best-fit $\Gamma$ values. The 68\%
confidence limits are reported for the free-$\Gamma$ fits.

The full-, soft-, and hard-band fluxes were measured by setting the
\texttt{cflux} integration limits to 0.5--7, 0.5--2, and 2--7~keV,
respectively, while retaining the 0.5--7~keV fitting range. We report best-fit fluxes and their associated $1\sigma$ uncertainties for detected bands and 90\% confidence-level upper limits for undetected ones.
The median full-, soft-, and hard-band fluxes of the detected samples are
$1.58\times10^{-15}$, $7.86\times10^{-16}$, and
$1.28\times10^{-15}~\mathrm{erg~cm^{-2}~s^{-1}}$, respectively. These
distributions are shown in the left column of
Figure~\ref{fig:photometry_distributions}.

\subsection{Positional Uncertainties}
\label{sec:positional_uncertainties}

We calibrated the positional uncertainties using the X-ray--KiDS matches
described in Section~\ref{sec:simulation_positions}, while using direct
Gaia matches for an independent test. We matched the positions from the
PSF fits of the 327 catalog sources  to Gaia DR3
\citep{GaiaCollaboration2023} within $1\arcsec$, requiring the associations to
be one-to-one, and obtained 20 X-ray--Gaia pairs. Among these sources, 18 are
also present in the sample of 111 X-ray--KiDS matches. We removed
these 18 sources from the coefficient fit, leaving 93 KiDS sources for the
calibration. 

\begin{figure*}[t]
    \centering
    \includegraphics[width=\textwidth]{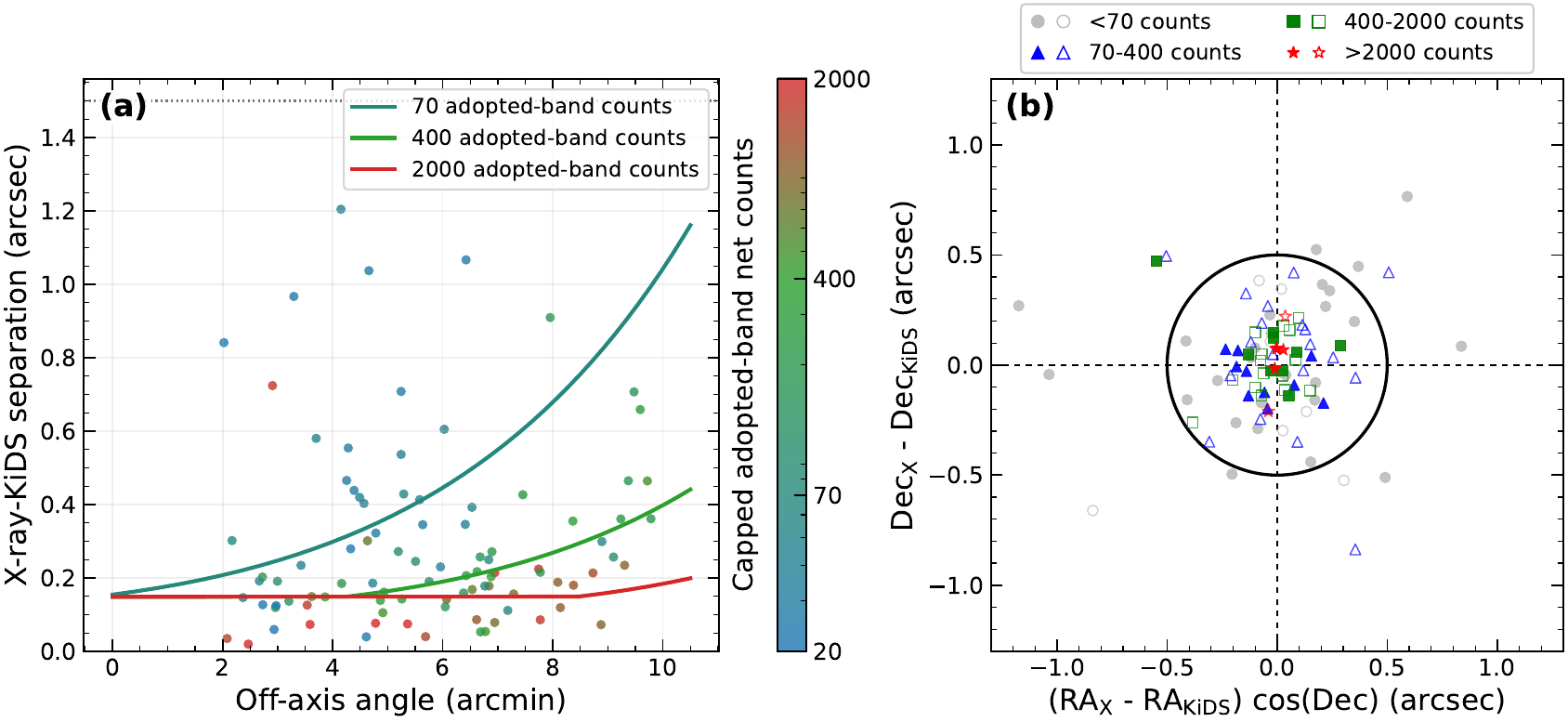}
    \caption{(a) X-ray--KiDS separation as a function of off-axis angle for
    the 93 sources used to calibrate the positional uncertainty relation.
    Points are color coded by the raw net counts in the band used to determine
    the X-ray position; the color scale is capped at 2000 counts for display,
    although the measured counts without a cap are used in the fit. The curves
    show the quadrature sum of the fitted X-ray uncertainty and the adopted
    $0.10\arcsec$ KiDS uncertainty for 70, 400, and 2000 counts; the
    $0.11\arcsec$ lower limit on the X-ray uncertainty is included. The
    horizontal dotted line marks the $1.5\arcsec$ matching radius. (b)
    X-ray-minus-KiDS positional offsets for the same 93 sources. The symbol
    shapes and colors indicate the adopted-band count ranges. Filled and open
    symbols denote sources at off-axis angles of $\leq6\arcmin$ and
    $>6\arcmin$, respectively. The solid circle has a radius of $0.5\arcsec$,
    and the dashed lines mark zero offset.}
    \label{fig:position_err}
\end{figure*}

The X-ray--KiDS separations depend on both the off-axis angle and source
counts, as shown in Figure~\ref{fig:position_err}. The former arises from the
broader Chandra PSF at larger off-axis angles, while the latter reflects the
greater difficulty of determining the positions of faint sources. Following
previous CDF catalogs \citep[e.g.,][]{Xue2011,Xue2016,Luo2017}, we adopted the
basic functional form of \citet{Kim2007}:
\begin{equation}
\label{eq:poserr_model}
\log_{10}\left(\frac{\sigma_{\rm X}}{\mathrm{arcsec}}\right)
=a\theta+b\log_{10}C+c,
\end{equation}
where $\sigma_{\rm X}$ is the nominal 68\% X-ray positional uncertainty in
arcseconds, $\theta$ is the off-axis angle in arcminutes, and $C$ is the net counts in the band used to determine the source
position. This band is selected from the PSF fits as the band with the largest
$\Delta C$
(Section~\ref{sec:simulation_positions}). The coefficients of Equation~\ref{eq:poserr_model} were determined so that for a given sample of X-ray--$K_s$
matches, the fraction of sources having positional offsets smaller than expectations 
($\sqrt{\sigma_{\rm X}^2+\sigma_{Ks}^2}$, where $\sigma_{K_s}=0\farcs1$ is the adopted KiDS source positional
uncertainty) is $\approx68\%$. The best-fit coefficients are $a=0.0945$, $b=-0.5685$, and
$c=0.1196$. Their corresponding 16th--84th percentile intervals
from 5000 source bootstrap resamples are $0.0561$--$0.1186$,
$-0.6286$--$-0.4447$, and $-0.0210$--$0.2443$, respectively.

The 7~Ms CDF-S catalog capped the count term at $C=2000$
\citep{Luo2017}. We also tested this choice for our calibration. Only six of
the 93 X-ray sources with KiDS matches have $C>2000$. The capped fit gives
$(a,b,c)=(0.0954,-0.5695,0.1174)$, compared with
$(0.0945,-0.5685,0.1196)$ without the cap, and both fits enclose 12 of the 20
Gaia validation sources (see below). The difference is much smaller than the
uncertainties in the coefficients, indicating that the choice of the cap has a negligible effect
on our results. We therefore use the measured counts without a cap. We retain
a lower limit of $0.11\arcsec$ for the cataloged positional uncertainties,
following \citet{Luo2017}, to avoid extrapolating the empirical relation to
unrealistically small radii. A total of 34 sources are assigned this lower
limit. The cataloged positional uncertainties have a median of
$0.37\arcsec$ and span $0.11$--$2.08\arcsec$.

For the independent Gaia test, 12 of the 20 direct counterparts lie within
their cataloged positional uncertainty radius, giving an enclosure fraction
of 60\%. The 68.27\% Jeffreys binomial interval is 48.8\%--70.2\%. We also
shifted the X-ray positions by $10\arcsec$ in eight directions and repeated
the matching, which predicts 0.4 chance associations among the 20 matches.
The independent enclosure fraction is therefore statistically consistent
with the nominal 68\% value, although the Gaia sample size is limited for a statistically accurate test.
For comparison, the main catalogs of the 2~Ms CDF-N and 7~Ms CDF-S both
report a median $1\sigma$ positional uncertainty of $0.47\arcsec$
\citep{Xue2016,Luo2017}.

{We also tested a calibration using simulations, which predicts
a median positional uncertainty of $0.26\arcsec$ for our catalog,
compared to $0.37\arcsec$ from the empirical calibration.
The simulation calibration also gives slightly lower enclosure fractions for
the observed X-ray--KiDS and X-ray--Gaia offsets (50.5\% and 45.0\%,
respectively) than for the empirical calibration (68.8\% and 60.0\%,
respectively). This small difference may
reflect residual astrometric errors or deviations from the assumed
PSF that are not fully reproduced by the simulations. We therefore
adopt the empirical calibration, which better represents the observed
offsets relevant to subsequent counterpart matching.}

\subsection{Data Release}
\label{sec: release}
We will release our results on the github repository. Before the acceptance of
this work, we temporarily release the X-ray catalog at \url{https://github.com/sywang0302/CLUES}. The catalog
contains 327 sources and 37 columns. Table~\ref{tab:catalog_excerpt} presents
the first five rows and the first 15 columns; the complete table will be
provided in machine-readable form.

\begin{deluxetable*}{ccccccccccccccc}
\tabletypesize{\scriptsize}
\tablewidth{\textwidth}
\tablecaption{Excerpt from the Main Chandra Source Catalog
\label{tab:catalog_excerpt}}
\tablehead{
\colhead{XID} &
\colhead{R.A.} &
\colhead{Decl.} &
\colhead{$\theta$} &
\colhead{$\sigma_{\rm X}$} &
\colhead{ICM} &
\colhead{$\Delta C_{\rm F}$} &
\colhead{$\Delta C_{\rm S}$} &
\colhead{$\Delta C_{\rm H}$} &
\colhead{$D_{\rm F}$} &
\colhead{$D_{\rm S}$} &
\colhead{$D_{\rm H}$} &
\colhead{$C_{\rm F}$} &
\colhead{$C_{\rm F,lo}$} &
\colhead{$C_{\rm F,hi}$} \\
\colhead{} &
\colhead{(deg)} &
\colhead{(deg)} &
\colhead{(arcmin)} &
\colhead{(arcsec)} &
\colhead{(flag)} &
\colhead{} &
\colhead{} &
\colhead{} &
\colhead{(flag)} &
\colhead{(flag)} &
\colhead{(flag)} &
\colhead{(counts)} &
\colhead{(counts)} &
\colhead{(counts)}
}
\startdata
1 & 3.397445 & $-30.353934$ & 10.374 & 2.079 & 0 & 89.51 & 34.76 & 96.20 & 1 & 1 & 1 & 42.43 & 35.55 & 49.53 \\
2 & 3.407472 & $-30.364199$ & 9.719 & 0.320 & 0 & 1297.55 & 527.93 & 849.25 & 1 & 1 & 1 & 498.94 & 475.70 & 521.33 \\
3 & 3.413048 & $-30.403893$ & 9.186 & 0.126 & 0 & 10881.02 & 6958.20 & 4469.50 & 1 & 1 & 1 & 2102.80 & 2056.16 & 2148.85 \\
4 & 3.413188 & $-30.375079$ & 9.302 & 0.421 & 0 & 290.93 & 66.74 & 241.66 & 1 & 1 & 1 & 261.87 & 244.86 & 278.17 \\
5 & 3.423558 & $-30.366463$ & 8.877 & 0.192 & 0 & 1810.53 & 67.20 & 1656.07 & 1 & 1 & 1 & 886.82 & 856.25 & 916.77 \\
\enddata
\tablecomments{The table shows the first 15 catalog columns. The ICM flag is
1 for sources within the adopted $3\arcmin$ ICM region and 0 otherwise.
$D_{\rm F}$, $D_{\rm S}$, and $D_{\rm H}$ are the full-, soft-, and hard-band
detection flags. $C_{\rm F}$ gives the full-band net counts, while
$C_{\rm F,lo}$ and $C_{\rm F,hi}$ are its 68\% lower and upper bounds. The
complete machine-readable catalog contains all 327 sources and 37 columns.}
\end{deluxetable*}

The catalog columns are defined as follows.
\begin{enumerate}
\item \textit{Column (1)}, \texttt{XID}. Unique X-ray source identifier.
\item \textit{Columns (2)--(3)}, \texttt{RA} and \texttt{DEC}. ICRS X-ray
R.A. and decl. in units of degrees.
\item \textit{Column (4)}, \texttt{OFFAXIS\_ARCMIN}. Off-axis angle in units of
arcminutes relative to the exposure-weighted mean aim point.
\item \textit{Column (5)}, \texttt{POS\_ERR\_ARCSEC}. The $1\sigma$ X-ray
positional uncertainty in units of arcseconds. See
Section~\ref{sec:positional_uncertainties}.
\item \textit{Column (6)}, \texttt{FLAG\_IN\_ICM}. Flag identifying a source
within $3\arcmin$ of the adopted ICM center. Our main catalog contains 51 sources within the ICM region and 276 outside it.
\item \textit{Columns (7)--(9)}, \texttt{DELTA\_C\_FULL},
\texttt{DELTA\_C\_SOFT}, and \texttt{DELTA\_C\_HARD}. Source-specific
$\Delta C_b$ values in the full, soft, and hard bands.
\item \textit{Columns (10)--(12)}, \texttt{F\_DET}, \texttt{S\_DET}, and
\texttt{H\_DET}. Band detection flags indicating $\Delta C_b\geq25$.
\item \textit{Columns (13)--(21)}, \texttt{F\_NET\_COUNTS},
\texttt{F\_NET\_LO68}, \texttt{F\_NET\_HI68}, and the corresponding soft-
and hard-band columns. For a detected band, these give the net counts and their
68\% lower and upper bounds. For an undetected band,
\texttt{NET\_COUNTS} contains the 90\% confidence-level upper limit, and the
two bound columns are blank.
\item \textit{Columns (22)--(24)}, \texttt{HR}, \texttt{HR\_LO68}, and
\texttt{HR\_HI68}. Hardness ratio and its 68\% lower and upper bounds.
\item \textit{Columns (25)--(27)}, \texttt{GAMMA}, \texttt{GAMMA\_LO68}, and
\texttt{GAMMA\_HI68}. Effective photon index and its 68\% lower and upper
bounds. The bounds are blank when $\Gamma=1.4$ is fixed or when the corresponding confidence bound is unconstrained.
\item \textit{Columns (28)--(36)}, \texttt{F\_FLUX},
\texttt{F\_FLUX\_LO68}, \texttt{F\_FLUX\_HI68}, and the corresponding soft-
and hard-band columns. Fluxes are in
$\mathrm{erg~cm^{-2}~s^{-1}}$. For a detected band, these give the flux and
its 68\% lower and upper bounds. For an undetected band,
\texttt{FLUX} contains the 90\% confidence-level upper limit, and the two
bound columns are blank.
\item \textit{Column (37)}, \texttt{W/DOF}. W-statistics divided by
the number of degrees of freedom in our spectral fitting with an absorbed power-law model.
\end{enumerate}
In addition to the catalog, we will also provide the final X-ray images, exposure
maps, background maps, and sensitivity maps (Section~\ref{sec:sensimap}) after
this paper is accepted.

\section{Sensitivity Map and Sky Coverage}
\label{sec:sensimap}

\subsection{Construction of Sensitivity Maps}
\label{sec:sensimap_cons}
We constructed sensitivity maps in the full, soft, and hard bands by calculating the minimum source counts required to satisfy our adopted threshold ($\Delta C_b\geq25$) at each pixel. The valid footprint contains approximately $5\times10^{6}$ pixels in each band, making it computationally prohibitive to construct a PSF from all 101 observations at every pixel. We therefore divided the image into cells of $20\times20$ ACIS pixels ($9.84\arcsec\times9.84\arcsec$). At the center of each cell, we constructed a stacked PSF by combining the PSFs from the 101 individual observations, weighted by their exposure times. This method assumes that the Chandra PSF does not change significantly on a scale of $\sim10\arcsec$. For every pixel within a cell, we adopted the stacked PSF calculated at the cell center and shifted it to the exact trial position. Thus, the PSF shape was sampled only once per grid cell, whereas the limiting source normalization was determined independently at every pixel.

At each pixel, we measured the local background, $B_{90}$, from the background map within the $R_{90}$ aperture. The fourth-order polynomial background model described in Section~\ref{sec:local_psf_background} was then used to describe the spatial distribution of the expected background counts, $b_i$, within the aperture while preserving the total normalization $B_{90}$.

Denoting $p_i$ as the fraction of the source counts in pixel $i$ predicted by the stacked PSF, the expected total counts in pixel $i$ are $b_i+Sp_i$, where $S$ is the total source counts. We calculated the expected likelihood improvement relative to the background-only model as
\begin{equation}
\Delta C(S)=
2\sum_{i\in R_{90}}
\left[
(b_i+Sp_i)\ln\left(1+\frac{Sp_i}{b_i}\right)-Sp_i
\right]
\label{eq:sensitivity_delta_c}
\end{equation}
 We numerically solved for the minimum source counts $S_{\rm lim}$ satisfying $\Delta C(S_{\rm lim})=25$, same as our adopted detection threshold.

We then converted $S_{\rm lim}$ to flux using the exposure maps, assuming a power-law spectrum with $\Gamma=1.4$ and Galactic absorption of $N_{\rm H}=1.34\times10^{20}~\mathrm{cm}^{-2}$. We retained pixels for which the exposure exceeds 10\% of its maximum value. The resulting sensitivity maps account for the spatial variations in the PSF, effective exposure due to effects such as vignetting and CCD gaps, and background across the field.
The deepest contiguous $\approx1~\mathrm{arcmin}^{2}$ regions have mean
limiting fluxes of $1.6\times10^{-16}$, $8.7\times10^{-17}$, and
$1.7\times10^{-16}~\mathrm{erg~cm^{-2}~s^{-1}}$ in the full, soft, and hard
bands, respectively. Unlike a blank field, the region near the mean aim point does not necessarily provide the greatest sensitivity because the advantage of its high effective exposure is reduced by the bright ICM background.

\begin{figure*}[t]
\centering
\includegraphics[width=0.72\textwidth]
{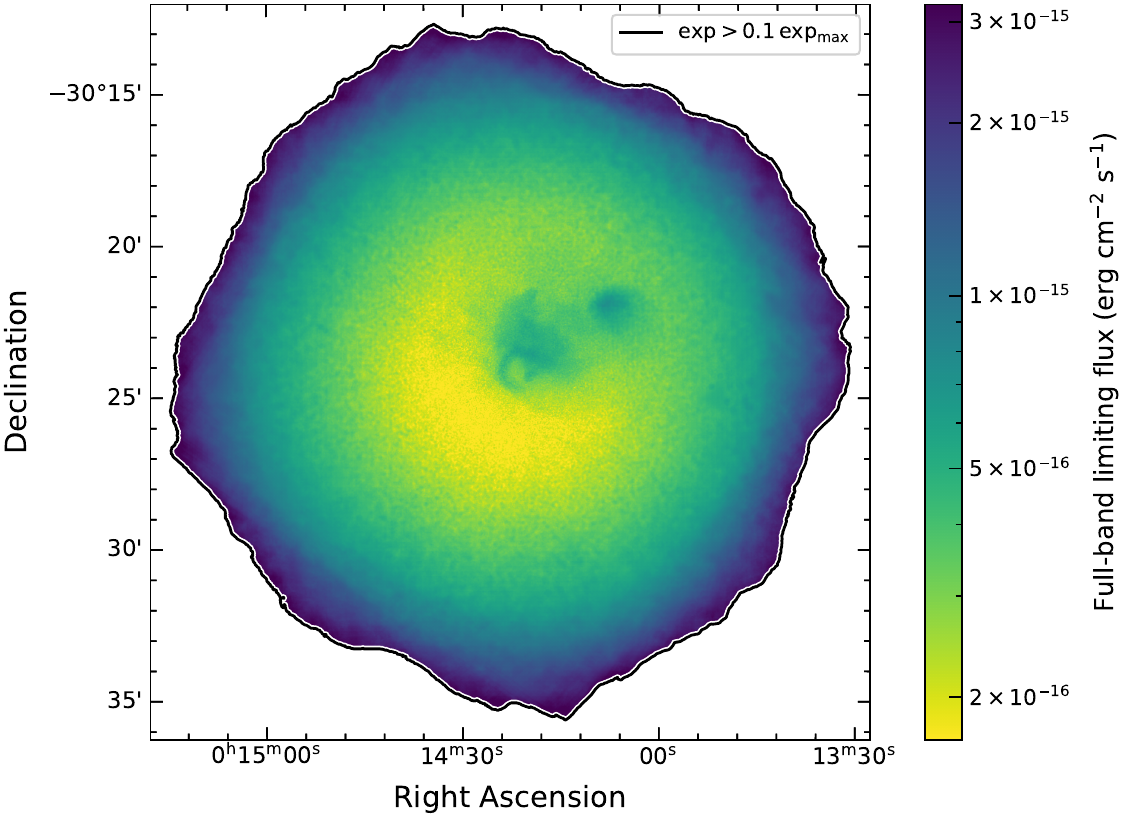}
\caption{
Full-band (0.5--7~keV)  sensitivity map derived from PSF likelihood calculation at each pixel.  Fluxes were calculated for a power-law spectrum with $\Gamma=1.4$. The black-and-white contour
encloses the full-band detection footprint, defined by exposure values greater than 10\% of the maximum full-band exposure. Lower limiting fluxes indicate greater point-source
sensitivity. No correction for gravitational lensing is applied.
}
\label{fig:full-sensimap}
\end{figure*}

The sensitivity map describes the nominal, first-order approximation of the limiting flux at a fixed position. It does not include the initial
\texttt{wavdetect} selection or other effects in the complete source-detection
procedure; these are included in the simulation-derived sky coverage in
Section~\ref{sec:simulation_sky_coverage}. Figure~\ref{fig:full-sensimap}
shows the resulting full-band sensitivity map. The limiting flux is lowest in regions with high effective exposure and low local background. The sensitivity becomes worse in regions with bright ICM emission, as well as at large off-axis angles. Near the cluster core, the complex sensitivity pattern results from the combination of deep on-axis exposure and spatial variations in the ICM background.

\subsection{Simulation-derived Sky Coverage}
\label{sec:simulation_sky_coverage}

We further measure the sky coverage as a function of flux in this subsection. To be accurate, we will rely on simulations instead of the nominal sensitivity maps derived in Section~\ref{sec:sensimap_cons}. The latter, although conventionally used in \mbox{X-ray} surveys, do not account for the full detection complexity including Poisson fluctuations, the initial \texttt{wavdetect} selection, or source deblending. These effects are naturally included in the simulations, which therefore provide a more realistic estimate of the sky coverage. 

\begin{figure}[t]
\centering
\includegraphics[width=\hsize]
{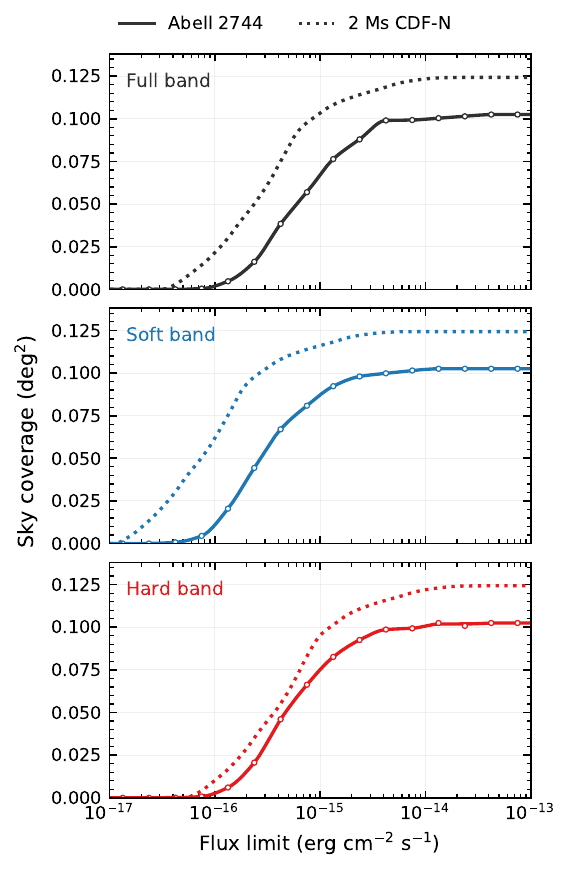}
\caption{
Sky coverage as a function of observed flux limit in the full
(0.5--7~keV; top), soft (0.5--2~keV; middle), and hard (2--7~keV; bottom)
bands. Black, blue, and red denote the full, soft, and hard bands, respectively. The three Abell~2744 curves share the same $369~\mathrm{arcmin^2}$ survey area.
Open circles show the binned sky coverage derived from ten source-injection
simulations at $\Delta C_b\geq25$ using
Equation~\ref{eq:simulation_sky_coverage}.
Dotted curves show the sky coverage of the 2~Ms CDF-N derived from the
sensitivity maps of \citet{Xue2016}. No correction for gravitational lensing
is applied.
}
\label{fig:skycoverage}
\end{figure}

For band $b$ and input flux $f$, we calculate the simulation-derived sky
coverage as
\begin{equation}
\Omega_{{\rm }b}(f)=
\Omega_{\rm survey}
\frac{N_{{\rm rec},b}(f)}{N_{{\rm in},b}(f)},
\label{eq:simulation_sky_coverage}
\end{equation}
where $\Omega_{\rm survey}=369.003~\mathrm{arcmin^2}$ is the area of the three-band union. The quantity $N_{{\rm in},b}(f)$ counts input sources whose true positions lie within this footprint in the corresponding band and flux bin, and $N_{{\rm rec},b}(f)$ counts those recovered in the same band with $\Delta C\geq25$ and fitted positions within the footprint. We pooled the
counts from all ten simulations before calculating the recovery fraction.
Because the recovery fractions are measured only in discrete input flux bins,
we constructed a continuous sky coverage curve for use at intermediate fluxes.
We first fitted the binned recovery fractions with a weighted monotonic regression,
using $N_{{\rm in},b}$ as the weights, to retain the expected nondecreasing
dependence on flux, and then applied a shape-preserving interpolation in $\log f$
between the fitted values. As shown in Figure~\ref{fig:skycoverage}, the fitted curves agree closely with
the completeness values measured directly in each input flux bin of the
simulations, indicating that the fitting procedure reliably represents the
simulation results.

Table~\ref{tab:flux_completeness} lists the flux limits corresponding to 20\%, 50\%, 80\%, and 90\% completeness in each band, measured from the binned completeness curves before smoothing.

To place the depth of the Abell~2744 field in context, we use the currently second and third deepest \mbox{X-ray} fields, 2~Ms CDF-N
\citep{Xue2016} and AEGIS-XD \citep{Nandra2015}, as comparisons.
Figure~\ref{fig:skycoverage} shows the comparison with the CDF-N, while for
AEGIS-XD we use its published 50\%-completeness flux limits. At 50\% completeness, the Abell~2744 flux limits are $6.3\times10^{-16}$,
$2.8\times10^{-16}$, and
$4.9\times10^{-16}~\mathrm{erg~cm^{-2}~s^{-1}}$ in the full, soft, and hard
bands, respectively. The corresponding limits are
$3.2\times10^{-16}$, $1.0\times10^{-16}$, and
$5.0\times10^{-16}~\mathrm{erg~cm^{-2}~s^{-1}}$ for the CDF-N, and
$6.3\times10^{-16}$, $2.0\times10^{-16}$, and
$9.4\times10^{-16}~\mathrm{erg~cm^{-2}~s^{-1}}$ for AEGIS-XD, after
converting its full- and hard-band limits to our energy ranges assuming
$\Gamma=1.4$. Abell~2744 therefore shares a similar hard-band depth with the CDF-N while becoming shallower in the full and soft bands. Compared with
AEGIS-XD, Abell~2744 has a similar full-band depth, is approximately 1.4 times
shallower in the soft band, and is approximately 1.9 times deeper in the hard
band. Lastly, we emphasize that, due to the lensing magnification by Abell~2744, sources intrinsically fainter than the detection thresholds could be magnified and become detectable, and thus the intrinsic, de-magnified limits could be improved in the central region, as discussed below in Section~\ref{sec:lensing_sensitivity}.

The relatively weaker full- and soft-band depths likely result from the
combined effects of the elevated cluster background and the long-term
reduction of the low-energy ACIS response. The CDF-N observations were
obtained in 1999--2002, whereas most of the Abell~2744 exposure was obtained
in 2022--2024. Contamination accumulated on the ACIS optical-blocking filters
over this interval has reduced the low-energy effective area, with the
strongest effect in the soft band and a smaller effect in the hard band
\citep{Plucinsky2018,Grant2024}. In addition, the bright ICM increases the
background in the deepest central region of Abell~2744, while the outer field
with little ICM emission has a lower effective exposure. These effects can
explain why the Abell~2744 full- and soft-band depths are relatively
shallower, while its hard-band depth remains competitive with the CDF-N. Differences in
pointing geometry, source selection, and background estimation may also
contribute to the remaining secondary differences among the three fields.

\subsection{Effect of Gravitational Lensing}
\label{sec:lensing_sensitivity}

Compared with traditional blank-field surveys, the bright and structured ICM of Abell~2744 raises the X-ray background, while the strong gravitational lensing of the cluster provides an opportunity to detect intrinsically less-luminous sources. The sensitivity map and sky coverage described above are defined in the image plane and do not include gravitational magnification. For a source at image-plane position $\boldsymbol{\theta}$ and redshift $z$, we denote the corresponding source-plane position as $\boldsymbol{\beta}$. The limiting flux and solid angle in the two planes are related by
\begin{equation}
\begin{split}
f_{\rm lim,int}(\boldsymbol{\beta},z)
&=\frac{f_{\rm lim,obs}(\boldsymbol{\theta})}
{|\mu(\boldsymbol{\theta},z)|},\\
d\Omega_{\rm src}(\boldsymbol{\beta},z)
&=\frac{d\Omega_{\rm img}(\boldsymbol{\theta})}
{|\mu(\boldsymbol{\theta},z)|}.
\end{split}
\label{eq:lensing_image_source_plane}
\end{equation}
Here, $\mu(\boldsymbol{\theta},z)$ denotes the signed gravitational-lensing magnification, with $|\mu|$ giving both the ratio of the observed to intrinsic flux and the ratio of the image-plane to source-plane solid angle. Lensing therefore allows us to probe intrinsically fainter sources, while mapping a smaller source-plane area onto a larger image-plane area and thereby diluting the surface density of background sources in the image plane.

The magnification depends strongly on source redshift and is unity for sources in front of the Abell~2744 cluster at $z\approx0.3$. Since the redshift distributions of sources in deep X-ray surveys generally peak at $z\approx1$--2 \citep[e.g.,][]{Luo2017,Ni2021}, we adopt $z=2$ as a representative redshift to illustrate the effect of lensing on the sensitivity. We constructed a magnification map using the publicly available lens model of \citet{Bergamini2023}, which incorporates 149 multiple images identified with GLASS-JWST, UNCOVER, and MUSE observations. We find that regions with $|\mu|>3$ cover a total image-plane area of approximately $6~\mathrm{arcmin^2}$ and are concentrated near the center of the Chandra field. Because these regions are spatially irregular and the local sensitivity varies across them, we adopt the full-band 20\% completeness limit of $2.7\times10^{-16}~\mathrm{erg~cm^{-2}~s^{-1}}$ as a representative value throughout these regions. Equation~\ref{eq:lensing_image_source_plane} then implies an intrinsic limiting flux below $9.0\times10^{-17}~\mathrm{erg~cm^{-2}~s^{-1}}$ and a total source-plane area smaller than $2~\mathrm{arcmin^2}$. Using the empirical full-band number-count relation of \citet{Georgakakis2008}, we expect approximately five background sources above the corresponding intrinsic flux limit within the $\approx2~\mathrm{arcmin^2}$ source-plane area. In addition, the mass distribution of Abell~2744 is spatially offset from the ICM emission \citep[e.g.,][]{Chadayammuri2024}, leaving regions where high magnification coincides with a relatively low ICM background. Together, these effects can produce intrinsic limiting fluxes in some regions below those reached by the 2~Ms CDF-N \citep{Xue2016}.

We do not present source-plane sensitivity maps or sky-coverage curves here because a self-consistent calculation depends on both source position and redshift and must also account for multiple images of the same physical source. The $z=2$ calculation above is intended only as a representative example.

\section{Number Counts}
\label{sec:number_counts}
We computed the cumulative number density of
sources, \hbox{$N(>S)$}, brighter than a given  flux, $S$.
For each band $b$, the surface density above an observed flux $S$ is calculated as
\begin{equation}
\begin{split}
N_b(>S) &= \sum_{S_{i,b}>S}
\frac{1}{\Omega_{{\rm }b}(S_{i,b})},\\
\end{split}
\label{eq:number_counts}
\end{equation}
where $S_{i,b}$ is the catalog flux of source $i$ in band $b$, and
$\Omega_b(S_{i,b})$ is the sky coverage evaluated at that flux, i.e., the effective
 area over which a source with flux $S_{i,b}$ can be detected, as measured in Section~\ref{sec:simulation_sky_coverage}.

\begin{figure}[t]
\centering
\includegraphics[width=\columnwidth]
{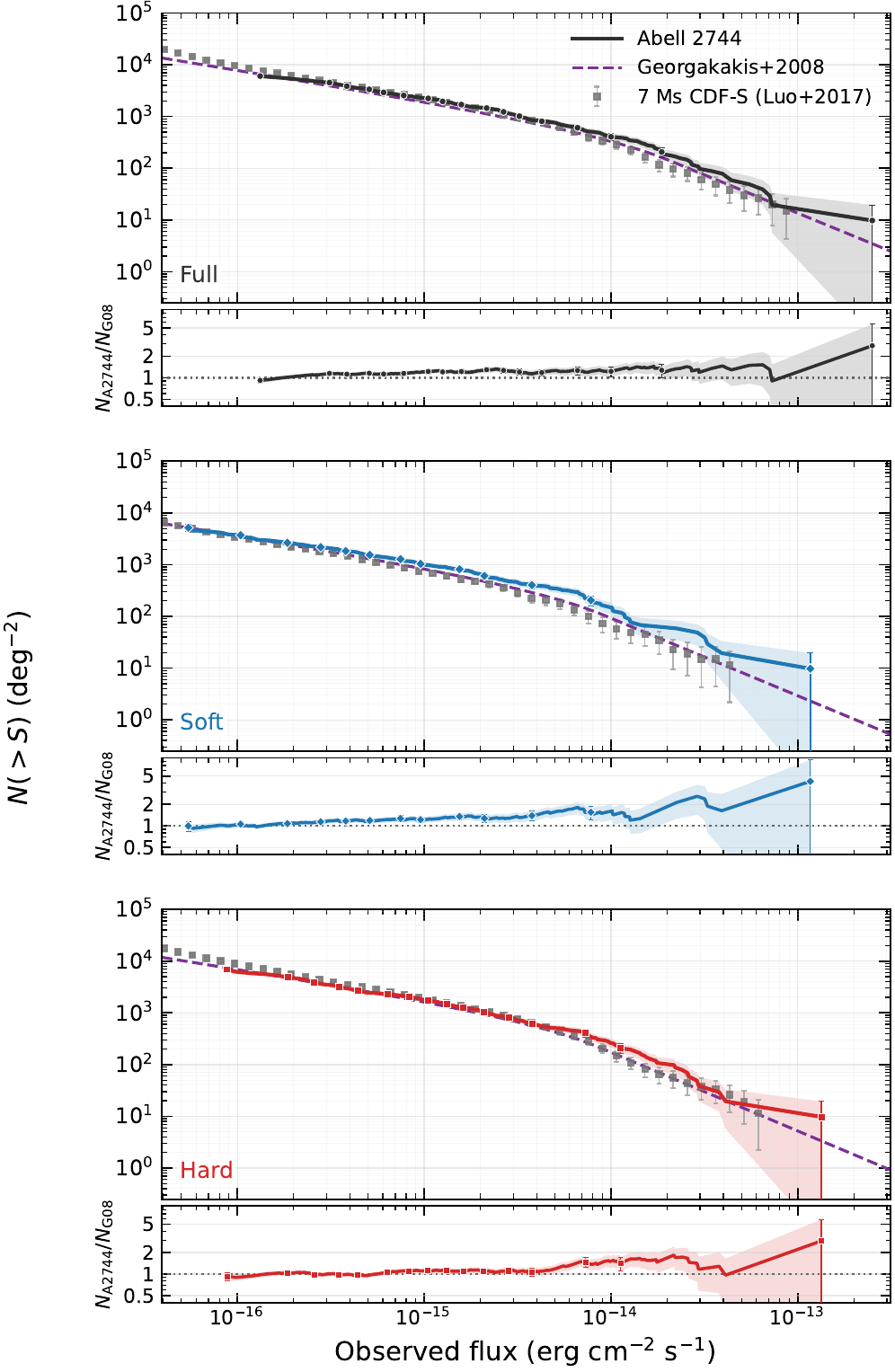}
\caption{
Cumulative number density of sources brighter than a given  flux for the main source catalog in the full (top), soft
(middle), and hard (bottom) bands. Only sources with $\Delta C_b\geq25$ in
the corresponding band are included. The black, blue, and red curves show the
Abell~2744 measurements, and the shaded regions show the 1$\sigma$
uncertainties. For clarity, explicit error bars are shown at intervals of 20 sources, and the brightest and faintest sources are included in each band. Gray squares show the
7~Ms CDF-S source number densities from \citet{Luo2017}, and the purple dashed curves
show the empirical relations of \citet{Georgakakis2008}. The latter were
converted from 0.5--10 to 0.5--7~keV in the full band and from 2--10 to
2--7~keV in the hard band assuming $\Gamma=1.4$.
The smaller panels show $N_{\rm A2744}/N_{\rm G08}$, the ratio to the
\citet{Georgakakis2008} relations. No
gravitational-lensing correction is applied to the Abell~2744 measurements.
}
\label{fig:number_counts}
\end{figure}

Figure~\ref{fig:number_counts} compares the resulting $N(>S)$ with the measurements from the 7~Ms CDF-S \citep{Luo2017} and the empirical broken
power-law relations of \citet{Georgakakis2008}. Our measurements are largely consistent with theirs at a first-order level. However, subtle differences exist especially above
$S\approx10^{-14}~\mathrm{erg~cm^{-2}~s^{-1}}$. At
$S=10^{-14}~\mathrm{erg~cm^{-2}~s^{-1}}$, the ratios of the Abell~2744
counts to the 7~Ms CDF-S counts are 1.33, 2.34, and 1.52 in the full, soft,
and hard bands, respectively, while the corresponding ratios to the
\citet{Georgakakis2008} relations are 1.23, 1.58, and 1.48. The largest
difference is therefore seen in the soft band. Such differences are likely physical because our field is not blank, as discussed below.

First, Abell~2744 itself may host an excess of X-ray sources among its galaxy members. Previous Chandra studies have shown that massive cluster
fields can contain an excess of bright X-ray point sources relative to blank
fields. \citet{Branchesi2007} found an approximately $2\sigma$ bright-end
excess in a sample of 18 distant clusters, with the excess concentrated within
0.5~Mpc of the cluster centers. Using a sample of 43 massive clusters,
\citet{Ehlert2013} measured $1.1\pm0.4$ excess sources per cluster
within $r_{500}$.  These results suggest that
cluster-associated sources may contribute to the number-count excess observed toward Abell~2744. 

Second, the overall $N(>S)$ is affected by the gravitational lensing exerted by Abell~2744 in a complex way.
For a background source, lensing magnification increases its observed flux,
moving it toward the bright end of the $\log N$--$\log S$ relation and
allowing an intrinsically fainter source to enter the catalog. At the same
time, Equation~\ref{eq:lensing_image_source_plane} shows that the corresponding
image-plane area is larger than the source-plane area, which dilutes the surface density of background sources in the image plane. If multiple imaging is ignored,
correcting both the source fluxes and survey area for lensing would therefore
move the X-ray $\log N$--$\log S$ relation toward the upper left in Figure~\ref{fig:number_counts}, corresponding
to lower intrinsic fluxes and higher source-plane surface densities. This is
not a uniform shift because the magnification depends on both source position
and redshift. Strong lensing can also produce multiple images of a single
background source; these images appear as separate catalog entries but
correspond to the same physical source. Thus the present observed-frame $\log N$--$\log S$ relation
cannot yet be interpreted as the intrinsic number counts of the background
population.

In the second paper of this series, we
will identify the multiwavelength counterparts of the X-ray sources and
determine their photometric or spectroscopic redshifts. We will then separate
cluster members from foreground and background sources, identify possible
multiple-image systems, and calculate the magnification of each background source. After accounting for both cluster members and
gravitational lensing, the intrinsic $\log N$--$\log S$ relation will be recovered. Nevertheless, the consistency within a factor of $\approx2$ between our observed $\log N$--$\log S$ relation with the blank-field ones suggests that the likely corrections would be largely secondary over the whole field.

\section{Summary}
\label{sec: summary}
As the first paper of the Chandra CLUES series, this work presents a new ultra-deep extragalactic field in \mbox{X-rays}, Abell~2744. Combining the long archival Chandra exposure and the strong-lensing magnification, this field represents the third deepest extragalactic \mbox{X-ray} survey, approaching the CDF-N depth. We summarize this work as follows.
\begin{enumerate}
\item The Abell~2744 field has been targeted by 103 Chandra ACIS-I observations, adding to over 2~Ms exposures in total. We systematically reduced 101 observations that share similar aim points; their total exposure is 2.102~Ms, while the cleaned exposure is 2.094~Ms. See Section~\ref{sec: chandra}.
\item We performed point-source detections in the full ($0.5-7$~keV), soft ($0.5-2$~keV), and hard ($2-7$~keV) bands. Our detections involve an initial wavelet detection and subsequent background and local PSF fitting to optimize the point-source significance. In total, we detected and cataloged 327 \mbox{X-ray} sources with $\Delta C\ge25$ in at least one band, among which 320, 226, and 291 are detected in the full, soft, and hard bands, respectively. See Sections~\ref{sec:candidate_detection} and \ref{sec:psf_modeling}.
\item We conducted end-to-end simulations of our detection procedures. Based on these simulations, we calculated the reliability and completeness versus the detection threshold for $\Delta C$. At our chosen threshold of $\Delta C=25$, the reliability is high ($98.0\%-99.2\%$) outside the central $3'$ ICM region but slightly degrades to $88.5\%-98.9\%$ inside the ICM region because small-scale fluctuations near steep surface-brightness gradients can be difficult to distinguish from real sources using \mbox{X-ray} data alone. The expected 50\% completeness flux is $6.3\times10^{-16}$, $2.8\times10^{-16}$, and $4.9\times10^{-16}~\mathrm{erg~cm^{-2}~s^{-1}}$ in the full, soft, and hard bands, respectively. Our analyses also suggest that our PSF-fitted positions are more accurate than the initial \texttt{wavdetect}-based positions. See Section~\ref{sec:simulations}.
\item We derived the \mbox{X-ray} photometry and provided basic spectral analyses of our cataloged sources. For the full, soft, and hard bands, the median net counts are 103.5, 58.0, 79.2, and the median fluxes are $1.6\times10^{-15}$, $7.9\times10^{-16}$, and $1.3\times10^{-15}~\mathrm{erg~cm^{-2}~s^{-1}}$, respectively. The hardness ratios and effective power-law photon indices are also cataloged. See Section~\ref{sec:xray_photometry}.
\item We quantified the \mbox{X-ray} positional accuracies by comparing the \mbox{X-ray} positions with the optical positions for sources brighter than $K_s=21$~mag. Our positions have been optimized thanks to the PSF fitting. An empirical relation is given to depict the positional uncertainty with off-axis angles and net counts. Most of the positional uncertainties are well within the sub-arcsec level, with a median value of $0.37''$, although a small fraction of off-axis sources can reach $2''$ positional uncertainties. See Section~\ref{sec:positional_uncertainties}.
\item Sensitivity maps are constructed after masking detected sources. We further derived the effective area as a function of limiting fluxes based on the simulated completeness. Our \mbox{X-ray} depth at a 50\% completeness level is similar to that of the CDF-N in the hard band but is slightly shallower by a factor of $2-3$ in the soft and full bands. However, our actual intrinsic sensitivities could be deeper in the central region after considering the strong-lensing effect. We also measured the cumulative number density of our survey, revealing a roughly consistent $\log N-\log S$ relation with previous blank-field results within a factor of 2, but an excess exists primarily in the bright end possibly due to galaxy-cluster members and gravitational lensing. See Sections~\ref{sec:sensimap} and \ref{sec:number_counts}.
\end{enumerate}
Overall, our ultra-deep \mbox{X-ray} catalog is an important resource in this unique strong-lensing field, especially considering its synergy with the archival/ongoing HST/JWST observations. These \mbox{X-ray} data products are publicly available; see Section~\ref{sec: release} for the release notes.\par
In the second of this series, we will focus on the multiwavelength data in our field and characterize the properties of our detected \mbox{X-ray} sources. We will identify their counterparts consistently across the entire field, regardless of whether they are within the JWST footprint. We will then classify the nature of the \mbox{X-ray} sources into three categories -- stars, galaxies, and AGNs -- and most of our sources are expected to be AGNs given the \mbox{X-ray} depth. Host-galaxy properties for galaxies and AGNs will also be derived through SED fitting.

\facilities{CXO (ACIS), HST, JWST, Gaia, VISTA (VIRCAM), VST (OmegaCAM)}

\software{
CIAO v4.18 \citep{CIAO2006},
CALDB v4.12.4,
Sherpa \citep{Freeman2001, Siemiginowska2024},
Astropy \citep{Astropy2022},
NumPy \citep{Harris2020},
SciPy \citep{Virtanen2020},
Matplotlib \citep{Hunter2007},
FastHR \citep{Zou2023}
}

\begin{acknowledgments}
S.W. and X.-B.W. acknowledge support from the National Key R\&D Program of China (grant Nos. 2025YFA1614100 and 2025YFA1614101) and the National Natural Science Foundation of China (grant No. 12133001). WX acknowledges the support of National Nature Science Foundation of China (No 12588202, 12203063), the science research grants from the China Manned Space Program (No. CMS-CSST-2025-A03, CMS-CSST-2025-A20), CAS Project for Young Scientists in Basic Research (No. YSBR-062).
The Chandra data covering this field were obtained under programs led by \'Akos Bogd\'an, Laurence David, Gordon Garmire, and Joshua Kempner, and retrieved from the public Chandra Data Archive; most of these observations were from the 2~Ms Chandra Cycle~23 Very Large Program (PI: \'Akos Bogd\'an). We acknowledge these investigators as the originators of the observations on which this work depends. This work made use of data from the ESO Science Archive Facility
(DOIs: \dataset[10.18727/archive/37]{https://doi.org/10.18727/archive/37} and \dataset[10.18727/archive/59]{https://doi.org/10.18727/archive/59}) and of data
products generated by the Kilo-Degree Survey (KiDS) Consortium.
The KiDS data-production effort received support from the Deutsche
Forschungsgemeinschaft, the European Research Council, NOVA, NWO-M
grants, Target, the University of Padova, and the University of
Naples Federico II. Some of the data products presented herein were retrieved from the Dawn JWST Archive (DJA). DJA is an initiative of the Cosmic Dawn Center (DAWN), which is funded by the Danish National Research Foundation under grant DNRF140.
\end{acknowledgments}

\bibliography{citations}{}

@ARTICLE{Allen2011,
       author = {{Allen}, Steven W. and {Evrard}, August E. and {Mantz}, Adam B.},
        title = "{Cosmological Parameters from Observations of Galaxy Clusters}",
      journal = {\araa},
         year = 2011,
        month = sep,
       volume = {49},
       number = {1},
        pages = {409-470},
          doi = {10.1146/annurev-astro-081710-102514},
archivePrefix = {arXiv},
       eprint = {1103.4829},
 primaryClass = {astro-ph.CO},
       adsurl = {https://ui.adsabs.harvard.edu/abs/2011ARA&A..49..409A}
}

@ARTICLE{Alvarez-Marquez2026,
       author = {{{\'A}lvarez-M{\'a}rquez}, J. and {Colina}, L. and {Crespo-Gomez}, A. and {Kendrew}, S. and {Zavala}, J. and {Marques-Chaves}, R. and {Prieto-Jim{\'e}nez}, C. and {Abdurro'uf} and {Blanco-Prieto}, C. and {Boogaard}, L.~A. and {Castellano}, M. and {Fontana}, A. and {Fudamoto}, Y. and {Fujimoto}, S. and {Garc{\'\i}a-Mar{\'\i}n}, M. and {Harikane}, Y. and {Harish}, S. and {Hashimoto}, T. and {Hsiao}, T. and {Iani}, E. and {Inoue}, A.~K. and {Langeroodi}, D. and {Lin}, R. and {Melinder}, J. and {Napolitano}, L. and {Ostlin}, G. and {P{\'e}rez-Gonz{\'a}lez}, P.~G. and {Rinaldi}, P. and {Rodr{\'\i}guez Del Pino}, B. and {Santini}, P. and {Sugahara}, Y. and {Treu}, T. and {Varo-O'ferral}, A. and {Wright}, G.},
        title = "{PRISMS. UNCOVER-26185, a metal-poor SFG at z=10.05 with no evidence for a X-ray-luminous AGN}",
      journal = {arXiv e-prints},
         year = 2026,
        month = feb,
          eid = {arXiv:2602.02323},
        pages = {arXiv:2602.02323},
          doi = {10.48550/arXiv.2602.02323},
archivePrefix = {arXiv},
       eprint = {2602.02323},
 primaryClass = {astro-ph.GA},
       adsurl = {https://ui.adsabs.harvard.edu/abs/2026arXiv260202323A}
}

@ARTICLE{Atek2025,
       author = {{Atek}, Hakim and {Chisholm}, John and {Kokorev}, Vasily and {Endsley}, Ryan and {Pan}, Richard and {Furtak}, Lukas and {Chemerynska}, Iryna and {Richard}, Johan and {Claeyssens}, Ad{\'e}la{\"\i}de and {Oesch}, Pascal and {Fujimoto}, Seiji and {Naidu}, Rohan and {Korber}, Damien and {Schaerer}, Daniel and {Blaizot}, Jeremy and {Rosdahl}, Joki and {Adamo}, Angela and {Asada}, Yoshihisa and {Basu}, Arghyadeep and {Beauchesne}, Benjamin and {Berg}, Danielle and {Bezanson}, Rachel and {Bouwens}, Rychard and {Brammer}, Gabriel and {Dessauges-Zavadsky}, Miroslava and {Ellien}, Ama{\"e}l and {Ezziati}, Meriam and {Fei}, Qinyue and {Goovaerts}, Ilias and {Heurtier}, Sylvain and {Hsiao}, Tiger Yu-Yang and {Jecmen}, Michelle and {Khullar}, Gourav and {Kneib}, Jean-Paul and {Labb{\'e}}, Ivo and {Leclercq}, Floriane and {Marques-Chaves}, Rui and {Mason}, Charlotte and {McQuinn}, Kristen B.~W. and {Mu{\~n}oz}, Julian B. and {Natarajan}, Priyamvada and {Saldana-Lopez}, Alberto and {Stephenson}, Mabel G. and {Trebitsch}, Maxime and {Volonteri}, Marta and {Weibel}, Andrea and {Zitrin}, Adi},
        title = "{JWST's GLIMPSE: an overview of the deepest probe of early galaxy formation and cosmic reionization}",
      journal = {arXiv e-prints},
         year = 2025,
        month = nov,
          eid = {arXiv:2511.07542},
        pages = {arXiv:2511.07542},
          doi = {10.48550/arXiv.2511.07542},
archivePrefix = {arXiv},
       eprint = {2511.07542},
 primaryClass = {astro-ph.GA},
       adsurl = {https://ui.adsabs.harvard.edu/abs/2025arXiv251107542A}
}

@ARTICLE{Bezanson2024,
       author = {{Bezanson}, Rachel and {Labbe}, Ivo and {Whitaker}, Katherine E. and {Leja}, Joel and {Price}, Sedona H. and {Franx}, Marijn and {Brammer}, Gabriel and {Marchesini}, Danilo and {Zitrin}, Adi and {Wang}, Bingjie and {Weaver}, John R. and {Furtak}, Lukas J. and {Atek}, Hakim and {Coe}, Dan and {Cutler}, Sam E. and {Dayal}, Pratika and {van Dokkum}, Pieter and {Feldmann}, Robert and {F{\"o}rster Schreiber}, Natascha M. and {Fujimoto}, Seiji and {Geha}, Marla and {Glazebrook}, Karl and {de Graaff}, Anna and {Greene}, Jenny E. and {Juneau}, St{\'e}phanie and {Kassin}, Susan and {Kriek}, Mariska and {Khullar}, Gourav and {Maseda}, Michael and {Mowla}, Lamiya A. and {Muzzin}, Adam and {Nanayakkara}, Themiya and {Nelson}, Erica J. and {Oesch}, Pascal A. and {Pacifici}, Camilla and {Pan}, Richard and {Papovich}, Casey and {Setton}, David J. and {Shapley}, Alice E. and {Smit}, Renske and {Stefanon}, Mauro and {Taylor}, Edward N. and {Williams}, Christina C.},
        title = "{The JWST UNCOVER Treasury Survey: Ultradeep NIRSpec and NIRCam Observations before the Epoch of Reionization}",
      journal = {\apj},
         year = 2024,
        month = oct,
       volume = {974},
       number = {1},
          eid = {92},
        pages = {92},
          doi = {10.3847/1538-4357/ad66cf},
archivePrefix = {arXiv},
       eprint = {2212.04026},
 primaryClass = {astro-ph.GA},
       adsurl = {https://ui.adsabs.harvard.edu/abs/2024ApJ...974...92B}
}

@ARTICLE{Bogdan2024,
       author = {{Bogd{\'a}n}, {\'A}kos and {Goulding}, Andy D. and {Natarajan}, Priyamvada and {Kov{\'a}cs}, Orsolya E. and {Tremblay}, Grant R. and {Chadayammuri}, Urmila and {Volonteri}, Marta and {Kraft}, Ralph P. and {Forman}, William R. and {Jones}, Christine and {Churazov}, Eugene and {Zhuravleva}, Irina},
        title = "{Evidence for heavy-seed origin of early supermassive black holes from a z ≍ 10 X-ray quasar}",
      journal = {Nature Astronomy},
         year = 2024,
        month = jan,
       volume = {8},
       number = {1},
        pages = {126-133},
          doi = {10.1038/s41550-023-02111-9},
archivePrefix = {arXiv},
       eprint = {2305.15458},
 primaryClass = {astro-ph.GA},
       adsurl = {https://ui.adsabs.harvard.edu/abs/2024NatAs...8..126B}
}

@ARTICLE{Boyett2024,
       author = {{Boyett}, Kristan and {Trenti}, Michele and {Leethochawalit}, Nicha and {Calabr{\'o}}, Antonello and {Metha}, Benjamin and {Roberts-Borsani}, Guido and {Dalmasso}, Nicol{\'o} and {Yang}, Lilan and {Santini}, Paola and {Treu}, Tommaso and {Jones}, Tucker and {Henry}, Alaina and {Mason}, Charlotte A. and {Morishita}, Takahiro and {Nanayakkara}, Themiya and {Roy}, Namrata and {Wang}, Xin and {Fontana}, Adriano and {Merlin}, Emiliano and {Castellano}, Marco and {Paris}, Diego and {Brada{\v{c}}}, Maru{\v{s}}a and {Malkan}, Matt and {Marchesini}, Danilo and {Mascia}, Sara and {Glazebrook}, Karl and {Pentericci}, Laura and {Vanzella}, Eros and {Vulcani}, Benedetta},
        title = "{A massive interacting galaxy 510 million years after the Big Bang}",
      journal = {Nature Astronomy},
         year = 2024,
        month = may,
       volume = {8},
        pages = {657-672},
          doi = {10.1038/s41550-024-02218-7},
archivePrefix = {arXiv},
       eprint = {2303.00306},
 primaryClass = {astro-ph.GA},
       adsurl = {https://ui.adsabs.harvard.edu/abs/2024NatAs...8..657B}
}

@ARTICLE{Brandt2015,
       author = {{Brandt}, W.~N. and {Alexander}, D.~M.},
        title = "{Cosmic X-ray surveys of distant active galaxies. The demographics, physics, and ecology of growing supermassive black holes}",
      journal = {\aapr},
         year = 2015,
        month = jan,
       volume = {23},
          eid = {1},
        pages = {1},
          doi = {10.1007/s00159-014-0081-z},
archivePrefix = {arXiv},
       eprint = {1501.01982},
 primaryClass = {astro-ph.HE},
       adsurl = {https://ui.adsabs.harvard.edu/abs/2015A&ARv..23....1B}
}

@INCOLLECTION{Brandt2024,
       author = {{Brandt}, W.~N. and {Yang}, G.},
        title = "{Surveys of the Cosmic X-Ray Background}",
    booktitle = {Handbook of X-ray and Gamma-ray Astrophysics},
         year = 2024,
        month = jan,
        pages = {5233--5267},
    publisher = {Springer Nature},
          doi = {10.1007/978-981-19-6960-7_130}
}

@ARTICLE{Chadayammuri2024,
       author = {{Chadayammuri}, Urmila and {Bogd{\'a}n}, {\'A}kos and {Schellenberger}, Gerrit and {ZuHone}, John},
        title = "{Closing Pandora's Box -- The deepest X-ray observations of Abell 2744 and a multi-wavelength merger picture}",
      journal = {arXiv e-prints},
         year = 2024,
        month = jul,
          eid = {arXiv:2407.03142},
        pages = {arXiv:2407.03142},
          doi = {10.48550/arXiv.2407.03142},
archivePrefix = {arXiv},
       eprint = {2407.03142},
 primaryClass = {astro-ph.CO},
       adsurl = {https://ui.adsabs.harvard.edu/abs/2024arXiv240703142C}
}

@ARTICLE{Edge2013,
       author = {{Edge}, A. and {Sutherland}, W. and {Kuijken}, K. and {Driver}, S. and {McMahon}, R. and {Eales}, S. and {Emerson}, J.~P.},
        title = "{The VISTA Kilo-degree Infrared Galaxy (VIKING) Survey: Bridging the Gap between Low and High Redshift}",
      journal = {The Messenger},
         year = 2013,
        month = dec,
       volume = {154},
        pages = {32-34},
       adsurl = {https://ui.adsabs.harvard.edu/abs/2013Msngr.154...32E}
}

@ARTICLE{Ehlert2013,
       author = {{Ehlert}, S. and {Allen}, S.~W. and {Brandt}, W.~N. and {Xue}, Y.~Q. and {Luo}, B. and {von der Linden}, A. and {Mantz}, A. and {Morris}, R.~G.},
        title = "{X-ray bright active galactic nuclei in massive galaxy clusters - I. Number counts and spatial distribution}",
      journal = {\mnras},
         year = 2013,
        month = feb,
       volume = {428},
       number = {4},
        pages = {3509-3525},
          doi = {10.1093/mnras/sts288},
archivePrefix = {arXiv},
       eprint = {1209.2132},
 primaryClass = {astro-ph.CO},
       adsurl = {https://ui.adsabs.harvard.edu/abs/2013MNRAS.428.3509E}
}

@ARTICLE{Evans2024,
       author = {{Evans}, Ian N. and {Evans}, Janet D. and {Mart{\'\i}nez-Galarza}, J. Rafael and {Miller}, Joseph B. and {Primini}, Francis A. and {Azadi}, Mojegan and {Burke}, Douglas J. and {Civano}, Francesca M. and {D'Abrusco}, Raffaele and {Fabbiano}, Giuseppina and {Graessle}, Dale E. and {Grier}, John D. and {Houck}, John C. and {Lauer}, Jennifer and {McCollough}, Michael L. and {Nowak}, Michael A. and {Plummer}, David A. and {Rots}, Arnold H. and {Siemiginowska}, Aneta and {Tibbetts}, Michael S.},
        title = "{The Chandra Source Catalog Release 2 Series}",
      journal = {\apjs},
         year = 2024,
        month = oct,
       volume = {274},
       number = {2},
          eid = {22},
        pages = {22},
          doi = {10.3847/1538-4365/ad6319},
archivePrefix = {arXiv},
       eprint = {2407.10799},
 primaryClass = {astro-ph.HE},
       adsurl = {https://ui.adsabs.harvard.edu/abs/2024ApJS..274...22E}
}

@ARTICLE{Fu2025,
       author = {{Fu}, Shuqi and {Sun}, Fengwu and {Jiang}, Linhua and {Lin}, Xiaojing and {Diego}, Jose M. and {Furtak}, Lukas J. and {Jauzac}, Mathilde and {Koekemoer}, Anton M. and {Li}, Mingyu and {Oguri}, Masamune and {Patel}, Nency R. and {Willmer}, Christopher N.~A. and {Windhorst}, Rogier A. and {Zitrin}, Adi and {Bauer}, Franz E. and {Chen}, Chian-Chou and {Chen}, Wenlei and {Cheng}, Cheng and {Conselice}, Christopher J. and {Eisenstein}, Daniel J. and {Egami}, Eiichi and {Espada}, Daniel and {Fan}, Xiaohui and {Fujimoto}, Seiji and {Hsiao}, Tiger Yu-Yang and {Jin}, Xiangyu and {Kohno}, Kotaro and {Lagattuta}, David J. and {Li}, Zihao and {Liu}, Weizhe and {Miralda-Escud{\'e}}, Jordi and {Ning}, Yuanhang and {Tacchella}, Sandro and {Tee}, Wei Leong and {Umehata}, Hideki and {Wang}, Feige and {Yan}, Haojing and {Zhu}, Yongda},
        title = "{Medium-band Astrophysics with the Grism of NIRCam In Frontier Fields (MAGNIF): Spectroscopic Census of H{\ensuremath{\alpha}} Luminosity Functions and Cosmic Star Formation at z {\ensuremath{\sim}} 4.5 and 6.3}",
      journal = {\apj},
         year = 2025,
        month = jul,
       volume = {987},
       number = {2},
          eid = {186},
        pages = {186},
          doi = {10.3847/1538-4357/adddb1},
archivePrefix = {arXiv},
       eprint = {2503.03829},
 primaryClass = {astro-ph.GA},
       adsurl = {https://ui.adsabs.harvard.edu/abs/2025ApJ...987..186F}
}

@ARTICLE{Furtak2023,
       author = {{Furtak}, Lukas J. and {Zitrin}, Adi and {Weaver}, John R. and {Atek}, Hakim and {Bezanson}, Rachel and {Labb{\'e}}, Ivo and {Whitaker}, Katherine E. and {Leja}, Joel and {Price}, Sedona H. and {Brammer}, Gabriel B. and {Wang}, Bingjie and {Marchesini}, Danilo and {Pan}, Richard and {Dayal}, Pratika and {van Dokkum}, Pieter and {Feldmann}, Robert and {Fujimoto}, Seiji and {Franx}, Marijn and {Khullar}, Gourav and {Nelson}, Erica J. and {Mowla}, Lamiya A.},
        title = "{UNCOVERing the extended strong lensing structures of Abell 2744 with the deepest JWST imaging}",
      journal = {\mnras},
         year = 2023,
        month = aug,
       volume = {523},
       number = {3},
        pages = {4568-4582},
          doi = {10.1093/mnras/stad1627},
archivePrefix = {arXiv},
       eprint = {2212.04381},
 primaryClass = {astro-ph.GA},
       adsurl = {https://ui.adsabs.harvard.edu/abs/2023MNRAS.523.4568F}
}

@ARTICLE{GaiaCollaboration2023,
       author = {{Gaia Collaboration} and {Vallenari}, A. and {Brown}, A.~G.~A. and {Prusti}, T. and {de Bruijne}, J.~H.~J. and {Arenou}, F. and {Babusiaux}, C. and {Biermann}, M. and {Creevey}, O.~L. and {Ducourant}, C. and {Evans}, D.~W. and {Eyer}, L. and {Guerra}, R. and {Hutton}, A. and {Jordi}, C. and {Klioner}, S.~A. and {Lammers}, U.~L. and {Lindegren}, L. and {Luri}, X. and {Mignard}, F. and {Panem}, C. and {Pourbaix}, D. and {Randich}, S. and {Sartoretti}, P. and {Soubiran}, C. and {Tanga}, P. and {Walton}, N.~A. and {Bailer-Jones}, C.~A.~L. and {Bastian}, U. and {Drimmel}, R. and {Jansen}, F. and {Katz}, D. and {Lattanzi}, M.~G. and {van Leeuwen}, F. and {Bakker}, J. and {Cacciari}, C. and {Casta{\~n}eda}, J. and {De Angeli}, F. and {Fabricius}, C. and {Fouesneau}, M. and {Fr{\'e}mat}, Y. and {Galluccio}, L. and {Guerrier}, A. and {Heiter}, U. and {Masana}, E. and {Messineo}, R. and {Mowlavi}, N. and {Nicolas}, C. and {Nienartowicz}, K. and {Pailler}, F. and {Panuzzo}, P. and {Riclet}, F. and {Roux}, W. and {Seabroke}, G.~M. and {Sordo}, R. and {Th{\'e}venin}, F. and {Gracia-Abril}, G. and {Portell}, J. and {Teyssier}, D. and {Altmann}, M. and {Andrae}, R. and {Audard}, M. and {Bellas-Velidis}, I. and {Benson}, K. and {Berthier}, J. and {Blomme}, R. and {Burgess}, P.~W. and {Busonero}, D. and {Busso}, G. and {C{\'a}novas}, H. and {Carry}, B. and {Cellino}, A. and {Cheek}, N. and {Clementini}, G. and {Damerdji}, Y. and {Davidson}, M. and {de Teodoro}, P. and {Nu{\~n}ez Campos}, M. and {Delchambre}, L. and {Dell'Oro}, A. and {Esquej}, P. and {Fern{\'a}ndez-Hern{\'a}ndez}, J. and {Fraile}, E. and {Garabato}, D. and {Garc{\'\i}a-Lario}, P. and {Gosset}, E. and {Haigron}, R. and {Halbwachs}, J.-L. and {Hambly}, N.~C. and {Harrison}, D.~L. and {Hern{\'a}ndez}, J. and {Hestroffer}, D. and {Hodgkin}, S.~T. and {Holl}, B. and {Jan{\ss}en}, K. and {Jevardat de Fombelle}, G. and {Jordan}, S. and {Krone-Martins}, A. and {Lanzafame}, A.~C. and {L{\"o}ffler}, W. and {Marchal}, O. and {Marrese}, P.~M. and {Moitinho}, A. and {Muinonen}, K. and {Osborne}, P. and {Pancino}, E. and {Pauwels}, T. and {Recio-Blanco}, A. and {Reyl{\'e}}, C. and {Riello}, M. and {Rimoldini}, L. and {Roegiers}, T. and {Rybizki}, J. and {Sarro}, L.~M. and {Siopis}, C. and {Smith}, M. and {Sozzetti}, A. and {Utrilla}, E. and {van Leeuwen}, M. and {Abbas}, U. and {{\'A}brah{\'a}m}, P. and {Abreu Aramburu}, A. and {Aerts}, C. and {Aguado}, J.~J. and {Ajaj}, M. and {Aldea-Montero}, F. and {Altavilla}, G. and {{\'A}lvarez}, M.~A. and {Alves}, J. and {Anders}, F. and {Anderson}, R.~I. and {Anglada Varela}, E. and {Antoja}, T. and {Baines}, D. and {Baker}, S.~G. and {Balaguer-N{\'u}{\~n}ez}, L. and {Balbinot}, E. and {Balog}, Z. and {Barache}, C. and {Barbato}, D. and {Barros}, M. and {Barstow}, M.~A. and {Bartolom{\'e}}, S. and {Bassilana}, J.-L. and {Bauchet}, N. and {Becciani}, U. and {Bellazzini}, M. and {Berihuete}, A. and {Bernet}, M. and {Bertone}, S. and {Bianchi}, L. and {Binnenfeld}, A. and {Blanco-Cuaresma}, S. and {Blazere}, A. and {Boch}, T. and {Bombrun}, A. and {Bossini}, D. and {Bouquillon}, S. and {Bragaglia}, A. and {Bramante}, L. and {Breedt}, E. and {Bressan}, A. and {Brouillet}, N. and {Brugaletta}, E. and {Bucciarelli}, B. and {Burlacu}, A. and {Butkevich}, A.~G. and {Buzzi}, R. and {Caffau}, E. and {Cancelliere}, R. and {Cantat-Gaudin}, T. and {Carballo}, R. and {Carlucci}, T. and {Carnerero}, M.~I. and {Carrasco}, J.~M. and {Casamiquela}, L. and {Castellani}, M. and {Castro-Ginard}, A. and {Chaoul}, L. and {Charlot}, P. and {Chemin}, L. and {Chiaramida}, V. and {Chiavassa}, A. and {Chornay}, N. and {Comoretto}, G. and {Contursi}, G. and {Cooper}, W.~J. and {Cornez}, T. and {Cowell}, S. and {Crifo}, F. and {Cropper}, M. and {Crosta}, M. and {Crowley}, C. and {Dafonte}, C. and {Dapergolas}, A. and {David}, M. and {David}, P. and {de Laverny}, P. and {De Luise}, F. and {De March}, R.},
        title = "{Gaia Data Release 3. Summary of the content and survey properties}",
      journal = {\aap},
         year = 2023,
        month = jun,
       volume = {674},
          eid = {A1},
        pages = {A1},
          doi = {10.1051/0004-6361/202243940},
archivePrefix = {arXiv},
       eprint = {2208.00211},
 primaryClass = {astro-ph.GA},
       adsurl = {https://ui.adsabs.harvard.edu/abs/2023A&A...674A...1G}
}

@ARTICLE{HI4PICollaboration2016,
       author = {{HI4PI Collaboration} and {Ben Bekhti}, N. and {Fl{\"o}er}, L. and {Keller}, R. and {Kerp}, J. and {Lenz}, D. and {Winkel}, B. and {Bailin}, J. and {Calabretta}, M.~R. and {Dedes}, L. and {Ford}, H.~A. and {Gibson}, B.~K. and {Haud}, U. and {Janowiecki}, S. and {Kalberla}, P.~M.~W. and {Lockman}, F.~J. and {McClure-Griffiths}, N.~M. and {Murphy}, T. and {Nakanishi}, H. and {Pisano}, D.~J. and {Staveley-Smith}, L.},
        title = "{HI4PI: A full-sky H I survey based on EBHIS and GASS}",
      journal = {\aap},
         year = 2016,
        month = oct,
       volume = {594},
          eid = {A116},
        pages = {A116},
          doi = {10.1051/0004-6361/201629178},
archivePrefix = {arXiv},
       eprint = {1610.06175},
 primaryClass = {astro-ph.GA},
       adsurl = {https://ui.adsabs.harvard.edu/abs/2016A&A...594A.116H}
}

@ARTICLE{Kovacs2024,
       author = {{Kov{\'a}cs}, Orsolya E. and {Bogd{\'a}n}, {\'A}kos and {Natarajan}, Priyamvada and {Werner}, Norbert and {Azadi}, Mojegan and {Volonteri}, Marta and {Tremblay}, Grant R. and {Chadayammuri}, Urmila and {Forman}, William R. and {Jones}, Christine and {Kraft}, Ralph P.},
        title = "{A Candidate Supermassive Black Hole in a Gravitationally Lensed Galaxy at Z {\ensuremath{\approx}} 10}",
      journal = {\apjl},
         year = 2024,
        month = apr,
       volume = {965},
       number = {2},
          eid = {L21},
        pages = {L21},
          doi = {10.3847/2041-8213/ad391f},
archivePrefix = {arXiv},
       eprint = {2403.14745},
 primaryClass = {astro-ph.GA},
       adsurl = {https://ui.adsabs.harvard.edu/abs/2024ApJ...965L..21K}
}

@ARTICLE{Lehmer2016,
       author = {{Lehmer}, B.~D. and {Basu-Zych}, A.~R. and {Mineo}, S. and {Brandt}, W.~N. and {Eufrasio}, R.~T. and {Fragos}, T. and {Hornschemeier}, A.~E. and {Luo}, B. and {Xue}, Y.~Q. and {Bauer}, F.~E. and {Gilfanov}, M. and {Ranalli}, P. and {Schneider}, D.~P. and {Shemmer}, O. and {Tozzi}, P. and {Trump}, J.~R. and {Vignali}, C. and {Wang}, J.-X. and {Yukita}, M. and {Zezas}, A.},
        title = "{The Evolution of Normal Galaxy X-Ray Emission through Cosmic History: Constraints from the 6 MS Chandra Deep Field-South}",
      journal = {\apj},
         year = 2016,
        month = jul,
       volume = {825},
       number = {1},
          eid = {7},
        pages = {7},
          doi = {10.3847/0004-637X/825/1/7},
archivePrefix = {arXiv},
       eprint = {1604.06461},
 primaryClass = {astro-ph.GA},
       adsurl = {https://ui.adsabs.harvard.edu/abs/2016ApJ...825....7L}
}

@ARTICLE{Lotz2017,
       author = {{Lotz}, J.~M. and {Koekemoer}, A. and {Coe}, D. and {Grogin}, N. and {Capak}, P. and {Mack}, J. and {Anderson}, J. and {Avila}, R. and {Barker}, E.~A. and {Borncamp}, D. and {Brammer}, G. and {Durbin}, M. and {Gunning}, H. and {Hilbert}, B. and {Jenkner}, H. and {Khandrika}, H. and {Levay}, Z. and {Lucas}, R.~A. and {MacKenty}, J. and {Ogaz}, S. and {Porterfield}, B. and {Reid}, N. and {Robberto}, M. and {Royle}, P. and {Smith}, L.~J. and {Storrie-Lombardi}, L.~J. and {Sunnquist}, B. and {Surace}, J. and {Taylor}, D.~C. and {Williams}, R. and {Bullock}, J. and {Dickinson}, M. and {Finkelstein}, S. and {Natarajan}, P. and {Richard}, J. and {Robertson}, B. and {Tumlinson}, J. and {Zitrin}, A. and {Flanagan}, K. and {Sembach}, K. and {Soifer}, B.~T. and {Mountain}, M.},
        title = "{The Frontier Fields: Survey Design and Initial Results}",
      journal = {\apj},
         year = 2017,
        month = mar,
       volume = {837},
       number = {1},
          eid = {97},
        pages = {97},
          doi = {10.3847/1538-4357/837/1/97},
archivePrefix = {arXiv},
       eprint = {1605.06567},
 primaryClass = {astro-ph.GA},
       adsurl = {https://ui.adsabs.harvard.edu/abs/2017ApJ...837...97L}
}

@ARTICLE{Luo2017,
       author = {{Luo}, B. and {Brandt}, W.~N. and {Xue}, Y.~Q. and {Lehmer}, B. and {Alexander}, D.~M. and {Bauer}, F.~E. and {Vito}, F. and {Yang}, G. and {Basu-Zych}, A.~R. and {Comastri}, A. and {Gilli}, R. and {Gu}, Q.-S. and {Hornschemeier}, A.~E. and {Koekemoer}, A. and {Liu}, T. and {Mainieri}, V. and {Paolillo}, M. and {Ranalli}, P. and {Rosati}, P. and {Schneider}, D.~P. and {Shemmer}, O. and {Smail}, I. and {Sun}, M. and {Tozzi}, P. and {Vignali}, C. and {Wang}, J.-X.},
        title = "{The Chandra Deep Field-South Survey: 7 Ms Source Catalogs}",
      journal = {\apjs},
         year = 2017,
        month = jan,
       volume = {228},
       number = {1},
          eid = {2},
        pages = {2},
          doi = {10.3847/1538-4365/228/1/2},
archivePrefix = {arXiv},
       eprint = {1611.03501},
 primaryClass = {astro-ph.GA},
       adsurl = {https://ui.adsabs.harvard.edu/abs/2017ApJS..228....2L}
}

@ARTICLE{Nandra2015,
       author = {{Nandra}, K. and {Laird}, E.~S. and {Aird}, J.~A. and {Salvato}, M. and {Georgakakis}, A. and {Barro}, G. and {Perez-Gonzalez}, P.~G. and {Barmby}, P. and {Chary}, R.-R. and {Coil}, A. and {Cooper}, M.~C. and {Davis}, M. and {Dickinson}, M. and {Faber}, S.~M. and {Fazio}, G.~G. and {Guhathakurta}, P. and {Gwyn}, S. and {Hsu}, L.-T. and {Huang}, J.-S. and {Ivison}, R.~J. and {Koo}, D.~C. and {Newman}, J.~A. and {Rangel}, C. and {Yamada}, T. and {Willmer}, C.},
        title = "{AEGIS-X: Deep Chandra Imaging of the Central Groth Strip}",
      journal = {\apjs},
         year = 2015,
        month = sep,
       volume = {220},
       number = {1},
          eid = {10},
        pages = {10},
          doi = {10.1088/0067-0049/220/1/10},
archivePrefix = {arXiv},
       eprint = {1503.09078},
 primaryClass = {astro-ph.HE},
       adsurl = {https://ui.adsabs.harvard.edu/abs/2015ApJS..220...10N}
}

@ARTICLE{Ni2021,
       author = {{Ni}, Qingling and {Brandt}, W.~N. and {Chen}, Chien-Ting and {Luo}, Bin and {Nyland}, Kristina and {Yang}, Guang and {Zou}, Fan and {Aird}, James and {Alexander}, David M. and {Bauer}, Franz Erik and {Lacy}, Mark and {Lehmer}, Bret D. and {Mallick}, Labani and {Salvato}, Mara and {Schneider}, Donald P. and {Tozzi}, Paolo and {Traulsen}, Iris and {Vaccari}, Mattia and {Vignali}, Cristian and {Vito}, Fabio and {Xue}, Yongquan and {Banerji}, Manda and {Chow}, Kate and {Comastri}, Andrea and {Del Moro}, Agnese and {Gilli}, Roberto and {Mullaney}, James and {Paolillo}, Maurizio and {Schwope}, Axel and {Shemmer}, Ohad and {Sun}, Mouyuan and {Timlin}, III, John D. and {Trump}, Jonathan R.},
        title = "{The XMM-SERVS Survey: XMM-Newton Point-source Catalogs for the W-CDF-S and ELAIS-S1 Fields}",
      journal = {\apjs},
         year = 2021,
        month = sep,
       volume = {256},
       number = {1},
          eid = {21},
        pages = {21},
          doi = {10.3847/1538-4365/ac0dc6},
archivePrefix = {arXiv},
       eprint = {2106.10572},
 primaryClass = {astro-ph.GA},
       adsurl = {https://ui.adsabs.harvard.edu/abs/2021ApJS..256...21N}
}

@ARTICLE{PlanckCollaboration2020,
       author = {{Planck Collaboration} and {Aghanim}, N. and {Akrami}, Y. and {Ashdown}, M. and {Aumont}, J. and {Baccigalupi}, C. and {Ballardini}, M. and {Banday}, A.~J. and {Barreiro}, R.~B. and {Bartolo}, N. and {Basak}, S. and {Battye}, R. and {Benabed}, K. and {Bernard}, J.-P. and {Bersanelli}, M. and {Bielewicz}, P. and {Bock}, J.~J. and {Bond}, J.~R. and {Borrill}, J. and {Bouchet}, F.~R. and {Boulanger}, F. and {Bucher}, M. and {Burigana}, C. and {Butler}, R.~C. and {Calabrese}, E. and {Cardoso}, J.-F. and {Carron}, J. and {Challinor}, A. and {Chiang}, H.~C. and {Chluba}, J. and {Colombo}, L.~P.~L. and {Combet}, C. and {Contreras}, D. and {Crill}, B.~P. and {Cuttaia}, F. and {de Bernardis}, P. and {de Zotti}, G. and {Delabrouille}, J. and {Delouis}, J.-M. and {Di Valentino}, E. and {Diego}, J.~M. and {Dor{\'e}}, O. and {Douspis}, M. and {Ducout}, A. and {Dupac}, X. and {Dusini}, S. and {Efstathiou}, G. and {Elsner}, F. and {En{\ss}lin}, T.~A. and {Eriksen}, H.~K. and {Fantaye}, Y. and {Farhang}, M. and {Fergusson}, J. and {Fernandez-Cobos}, R. and {Finelli}, F. and {Forastieri}, F. and {Frailis}, M. and {Fraisse}, A.~A. and {Franceschi}, E. and {Frolov}, A. and {Galeotta}, S. and {Galli}, S. and {Ganga}, K. and {G{\'e}nova-Santos}, R.~T. and {Gerbino}, M. and {Ghosh}, T. and {Gonz{\'a}lez-Nuevo}, J. and {G{\'o}rski}, K.~M. and {Gratton}, S. and {Gruppuso}, A. and {Gudmundsson}, J.~E. and {Hamann}, J. and {Handley}, W. and {Hansen}, F.~K. and {Herranz}, D. and {Hildebrandt}, S.~R. and {Hivon}, E. and {Huang}, Z. and {Jaffe}, A.~H. and {Jones}, W.~C. and {Karakci}, A. and {Keih{\"a}nen}, E. and {Keskitalo}, R. and {Kiiveri}, K. and {Kim}, J. and {Kisner}, T.~S. and {Knox}, L. and {Krachmalnicoff}, N. and {Kunz}, M. and {Kurki-Suonio}, H. and {Lagache}, G. and {Lamarre}, J.-M. and {Lasenby}, A. and {Lattanzi}, M. and {Lawrence}, C.~R. and {Le Jeune}, M. and {Lemos}, P. and {Lesgourgues}, J. and {Levrier}, F. and {Lewis}, A. and {Liguori}, M. and {Lilje}, P.~B. and {Lilley}, M. and {Lindholm}, V. and {L{\'o}pez-Caniego}, M. and {Lubin}, P.~M. and {Ma}, Y.-Z. and {Mac{\'\i}as-P{\'e}rez}, J.~F. and {Maggio}, G. and {Maino}, D. and {Mandolesi}, N. and {Mangilli}, A. and {Marcos-Caballero}, A. and {Maris}, M. and {Martin}, P.~G. and {Martinelli}, M. and {Mart{\'\i}nez-Gonz{\'a}lez}, E. and {Matarrese}, S. and {Mauri}, N. and {McEwen}, J.~D. and {Meinhold}, P.~R. and {Melchiorri}, A. and {Mennella}, A. and {Migliaccio}, M. and {Millea}, M. and {Mitra}, S. and {Miville-Desch{\^e}nes}, M.-A. and {Molinari}, D. and {Montier}, L. and {Morgante}, G. and {Moss}, A. and {Natoli}, P. and {N{\o}rgaard-Nielsen}, H.~U. and {Pagano}, L. and {Paoletti}, D. and {Partridge}, B. and {Patanchon}, G. and {Peiris}, H.~V. and {Perrotta}, F. and {Pettorino}, V. and {Piacentini}, F. and {Polastri}, L. and {Polenta}, G. and {Puget}, J.-L. and {Rachen}, J.~P. and {Reinecke}, M. and {Remazeilles}, M. and {Renzi}, A. and {Rocha}, G. and {Rosset}, C. and {Roudier}, G. and {Rubi{\~n}o-Mart{\'\i}n}, J.~A. and {Ruiz-Granados}, B. and {Salvati}, L. and {Sandri}, M. and {Savelainen}, M. and {Scott}, D. and {Shellard}, E.~P.~S. and {Sirignano}, C. and {Sirri}, G. and {Spencer}, L.~D. and {Sunyaev}, R. and {Suur-Uski}, A.-S. and {Tauber}, J.~A. and {Tavagnacco}, D. and {Tenti}, M. and {Toffolatti}, L. and {Tomasi}, M. and {Trombetti}, T. and {Valenziano}, L. and {Valiviita}, J. and {Van Tent}, B. and {Vibert}, L. and {Vielva}, P. and {Villa}, F. and {Vittorio}, N. and {Wandelt}, B.~D. and {Wehus}, I.~K. and {White}, M. and {White}, S.~D.~M. and {Zacchei}, A. and {Zonca}, A.},
        title = "{Planck 2018 results. VI. Cosmological parameters}",
      journal = {\aap},
         year = 2020,
        month = sep,
       volume = {641},
          eid = {A6},
        pages = {A6},
          doi = {10.1051/0004-6361/201833910},
archivePrefix = {arXiv},
       eprint = {1807.06209},
 primaryClass = {astro-ph.CO},
       adsurl = {https://ui.adsabs.harvard.edu/abs/2020A&A...641A...6P}
}

@ARTICLE{Salvato2018,
       author = {{Salvato}, M. and {Buchner}, J. and {Budav{\'a}ri}, T. and {Dwelly}, T. and {Merloni}, A. and {Brusa}, M. and {Rau}, A. and {Fotopoulou}, S. and {Nandra}, K.},
        title = "{Finding counterparts for all-sky X-ray surveys with NWAY: a Bayesian algorithm for cross-matching multiple catalogues}",
      journal = {\mnras},
         year = 2018,
        month = feb,
       volume = {473},
       number = {4},
        pages = {4937-4955},
          doi = {10.1093/mnras/stx2651},
archivePrefix = {arXiv},
       eprint = {1705.10711},
 primaryClass = {astro-ph.GA},
       adsurl = {https://ui.adsabs.harvard.edu/abs/2018MNRAS.473.4937S}
}

@ARTICLE{Siemiginowska2024,
       author = {{Siemiginowska}, Aneta and {Burke}, Douglas and {G{\"u}nther}, Hans Moritz and {Lee}, Nicholas P. and {McLaughlin}, Warren and {Principe}, David A. and {Cheer}, Harlan and {Fruscione}, Antonella and {Laurino}, Omar and {McDowell}, Jonathan and {Terrell}, Marie},
        title = "{Sherpa: An Open-source Python Fitting Package}",
      journal = {\apjs},
         year = 2024,
        month = oct,
       volume = {274},
       number = {2},
          eid = {43},
        pages = {43},
          doi = {10.3847/1538-4365/ad7bab},
archivePrefix = {arXiv},
       eprint = {2409.10400},
 primaryClass = {astro-ph.IM},
       adsurl = {https://ui.adsabs.harvard.edu/abs/2024ApJS..274...43S}
}

@ARTICLE{Slane2025,
       author = {{Slane}, Patrick and {Bogd{\'a}n}, {\'A}kos and {Pooley}, David},
        title = "{25 years of groundbreaking discoveries with Chandra}",
      journal = {Nature Astronomy},
         year = 2025,
        month = oct,
       volume = {9},
        pages = {1431-1443},
          doi = {10.1038/s41550-025-02675-8},
archivePrefix = {arXiv},
       eprint = {2510.25873},
 primaryClass = {astro-ph.HE},
       adsurl = {https://ui.adsabs.harvard.edu/abs/2025NatAs...9.1431S}
}

@ARTICLE{Steinhardt2020,
       author = {{Steinhardt}, Charles L. and {Jauzac}, Mathilde and {Acebron}, Ana and {Atek}, Hakim and {Capak}, Peter and {Davidzon}, Iary and {Eckert}, Dominique and {Harvey}, David and {Koekemoer}, Anton M. and {Lagos}, Claudia D.~P. and {Mahler}, Guillaume and {Montes}, Mireia and {Niemiec}, Anna and {Nonino}, Mario and {Oesch}, P.~A. and {Richard}, Johan and {Rodney}, Steven A. and {Schaller}, Matthieu and {Sharon}, Keren and {Strolger}, Louis-Gregory and {Allingham}, Joseph and {Amara}, Adam and {Bah{\'e}}, Yannick and {B{\oe}hm}, C{\'e}line and {Bose}, Sownak and {Bouwens}, Rychard J. and {Bradley}, Larry D. and {Brammer}, Gabriel and {Broadhurst}, Tom and {Ca{\~n}as}, Rodrigo and {Cen}, Renyue and {Cl{\'e}ment}, Benjamin and {Clowe}, Douglas and {Coe}, Dan and {Connor}, Thomas and {Darvish}, Behnam and {Diego}, Jose M. and {Ebeling}, Harald and {Edge}, A.~C. and {Egami}, Eiichi and {Ettori}, Stefano and {Faisst}, Andreas L. and {Frye}, Brenda and {Furtak}, Lukas J. and {G{\'o}mez-Guijarro}, C. and {Remolina Gonz{\'a}lez}, J.~D. and {Gonzalez}, Anthony and {Graur}, Or and {Gruen}, Daniel and {Harvey}, David and {Hensley}, Hagan and {Hovis-Afflerbach}, Beryl and {Jablonka}, Pascale and {Jha}, Saurabh W. and {Jullo}, Eric and {Kneib}, Jean-Paul and {Kokorev}, Vasily and {Lagattuta}, David J. and {Limousin}, Marceau and {von der Linden}, Anja and {Linzer}, Nora B. and {Lopez}, Adrian and {Magdis}, Georgios E. and {Massey}, Richard and {Masters}, Daniel C. and {Maturi}, Matteo and {McCully}, Curtis and {McGee}, Sean L. and {Meneghetti}, Massimo and {Mobasher}, Bahram and {Moustakas}, Leonidas A. and {Murphy}, Eric J. and {Natarajan}, Priyamvada and {Neyrinck}, Mark and {O'Connor}, Kyle and {Oguri}, Masamune and {Pagul}, Amanda and {Rhodes}, Jason and {Rich}, R. Michael and {Robertson}, Andrew and {Sereno}, Mauro and {Shan}, Huanyuan and {Smith}, Graham P. and {Sneppen}, Albert and {Squires}, Gordon K. and {Tam}, Sut-Ieng and {Tchernin}, C{\'e}line and {Toft}, Sune and {Umetsu}, Keiichi and {Weaver}, John R. and {van Weeren}, R.~J. and {Williams}, Liliya L.~R. and {Wilson}, Tom J. and {Yan}, Lin and {Zitrin}, Adi},
        title = "{The BUFFALO HST Survey}",
      journal = {\apjs},
         year = 2020,
        month = apr,
       volume = {247},
       number = {2},
          eid = {64},
        pages = {64},
          doi = {10.3847/1538-4365/ab75ed},
archivePrefix = {arXiv},
       eprint = {2001.09999},
 primaryClass = {astro-ph.GA},
       adsurl = {https://ui.adsabs.harvard.edu/abs/2020ApJS..247...64S}
}

@ARTICLE{Treu2010,
       author = {{Treu}, Tommaso},
        title = "{Strong Lensing by Galaxies}",
      journal = {\araa},
         year = 2010,
        month = sep,
       volume = {48},
        pages = {87-125},
          doi = {10.1146/annurev-astro-081309-130924},
archivePrefix = {arXiv},
       eprint = {1003.5567},
 primaryClass = {astro-ph.CO},
       adsurl = {https://ui.adsabs.harvard.edu/abs/2010ARA&A..48...87T}
}

@ARTICLE{Treu2022,
       author = {{Treu}, T. and {Roberts-Borsani}, G. and {Bradac}, M. and {Brammer}, G. and {Fontana}, A. and {Henry}, A. and {Mason}, C. and {Morishita}, T. and {Pentericci}, L. and {Wang}, X. and {Acebron}, A. and {Bagley}, M. and {Bergamini}, P. and {Belfiori}, D. and {Bonchi}, A. and {Boyett}, K. and {Boutsia}, K. and {Calabr{\'o}}, A. and {Caminha}, G.~B. and {Castellano}, M. and {Dressler}, A. and {Glazebrook}, K. and {Grillo}, C. and {Jacobs}, C. and {Jones}, T. and {Kelly}, P.~L. and {Leethochawalit}, N. and {Malkan}, M.~A. and {Marchesini}, D. and {Mascia}, S. and {Mercurio}, A. and {Merlin}, E. and {Nanayakkara}, T. and {Nonino}, M. and {Paris}, D. and {Poggianti}, B. and {Rosati}, P. and {Santini}, P. and {Scarlata}, C. and {Shipley}, H.~V. and {Strait}, V. and {Trenti}, M. and {Tubthong}, C. and {Vanzella}, E. and {Vulcani}, B. and {Yang}, L.},
        title = "{The GLASS-JWST Early Release Science Program. I. Survey Design and Release Plans}",
      journal = {\apj},
         year = 2022,
        month = aug,
       volume = {935},
       number = {2},
          eid = {110},
        pages = {110},
          doi = {10.3847/1538-4357/ac8158},
archivePrefix = {arXiv},
       eprint = {2206.07978},
 primaryClass = {astro-ph.GA},
       adsurl = {https://ui.adsabs.harvard.edu/abs/2022ApJ...935..110T}
}

@ARTICLE{Weisskopf2007,
       author = {{Weisskopf}, Martin C. and {Wu}, Kinwah and {Trimble}, Virginia and {O'Dell}, Stephen L. and {Elsner}, Ronald F. and {Zavlin}, Vyacheslav E. and {Kouveliotou}, Chryssa},
        title = "{A Chandra Search for Coronal X-Rays from the Cool White Dwarf GD 356}",
      journal = {\apj},
         year = 2007,
        month = mar,
       volume = {657},
       number = {2},
        pages = {1026-1036},
          doi = {10.1086/510776},
archivePrefix = {arXiv},
       eprint = {astro-ph/0609585},
 primaryClass = {astro-ph},
       adsurl = {https://ui.adsabs.harvard.edu/abs/2007ApJ...657.1026W}
}

@ARTICLE{Xue2011,
       author = {{Xue}, Y.~Q. and {Luo}, B. and {Brandt}, W.~N. and others},
        title = "{The Chandra Deep Field-South Survey: 4 Ms Source Catalogs}",
      journal = {\apjs},
         year = 2011,
        month = jul,
       volume = {195},
       number = {1},
          eid = {10},
        pages = {10},
          doi = {10.1088/0067-0049/195/1/10},
archivePrefix = {arXiv},
       eprint = {1105.5643},
 primaryClass = {astro-ph.CO},
       adsurl = {https://ui.adsabs.harvard.edu/abs/2011ApJS..195...10X}
}

@ARTICLE{Xue2016,
       author = {{Xue}, Y.~Q. and {Luo}, B. and {Brandt}, W.~N. and others},
        title = "{The 2 Ms Chandra Deep Field-North Survey and the 250 ks Extended Chandra Deep Field-South Survey: Improved Point-source Catalogs}",
      journal = {\apjs},
         year = 2016,
        month = jun,
       volume = {224},
       number = {2},
          eid = {15},
        pages = {15},
          doi = {10.3847/0067-0049/224/2/15},
archivePrefix = {arXiv},
       eprint = {1602.06299},
 primaryClass = {astro-ph.GA},
       adsurl = {https://ui.adsabs.harvard.edu/abs/2016ApJS..224...15X}
}

@ARTICLE{Zou2026,
       author = {{Zou}, Fan and {Gallo}, Elena and {Zuo}, Zihao and {Hodges-Kluck}, Edmund and {Nguyen}, Dieu D. and {Roberts-Borsani}, Guido and {Madau}, Piero and {Pacucci}, Fabio and {Seth}, Anil C. and {Treu}, Tommaso and {Wang}, Shouyi},
        title = "{Revisiting the Claim for a Direct-Collapse Black Hole in UHZ1 at $z=10.05$}",
      journal = {arXiv e-prints},
         year = 2026,
        month = mar,
          eid = {arXiv:2603.24893},
        pages = {arXiv:2603.24893},
          doi = {10.48550/arXiv.2603.24893},
archivePrefix = {arXiv},
       eprint = {2603.24893},
 primaryClass = {astro-ph.GA},
       adsurl = {https://ui.adsabs.harvard.edu/abs/2026arXiv260324893Z}
}

@INPROCEEDINGS{CIAO2006,
       author = {{Fruscione}, Antonella and {McDowell}, Jonathan C. and {Allen}, Glenn E. and {Brickhouse}, Nancy S. and {Burke}, Douglas J. and {Davis}, John E. and {Durham}, Nick and {Elvis}, Martin and {Galle}, Elizabeth C. and {Harris}, Daniel E. and {Huenemoerder}, David P. and {Houck}, John C. and {Ishibashi}, Bish and {Karovska}, Margarita and {Nicastro}, Fabrizio and {Noble}, Michael S. and {Nowak}, Michael A. and {Primini}, Frank A. and {Siemiginowska}, Aneta and {Smith}, Randall K. and {Wise}, Michael},
        title = "{CIAO: Chandra's data analysis system}",
    booktitle = {Observatory Operations: Strategies, Processes, and Systems},
         year = 2006,
       editor = {{Silva}, David R. and {Doxsey}, Rodger E.},
       series = {Society of Photo-Optical Instrumentation Engineers (SPIE) Conference Series},
       volume = {6270},
        month = jun,
          eid = {62701V},
        pages = {62701V},
          doi = {10.1117/12.671760},
       adsurl = {https://ui.adsabs.harvard.edu/abs/2006SPIE.6270E..1VF}
}

@ARTICLE{Dey2019,
       author = {{Dey}, Arjun and {Schlegel}, David J. and {Lang}, Dustin and {Blum}, Robert and {Burleigh}, Kaylan and {Fan}, Xiaohui and {Findlay}, Joseph R. and {Finkbeiner}, Doug and {Herrera}, David and {Juneau}, St{\'e}phanie and {Landriau}, Martin and {Levi}, Michael and {McGreer}, Ian and {Meisner}, Aaron and {Myers}, Adam D. and {Moustakas}, John and {Nugent}, Peter and {Patej}, Anna and {Schlafly}, Edward F. and {Walker}, Alistair R. and {Valdes}, Francisco and {Weaver}, Benjamin A. and {Y{\`e}che}, Christophe and {Zou}, Hu and {Zhou}, Xu and {Abareshi}, Behzad and {Abbott}, T.~M.~C. and {Abolfathi}, Bela and {Aguilera}, C. and {Alam}, Shadab and {Allen}, Lori and {Alvarez}, A. and {Annis}, James and {Ansarinejad}, Behzad and {Aubert}, Marie and {Beechert}, Jacqueline and {Bell}, Eric F. and {BenZvi}, Segev Y. and {Beutler}, Florian and {Bielby}, Richard M. and {Bolton}, Adam S. and {Brice{\~n}o}, C{\'e}sar and {Buckley-Geer}, Elizabeth J. and {Butler}, Karen and {Calamida}, Annalisa and {Carlberg}, Raymond G. and {Carter}, Paul and {Casas}, Ricard and {Castander}, Francisco J. and {Choi}, Yumi and {Comparat}, Johan and {Cukanovaite}, Elena and {Delubac}, Timoth{\'e}e and {DeVries}, Kaitlin and {Dey}, Sharmila and {Dhungana}, Govinda and {Dickinson}, Mark and {Ding}, Zhejie and {Donaldson}, John B. and {Duan}, Yutong and {Duckworth}, Christopher J. and {Eftekharzadeh}, Sarah and {Eisenstein}, Daniel J. and {Etourneau}, Thomas and {Fagrelius}, Parker A. and {Farihi}, Jay and {Fitzpatrick}, Mike and {Font-Ribera}, Andreu and {Fulmer}, Leah and {G{\"a}nsicke}, Boris T. and {Gaztanaga}, Enrique and {George}, Koshy and {Gerdes}, David W. and {Gontcho}, Satya Gontcho A. and {Gorgoni}, Claudio and {Green}, Gregory and {Guy}, Julien and {Harmer}, Diane and {Hernandez}, M. and {Honscheid}, Klaus and {Huang}, Lijuan Wendy and {James}, David J. and {Jannuzi}, Buell T. and {Jiang}, Linhua and {Joyce}, Richard and {Karcher}, Armin and {Karkar}, Sonia and {Kehoe}, Robert and {Kneib}, Jean-Paul and {Kueter-Young}, Andrea and {Lan}, Ting-Wen and {Lauer}, Tod R. and {Le Guillou}, Laurent and {Le Van Suu}, Auguste and {Lee}, Jae Hyeon and {Lesser}, Michael and {Perreault Levasseur}, Laurence and {Li}, Ting S. and {Mann}, Justin L. and {Marshall}, Robert and {Mart{\'\i}nez-V{\'a}zquez}, C.~E. and {Martini}, Paul and {du Mas des Bourboux}, H{\'e}lion and {McManus}, Sean and {Meier}, Tobias Gabriel and {M{\'e}nard}, Brice and {Metcalfe}, Nigel and {Mu{\~n}oz-Guti{\'e}rrez}, Andrea and {Najita}, Joan and {Napier}, Kevin and {Narayan}, Gautham and {Newman}, Jeffrey A. and {Nie}, Jundan and {Nord}, Brian and {Norman}, Dara J. and {Olsen}, Knut A.~G. and {Paat}, Anthony and {Palanque-Delabrouille}, Nathalie and {Peng}, Xiyan and {Poppett}, Claire L. and {Poremba}, Megan R. and {Prakash}, Abhishek and {Rabinowitz}, David and {Raichoor}, Anand and {Rezaie}, Mehdi and {Robertson}, A.~N. and {Roe}, Natalie A. and {Ross}, Ashley J. and {Ross}, Nicholas P. and {Rudnick}, Gregory and {Safonova}, Sasha and {Saha}, Abhijit and {S{\'a}nchez}, F. Javier and {Savary}, Elodie and {Schweiker}, Heidi and {Scott}, Adam and {Seo}, Hee-Jong and {Shan}, Huanyuan and {Silva}, David R. and {Slepian}, Zachary and {Soto}, Christian and {Sprayberry}, David and {Staten}, Ryan and {Stillman}, Coley M. and {Stupak}, Robert J. and {Summers}, David L. and {Sien Tie}, Suk and {Tirado}, H. and {Vargas-Maga{\~n}a}, Mariana and {Vivas}, A. Katherina and {Wechsler}, Risa H. and {Williams}, Doug and {Yang}, Jinyi and {Yang}, Qian and {Yapici}, Tolga and {Zaritsky}, Dennis and {Zenteno}, A. and {Zhang}, Kai and {Zhang}, Tianmeng and {Zhou}, Rongpu and {Zhou}, Zhimin},
        title = "{Overview of the DESI Legacy Imaging Surveys}",
      journal = {\aj},
         year = 2019,
        month = may,
       volume = {157},
       number = {5},
          eid = {168},
        pages = {168},
          doi = {10.3847/1538-3881/ab089d},
archivePrefix = {arXiv},
       eprint = {1804.08657},
 primaryClass = {astro-ph.IM},
       adsurl = {https://ui.adsabs.harvard.edu/abs/2019AJ....157..168D}
}

@ARTICLE{Astropy2022,
       author = {{Astropy Collaboration} and {Price-Whelan}, Adrian M. and {Lim}, Pey Lian and {Earl}, Nicholas and {Starkman}, Nathaniel and {Bradley}, Larry and {Shupe}, David L. and {Patil}, Aarya A. and {Corrales}, Lia and {Brasseur}, C.~E. and {N{\"o}the}, Maximilian and {Donath}, Axel and {Tollerud}, Erik and {Morris}, Brett M. and {Ginsburg}, Adam and {Vaher}, Eero and {Weaver}, Benjamin A. and {Tocknell}, James and {Jamieson}, William and {van Kerkwijk}, Marten H. and {Robitaille}, Thomas P. and {Merry}, Bruce and {Bachetti}, Matteo and {G{\"u}nther}, H. Moritz and {Aldcroft}, Thomas L. and {Alvarado-Montes}, Jaime A. and {Archibald}, Anne M. and {B{\'o}di}, Attila and {Bapat}, Shreyas and {Barentsen}, Geert and {Baz{\'a}n}, Juanjo and {Biswas}, Manish and {Boquien}, M{\'e}d{\'e}ric and {Burke}, D.~J. and {Cara}, Daria and {Cara}, Mihai and {Conroy}, Kyle E. and {Conseil}, Simon and {Craig}, Matthew W. and {Cross}, Robert M. and {Cruz}, Kelle L. and {D'Eugenio}, Francesco and {Dencheva}, Nadia and {Devillepoix}, Hadrien A.~R. and {Dietrich}, J{\"o}rg P. and {Eigenbrot}, Arthur Davis and {Erben}, Thomas and {Ferreira}, Leonardo and {Foreman-Mackey}, Daniel and {Fox}, Ryan and {Freij}, Nabil and {Garg}, Suyog and {Geda}, Robel and {Glattly}, Lauren and {Gondhalekar}, Yash and {Gordon}, Karl D. and {Grant}, David and {Greenfield}, Perry and {Groener}, Austen M. and {Guest}, Steve and {Gurovich}, Sebastian and {Handberg}, Rasmus and {Hart}, Akeem and {Hatfield-Dodds}, Zac and {Homeier}, Derek and {Hosseinzadeh}, Griffin and {Jenness}, Tim and {Jones}, Craig K. and {Joseph}, Prajwel and {Kalmbach}, J. Bryce and {Karamehmetoglu}, Emir and {Ka{\l}uszy{\'n}ski}, Miko{\l}aj and {Kelley}, Michael S.~P. and {Kern}, Nicholas and {Kerzendorf}, Wolfgang E. and {Koch}, Eric W. and {Kulumani}, Shankar and {Lee}, Antony and {Ly}, Chun and {Ma}, Zhiyuan and {MacBride}, Conor and {Maljaars}, Jakob M. and {Muna}, Demitri and {Murphy}, N.~A. and {Norman}, Henrik and {O'Steen}, Richard and {Oman}, Kyle A. and {Pacifici}, Camilla and {Pascual}, Sergio and {Pascual-Granado}, J. and {Patil}, Rohit R. and {Perren}, Gabriel I. and {Pickering}, Timothy E. and {Rastogi}, Tanuj and {Roulston}, Benjamin R. and {Ryan}, Daniel F. and {Rykoff}, Eli S. and {Sabater}, Jose and {Sakurikar}, Parikshit and {Salgado}, Jes{\'u}s and {Sanghi}, Aniket and {Saunders}, Nicholas and {Savchenko}, Volodymyr and {Schwardt}, Ludwig and {Seifert-Eckert}, Michael and {Shih}, Albert Y. and {Jain}, Anany Shrey and {Shukla}, Gyanendra and {Sick}, Jonathan and {Simpson}, Chris and {Singanamalla}, Sudheesh and {Singer}, Leo P. and {Singhal}, Jaladh and {Sinha}, Manodeep and {Sip{\H{o}}cz}, Brigitta M. and {Spitler}, Lee R. and {Stansby}, David and {Streicher}, Ole and {{\v{S}}umak}, Jani and {Swinbank}, John D. and {Taranu}, Dan S. and {Tewary}, Nikita and {Tremblay}, Grant R. and {de Val-Borro}, Miguel and {Van Kooten}, Samuel J. and {Vasovi{\'c}}, Zlatan and {Verma}, Shresth and {de Miranda Cardoso}, Jos{\'e} Vin{\'\i}cius and {Williams}, Peter K.~G. and {Wilson}, Tom J. and {Winkel}, Benjamin and {Wood-Vasey}, W.~M. and {Xue}, Rui and {Yoachim}, Peter and {Zhang}, Chen and {Zonca}, Andrea and {Astropy Project Contributors}},
        title = "{The Astropy Project: Sustaining and Growing a Community-oriented Open-source Project and the Latest Major Release (v5.0) of the Core Package}",
      journal = {\apj},
         year = 2022,
        month = aug,
       volume = {935},
       number = {2},
          eid = {167},
        pages = {167},
          doi = {10.3847/1538-4357/ac7c74},
archivePrefix = {arXiv},
       eprint = {2206.14220},
 primaryClass = {astro-ph.IM},
       adsurl = {https://ui.adsabs.harvard.edu/abs/2022ApJ...935..167A}
}

@ARTICLE{Valentino2023DJA,
       author = {{Valentino}, Francesco and {Brammer}, Gabriel and {Gould}, Katriona M.~L. and {Kokorev}, Vasily and {Fujimoto}, Seiji and {Jespersen}, Christian Kragh and {Vijayan}, Aswin P. and {Weaver}, John R. and {Ito}, Kei and {Tanaka}, Masayuki and {Ilbert}, Olivier and {Magdis}, Georgios E. and {Whitaker}, Katherine E. and {Faisst}, Andreas L. and {Gallazzi}, Anna and {Gillman}, Steven and {Gim{\'e}nez-Arteaga}, Clara and {G{\'o}mez-Guijarro}, Carlos and {Kubo}, Mariko and {Heintz}, Kasper E. and {Hirschmann}, Michaela and {Oesch}, Pascal and {Onodera}, Masato and {Rizzo}, Francesca and {Lee}, Minju and {Strait}, Victoria and {Toft}, Sune},
        title = "{An Atlas of Color-selected Quiescent Galaxies at z > 3 in Public JWST Fields}",
      journal = {\apj},
         year = 2023,
        month = apr,
       volume = {947},
       number = {1},
          eid = {20},
        pages = {20},
          doi = {10.3847/1538-4357/acbefa},
archivePrefix = {arXiv},
       eprint = {2302.10936},
 primaryClass = {astro-ph.GA},
       adsurl = {https://ui.adsabs.harvard.edu/abs/2023ApJ...947...20V}
}

@ARTICLE{Bergamini2023,
       author = {{Bergamini}, Pietro and {Acebron}, Ana and {Grillo}, Claudio and {Rosati}, Piero and {Caminha}, Gabriel Bartosch and {Mercurio}, Amata and {Vanzella}, Eros and {Mason}, Charlotte and {Treu}, Tommaso and {Angora}, Giuseppe and {Brammer}, Gabriel B. and {Meneghetti}, Massimo and {Nonino}, Mario and {Boyett}, Kristan and {Brada{\v{c}}}, Maru{\v{s}}a and {Castellano}, Marco and {Fontana}, Adriano and {Morishita}, Takahiro and {Paris}, Diego and {Prieto-Lyon}, Gonzalo and {Roberts-Borsani}, Guido and {Roy}, Namrata and {Santini}, Paola and {Vulcani}, Benedetta and {Wang}, Xin and {Yang}, Lilan},
        title = "{The {GLASS-JWST} Early Release Science Program.
                 III. Strong-lensing Model of Abell 2744 and Its
                 Infalling Regions}",
      journal = {\apj},
         year = 2023,
        month = jul,
       volume = {952},
          eid = {84},
        pages = {84},
          doi = {10.3847/1538-4357/acd643},
archivePrefix = {arXiv},
       eprint = {2303.10210},
 primaryClass = {astro-ph.CO},
       adsurl = {https://ui.adsabs.harvard.edu/abs/2023ApJ...952...84B}
}

@ARTICLE{Castellano2022,
       author = {{Castellano}, Marco and {Fontana}, Adriano and {Treu}, Tommaso and {Santini}, Paola and {Merlin}, Emiliano and {Leethochawalit}, Nicha and {Trenti}, Michele and {Vanzella}, Eros and {Mestric}, Uros and {Bonchi}, Andrea and {Belfiori}, Davide and {Nonino}, Mario and {Paris}, Diego and {Polenta}, Gianluca and {Roberts-Borsani}, Guido and {Boyett}, Kristan and {Brada{\v{c}}}, Maru{\v{s}}a and {Calabrò}, Antonello and {Glazebrook}, Karl and {Grillo}, Claudio and {Mascia}, Sara and {Mason}, Charlotte and {Mercurio}, Amata and {Morishita}, Takahiro and {Nanayakkara}, Themiya and {Pentericci}, Laura and {Rosati}, Piero and {Vulcani}, Benedetta and {Wang}, Xin and {Yang}, Lilan},
        title = "{Early Results from {GLASS-JWST}. III. Galaxy
                 Candidates at $z\sim9$--15}",
      journal = {\apjl},
         year = 2022,
        month = oct,
       volume = {938},
          eid = {L15},
        pages = {L15},
          doi = {10.3847/2041-8213/ac94d0},
archivePrefix = {arXiv},
       eprint = {2207.09436},
 primaryClass = {astro-ph.GA},
       adsurl = {https://ui.adsabs.harvard.edu/abs/2022ApJ...938L..15C}
}

@ARTICLE{Fujimoto2024,
       author = {{Fujimoto}, Seiji and {Wang}, Bingjie and {Weaver}, John R. and {Kokorev}, Vasily and {Atek}, Hakim and {Bezanson}, Rachel and {Labbe}, Ivo and {Brammer}, Gabriel and {Greene}, Jenny E. and {Chemerynska}, Iryna and {Dayal}, Pratika and {de Graaff}, Anna and {Furtak}, Lukas J. and {Oesch}, Pascal A. and {Setton}, David J. and {Price}, Sedona H. and {Miller}, Tim B. and {Williams}, Christina C. and {Whitaker}, Katherine E. and {Zitrin}, Adi and {Cutler}, Sam E. and {Leja}, Joel and {Pan}, Richard and {Coe}, Dan and {van Dokkum}, Pieter and {Feldmann}, Robert and {Fudamoto}, Yoshinobu and {Goulding}, Andy D. and {Khullar}, Gourav and {Marchesini}, Danilo and {Maseda}, Michael and {Nanayakkara}, Themiya and {Nelson}, Erica J. and {Smit}, Renske and {Stefanon}, Mauro and {Weibel}, Andrea},
        title = "{{UNCOVER}: A {NIRSpec} Census of Lensed Galaxies
                 at $z=8.50$--13.08 Probing a High-{AGN} Fraction
                 and Ionized Bubbles in the Shadow}",
      journal = {\apj},
         year = 2024,
        month = dec,
       volume = {977},
          eid = {250},
        pages = {250},
          doi = {10.3847/1538-4357/ad9027},
archivePrefix = {arXiv},
       eprint = {2308.11609},
 primaryClass = {astro-ph.GA},
       adsurl = {https://ui.adsabs.harvard.edu/abs/2024ApJ...977..250F}
}

@ARTICLE{Napolitano2025,
       author = {{Napolitano}, L. and {Castellano}, M. and {Pentericci}, L. and {Arrabal Haro}, P. and {Fontana}, A. and {Treu}, T. and {Bergamini}, P. and {Calabrò}, A. and {Mascia}, S. and {Morishita}, T. and {Roberts-Borsani}, G. and {Santini}, P. and {Vanzella}, E. and {Vulcani}, B. and {Zakharova}, D. and {Bakx}, T. and {Dickinson}, M. and {Grillo}, C. and {Leethochawalit}, N. and {Llerena}, M. and {Merlin}, E. and {Paris}, D. and {Rojas-Ruiz}, S. and {Rosati}, P. and {Wang}, X. and {Yoon}, I. and {Zavala}, J.},
        title = "{Seven Wonders of Cosmic Dawn: {JWST} Confirms
                 a High Abundance of Galaxies and {AGN} at
                 $z\simeq9$--11 in the {GLASS} Field}",
      journal = {\aap},
         year = 2025,
        month = jan,
       volume = {693},
          eid = {A50},
        pages = {A50},
          doi = {10.1051/0004-6361/202452090},
archivePrefix = {arXiv},
       eprint = {2410.10967},
 primaryClass = {astro-ph.GA},
       adsurl = {https://ui.adsabs.harvard.edu/abs/2025A&A...693A..50N}
}

@ARTICLE{RobertsBorsani2023,
       author = {{Roberts-Borsani}, Guido and {Treu}, Tommaso and {Chen}, Wenlei and {Morishita}, Takahiro and {Vanzella}, Eros and {Zitrin}, Adi and {Bergamini}, Pietro and {Castellano}, Marco and {Fontana}, Adriano and {Glazebrook}, Karl and {Grillo}, Claudio and {Kelly}, Patrick L. and {Merlin}, Emiliano and {Nanayakkara}, Themiya and {Paris}, Diego and {Rosati}, Piero and {Yang}, Lilan and {Acebron}, Ana and {Bonchi}, Andrea and {Boyett}, Kit and {Brada{\v{c}}}, Maru{\v{s}}a and {Brammer}, Gabriel and {Broadhurst}, Tom and {Calabr{\'o}}, Antonello and {Diego}, Jose M. and {Dressler}, Alan and {Furtak}, Lukas J. and {Filippenko}, Alexei V. and {Henry}, Alaina and {Koekemoer}, Anton M. and {Leethochawalit}, Nicha and {Malkan}, Matthew A. and {Mason}, Charlotte and {Mercurio}, Amata and {Metha}, Benjamin and {Pentericci}, Laura and {Pierel}, Justin and {Rieck}, Steven and {Roy}, Namrata and {Santini}, Paola and {Strait}, Victoria and {Strausbaugh}, Robert and {Trenti}, Michele and {Vulcani}, Benedetta and {Wang}, Lifan and {Wang}, Xin and {Windhorst}, Rogier A.},
        title = "{The Nature of an Ultrafaint Galaxy in the
                 Cosmic Dark Ages Seen with {JWST}}",
      journal = {Nature},
         year = 2023,
        month = jun,
       volume = {618},
       number = {7965},
        pages = {480--483},
          doi = {10.1038/s41586-023-05994-w},
archivePrefix = {arXiv},
       eprint = {2210.15639},
 primaryClass = {astro-ph.GA},
       adsurl = {https://ui.adsabs.harvard.edu/abs/2023Natur.618..480R}
}

@ARTICLE{RobertsBorsani2026,
       author = {{Roberts-Borsani}, Guido and {Oesch}, Pascal A and {Ellis}, Richard and {Weibel}, Andrea and {Giovinazzo}, Emma and {Bouwens}, Rychard and {Dayal}, Pratika and {Fontana}, Adriano and {Heintz}, Kasper E and {Matthee}, Jorryt and {Meyer}, Romain A and {Pentericci}, Laura and {Shapley}, Alice and {Tacchella}, Sandro and {Treu}, Tommaso and {Walter}, Fabian and {Atek}, Hakim and {Bose}, Sownak and {Castellano}, Marco and {Fudamoto}, Yoshinobu and {Morishita}, Takahiro and {Naidu}, Rohan P and {Sanders}, Ryan L and {van der Wel}, Arjen},
        title = "{{JWST} Spectroscopic Insights into the Diversity
                 of Galaxies in the First 500 Myr: Short-lived
                 Snapshots along a Common Evolutionary Pathway}",
      journal = {\mnras},
         year = 2026,
        month = apr,
       volume = {548},
       number = {3},
          eid = {stag701},
        pages = {stag701},
          doi = {10.1093/mnras/stag701},
archivePrefix = {arXiv},
       eprint = {2508.21708},
 primaryClass = {astro-ph.GA},
       adsurl = {https://ui.adsabs.harvard.edu/abs/2025arXiv250821708R}
}

@article{Cash1979ApJ...228..939C,
       author = {{Cash}, W.},
  title = {Parameter Estimation in Astronomy through Application of the Likelihood Ratio},
  journal = {\apj},
  year = {1979},
        month = mar,
  volume = {228},
  pages = {939--947},
  doi = {10.1086/156922},
       adsurl = {https://ui.adsabs.harvard.edu/abs/1979ApJ...228..939C}
}

@inproceedings{Davis2012SPIE.8443E..1AD,
       author = {{Davis}, John E. and {Bautz}, Marshall W. and {Dewey}, Daniel and {Heilmann}, Ralf K. and {Houck}, John C. and {Huenemoerder}, David P. and {Marshall}, Herman L. and {Nowak}, Michael A. and {Schattenburg}, Mark L. and {Schulz}, Norbert S. and {Smith}, Randall K.},
  title = {Raytracing with MARX: X-Ray Observatory Design, Calibration, and Support},
  booktitle = {Space Telescopes and Instrumentation 2012: Ultraviolet to Gamma Ray},
  series = {Proceedings of SPIE},
  year = {2012},
        month = sep,
  volume = {8443},
  eid = {84431A},
  doi = {10.1117/12.926937},
       adsurl = {https://ui.adsabs.harvard.edu/abs/2012SPIE.8443E..1AD}
}

@article{Freeman2002ApJS..138..185F,
       author = {{Freeman}, P. E. and {Kashyap}, V. and {Rosner}, R. and {Lamb}, D. Q.},
  title = {A Wavelet-Based Algorithm for the Spatial Analysis of Poisson Data},
  journal = {\apjs},
  year = {2002},
        month = jan,
  volume = {138},
  pages = {185--218},
  doi = {10.1086/324017},
       adsurl = {https://ui.adsabs.harvard.edu/abs/2002ApJS..138..185F}
}

@article{Puccetti2009ApJS..185..586P,
       author = {{Puccetti}, S. and {Vignali}, C. and {Cappelluti}, N. and {Fiore}, F. and {Zamorani}, G. and {Aldcroft}, T. L. and {Elvis}, M. and {Gilli}, R. and {Miyaji}, T. and {Brunner}, H. and {Brusa}, M. and {Civano}, F. and {Comastri}, A. and {Damiani}, F. and {Fruscione}, A. and {Finoguenov}, A. and {Koekemoer}, A. M. and {Mainieri}, V.},
  title = {The Chandra Survey of the COSMOS Field. II. Source Detection and Photometry},
  journal = {\apjs},
  year = {2009},
        month = dec,
  volume = {185},
  pages = {586--601},
  doi = {10.1088/0067-0049/185/2/586},
       adsurl = {https://ui.adsabs.harvard.edu/abs/2009ApJS..185..586P}
}

@ARTICLE{Kraft1991,
       author = {{Kraft}, Ralph P. and {Burrows}, David N. and {Nousek}, John A.},
        title = "{Determination of Confidence Limits for Experiments with Low Numbers of Counts}",
      journal = {\apj},
         year = 1991,
        month = jun,
       volume = {374},
        pages = {344},
          doi = {10.1086/170124},
       adsurl = {https://ui.adsabs.harvard.edu/abs/1991ApJ...374..344K}
}

@ARTICLE{Zou2023,
       author = {{Zou}, Fan and {Brandt}, W.~N. and {Ni}, Qingling and {Zhu}, Shifu and {Alexander}, David M. and {Bauer}, Franz E. and {Chen}, Chien-Ting J. and {Luo}, Bin and {Sun}, Mouyuan and {Vignali}, Cristian and {Vito}, Fabio and {Xue}, Yongquan and {Yan}, Wei},
        title = "{Identification and Characterization of a Large Sample of Distant Active Dwarf Galaxies in XMM-SERVS}",
      journal = {\apj},
         year = 2023,
        month = jun,
       volume = {950},
       number = {2},
          eid = {136},
        pages = {136},
          doi = {10.3847/1538-4357/acce39},
archivePrefix = {arXiv},
       eprint = {2304.09904},
 primaryClass = {astro-ph.GA},
       adsurl = {https://ui.adsabs.harvard.edu/abs/2023ApJ...950..136Z}
}

@INPROCEEDINGS{Freeman2001,
       author = {{Freeman}, Peter and {Doe}, Stephen and {Siemiginowska}, Aneta},
        title = "{Sherpa: a mission-independent data analysis application}",
    booktitle = {Astronomical Data Analysis},
         year = 2001,
       editor = {{Starck}, Jean-Luc and {Murtagh}, Fionn D.},
       series = {Society of Photo-Optical Instrumentation Engineers (SPIE) Conference Series},
       volume = {4477},
        month = nov,
        pages = {76-87},
          doi = {10.1117/12.447161},
       adsurl = {https://ui.adsabs.harvard.edu/abs/2001SPIE.4477...76F}
}

@ARTICLE{Wright2024,
       author = {{Wright}, Angus H. and {Kuijken}, Konrad and {Hildebrandt}, Hendrik and {Radovich}, Mario and {Bilicki}, Maciej and {Dvornik}, Andrej and {Getman}, Fedor and {Heymans}, Catherine and {Hoekstra}, Henk and {Li}, Shun-Sheng and {Miller}, Lance and {Napolitano}, Nicola R. and {Xia}, Qianli and {Asgari}, Marika and {Brescia}, Massimo and {Buddelmeijer}, Hugo and {Burger}, Pierre and {Castignani}, Gianluca and {Cavuoti}, Stefano and {de Jong}, Jelte and {Edge}, Alastair and {Giblin}, Benjamin and {Giocoli}, Carlo and {Harnois-D{\'e}raps}, Joachim and {Jalan}, Priyanka and {Joachimi}, Benjamin and {John William}, Anjitha and {Joudaki}, Shahab and {Kannawadi}, Arun and {Kaur}, Gursharanjit and {La Barbera}, Francesco and {Linke}, Laila and {Mahony}, Constance and {Maturi}, Matteo and {Moscardini}, Lauro and {Nakoneczny}, Szymon J. and {Paolillo}, Maurizio and {Porth}, Lucas and {Puddu}, Emanuella and {Reischke}, Robert and {Schneider}, Peter and {Sereno}, Mauro and {Shan}, HuanYuan and {Sif{\'o}n}, Crist{\'o}bal and {St{\"o}lzner}, Benjamin and {Tr{\"o}ster}, Tilman and {Valentijn}, Edwin and {van den Busch}, Jan Luca and {Verdoes Kleijn}, Gijs and {Wittje}, Anna and {Yan}, Ziang and {Yao}, Ji and {Yoon}, Mijin and {Zhang}, Yun-Hao},
        title = "{The fifth data release of the Kilo Degree Survey: Multi-epoch optical/NIR imaging covering wide and legacy-calibration fields}",
      journal = {\aap},
         year = 2024,
        month = jun,
       volume = {686},
          eid = {A170},
        pages = {A170},
          doi = {10.1051/0004-6361/202346730},
archivePrefix = {arXiv},
       eprint = {2503.19439},
 primaryClass = {astro-ph.GA},
       adsurl = {https://ui.adsabs.harvard.edu/abs/2024A&A...686A.170W}
}

@ARTICLE{Kim2007,
       author = {{Kim}, Minsun and {Kim}, Dong-Woo and {Wilkes}, Belinda J. and {Green}, Paul J. and {Kim}, Eunhyeuk and {Anderson}, Craig S. and {Barkhouse}, Wayne A. and {Evans}, Nancy R. and {Ivezi{\'c}}, {\v{Z}}eljko and {Karovska}, Margarita and {Kashyap}, Vinay L. and {Lee}, Myung Gyoon and {Maksym}, Peter and {Mossman}, Amy E. and {Silverman}, John D. and {Tananbaum}, Harvey D.},
        title = "{Chandra Multiwavelength Project X-Ray Point Source Catalog}",
      journal = {\apjs},
         year = 2007,
        month = apr,
       volume = {169},
       number = {2},
        pages = {401-429},
          doi = {10.1086/511634},
archivePrefix = {arXiv},
       eprint = {astro-ph/0611840},
 primaryClass = {astro-ph},
       adsurl = {https://ui.adsabs.harvard.edu/abs/2007ApJS..169..401K}
}

@INPROCEEDINGS{Grant2024,
       author = {{Grant}, Catherine E. and {Bautz}, Marshall W. and {Plucinsky}, Paul P. and {Ford}, Peter G.},
        title = "{The advanced CCD imaging spectrometer on the Chandra x-ray observatory: twenty-five years of on-orbit operation}",
    booktitle = {Space Telescopes and Instrumentation 2024: Ultraviolet to Gamma Ray},
         year = 2024,
       editor = {{den Herder}, Jan-Willem A. and {Nikzad}, Shouleh and {Nakazawa}, Kazuhiro},
       series = {Society of Photo-Optical Instrumentation Engineers (SPIE) Conference Series},
       volume = {13093},
        month = aug,
          eid = {130931E},
        pages = {130931E},
          doi = {10.1117/12.3018498},
archivePrefix = {arXiv},
       eprint = {2406.18395},
 primaryClass = {astro-ph.IM},
       adsurl = {https://ui.adsabs.harvard.edu/abs/2024SPIE13093E..1EG}
}

@INPROCEEDINGS{Plucinsky2018,
       author = {{Plucinsky}, Paul P. and {Bogd{\'a}n}, {\'A}kos and {Marshall}, Herman L. and {Tice}, Neil W.},
        title = "{The complicated evolution of the ACIS contamination layer over the mission life of the Chandra X-ray Observatory}",
    booktitle = {Space Telescopes and Instrumentation 2018: Ultraviolet to Gamma Ray},
         year = 2018,
       editor = {{den Herder}, Jan-Willem A. and {Nikzad}, Shouleh and {Nakazawa}, Kazuhiro},
       series = {Society of Photo-Optical Instrumentation Engineers (SPIE) Conference Series},
       volume = {10699},
        month = jul,
          eid = {106996B},
        pages = {106996B},
          doi = {10.1117/12.2312748},
archivePrefix = {arXiv},
       eprint = {1809.02225},
 primaryClass = {astro-ph.IM},
       adsurl = {https://ui.adsabs.harvard.edu/abs/2018SPIE10699E..6BP}
}

@ARTICLE{Watson2009,
  author        = {{Watson}, M. G. and {Schr{\"o}der}, A. C. and {Fyfe}, D. and
                   {Page}, C. G. and {Lamer}, G. and {Mateos}, S. and {Pye}, J. and
                   {Sakano}, M. and {Rosen}, S. and others},
  title         = "{The XMM-Newton Serendipitous Survey. V. The Second XMM-Newton Serendipitous Source Catalogue}",
  journal       = {\aap},
  year          = {2009},
  month         = jan,
  volume        = {493},
  number        = {1},
  pages         = {339--373},
  doi           = {10.1051/0004-6361:200810534},
  archivePrefix = {arXiv},
  eprint        = {0807.1067},
  primaryClass  = {astro-ph}
}

@ARTICLE{Brunner2022,
       author = {{Brunner}, H. and {Liu}, T. and {Lamer}, G. and
                 {Georgakakis}, A. and {Merloni}, A. and {Brusa}, M. and
                 {Bulbul}, E. and {Dennerl}, K. and {Friedrich}, S. and
                 others},
        title = "{The eROSITA Final Equatorial Depth Survey (eFEDS): X-ray catalogue}",
      journal = {\aap},
         year = 2022,
        month = may,
       volume = {661},
          eid = {A1},
        pages = {A1},
          doi = {10.1051/0004-6361/202141266},
       adsurl = {https://ui.adsabs.harvard.edu/abs/2022A&A...661A...1B}
}

@ARTICLE{Cappelluti2007,
  author = {{Cappelluti}, N. and {Hasinger}, G. and {Brusa}, M. and
            {Comastri}, A. and {Zamorani}, G. and {B{\"o}hringer}, H. and
            {Brunner}, H. and {Civano}, F. and {Finoguenov}, A. and
            {Fiore}, F. and {Gilli}, R. and {Griffiths}, R.~E. and
            {Mainieri}, V. and {Matute}, I. and {Miyaji}, T. and
            {Silverman}, J.},
  title = "{The XMM-Newton Wide-Field Survey in the COSMOS Field. II.
            X-Ray Data and the log N--log S Relations}",
  journal = {\apjs},
  year = {2007},
  volume = {172},
  number = {1},
  pages = {341--352},
  doi = {10.1086/516586},
  adsurl = {https://ui.adsabs.harvard.edu/abs/2007ApJS..172..341C}
}

@ARTICLE{Cappelluti2009,
  author = {{Cappelluti}, N. and {Brusa}, M. and {Hasinger}, G. and
            {Comastri}, A. and {Zamorani}, G. and {Finoguenov}, A. and
            {Gilli}, R. and {Puccetti}, S. and {Miyaji}, T. and
            {Salvato}, M. and {Vignali}, C. and {Aldcroft}, T.~L. and
            {B{\"o}hringer}, H. and {Brunner}, H. and {Civano}, F. and
            {Elvis}, M. and {Fiore}, F. and {Fruscione}, A. and
            {Griffiths}, R.~E. and {Guzzo}, L. and {Iovino}, A. and
            {Koekemoer}, A.~M. and {Mainieri}, V. and {Scoville}, N.~Z. and
            {Shopbell}, P. and {Silverman}, J. and {Urry}, C.~M.},
  title = "{The XMM-Newton wide-field survey in the COSMOS field.
            The point-like X-ray source catalogue}",
  journal = {\aap},
  year = {2009},
  volume = {497},
  pages = {635--648},
  doi = {10.1051/0004-6361/200810794},
  adsurl = {https://ui.adsabs.harvard.edu/abs/2009A&A...497..635C}
}

@article{Georgakakis2008,
  author = {{Georgakakis}, A. and {Nandra}, K. and {Laird}, E.~S. and
            {Aird}, J. and {Trichas}, M.},
  title = {A New Method for Determining the Sensitivity of X-Ray Imaging
           Observations and the X-Ray Number Counts},
  journal = {\mnras},
  year = {2008},
  volume = {388},
  pages = {1205--1213},
  doi = {10.1111/j.1365-2966.2008.13423.x}
}

@ARTICLE{1986ApJ...303..336G,
       author = {{Gehrels}, N.},
        title = "{Confidence Limits for Small Numbers of Events in Astrophysical Data}",
      journal = {\apj},
         year = 1986,
        month = apr,
       volume = {303},
        pages = {336},
          doi = {10.1086/164079},
       adsurl = {https://ui.adsabs.harvard.edu/abs/1986ApJ...303..336G}
}

@ARTICLE{Broos2010,
       author = {{Broos}, P.~S. and {Townsley}, L.~K. and
                 {Feigelson}, E.~D. and {Getman}, K.~V. and
                 {Bauer}, F.~E. and {Garmire}, G.~P.},
        title = "{Innovations in the Analysis of Chandra-ACIS Observations}",
      journal = {\apj},
         year = 2010,
        month = may,
       volume = {714},
       number = {2},
        pages = {1582--1605},
          doi = {10.1088/0004-637X/714/2/1582},
archivePrefix = {arXiv},
       eprint = {1003.2397},
 primaryClass = {astro-ph.HE},
       adsurl = {https://ui.adsabs.harvard.edu/abs/2010ApJ...714.1582B}
}

@MISC{Broos2012,
       author = {{Broos}, Patrick and {Townsley}, Leisa and
                 {Getman}, Konstantin and {Bauer}, Franz},
        title = "{AE: ACIS Extract}",
 howpublished = {Astrophysics Source Code Library, record ascl:1203.001},
         year = 2012,
        month = mar,
          eid = {ascl:1203.001},
archivePrefix = {ascl},
       eprint = {1203.001},
       adsurl = {https://ui.adsabs.harvard.edu/abs/2012ascl.soft03001B}
}

@article{Branchesi2007,
  author = {Branchesi, M. and Gioia, I. M. and Fanti, C. and Fanti, R. and Cappelluti, N.},
  title = {{Chandra} Point-Source Counts in Distant Galaxy Clusters},
  journal = {Astronomy \& Astrophysics},
  year = {2007},
  volume = {462},
  pages = {449--458},
  doi = {10.1051/0004-6361:20066196}
}

@ARTICLE{Harris2020,
  author  = {{Harris}, Charles R. and {Millman}, K. Jarrod and
             {van der Walt}, St{\'e}fan J. and {Gommers}, Ralf and
             {Virtanen}, Pauli and {Cournapeau}, David and {Wieser}, Eric and
             {Taylor}, Julian and {Berg}, Sebastian and {Smith}, Nathaniel J. and
             {Kern}, Robert and {Picus}, Matti and {Hoyer}, Stephan and
             {van Kerkwijk}, Marten H. and {Brett}, Matthew and {Haldane}, Allan and
             {Fern{\'a}ndez del R{\'i}o}, Jaime and {Wiebe}, Mark and
             {Peterson}, Pearu and {G{\'e}rard-Marchant}, Pierre and
             {Sheppard}, Kevin and {Reddy}, Tyler and {Weckesser}, Warren and
             {Abbasi}, Hameer and {Gohlke}, Christoph and {Oliphant}, Travis E.},
  title   = {Array programming with {NumPy}},
  journal = {Nature},
  year    = {2020},
  volume  = {585},
  number  = {7825},
  pages   = {357--362},
  doi     = {10.1038/s41586-020-2649-2}
}

@ARTICLE{Virtanen2020,
  author  = {{Virtanen}, Pauli and {Gommers}, Ralf and {Oliphant}, Travis E. and
             {Haberland}, Matt and {Reddy}, Tyler and {Cournapeau}, David and
             {Burovski}, Evgeni and {Peterson}, Pearu and {Weckesser}, Warren and
             {Bright}, Jonathan and {van der Walt}, St{\'e}fan J. and
             {Brett}, Matthew and {Wilson}, Joshua and {Millman}, K. Jarrod and
             {Mayorov}, Nikolay and {Nelson}, Andrew R. J. and {Jones}, Eric and
             {Kern}, Robert and {Larson}, Eric and {Carey}, C J and
             {Polat}, {\.I}lhan and {Feng}, Yu and {Moore}, Eric W. and
             {VanderPlas}, Jake and {Laxalde}, Denis and {Perktold}, Josef and
             {Cimrman}, Robert and {Henriksen}, Ian and {Quintero}, E. A. and
             {Harris}, Charles R. and {Archibald}, Anne M. and
             {Ribeiro}, Ant{\^o}nio H. and {Pedregosa}, Fabian and
             {van Mulbregt}, Paul and {{SciPy} 1.0 Contributors}},
  title   = {{SciPy} 1.0: Fundamental Algorithms for Scientific Computing in {Python}},
  journal = {Nature Methods},
  year    = {2020},
  volume  = {17},
  number  = {3},
  pages   = {261--272},
  doi     = {10.1038/s41592-019-0686-2}
}

@ARTICLE{Hunter2007,
  author  = {{Hunter}, John D.},
  title   = {{Matplotlib}: A {2D} Graphics Environment},
  journal = {Computing in Science \& Engineering},
  year    = {2007},
  volume  = {9},
  number  = {3},
  pages   = {90--95},
  doi     = {10.1109/MCSE.2007.55}
}

@ARTICLE{Fujimoto2025DUALZ,
       author = {{Fujimoto}, Seiji and {Bezanson}, Rachel and {Labb{\'e}}, Ivo and
                 {Brammer}, Gabriel and {Price}, Sedona H. and {Wang}, Bingjie and
                 {Weaver}, John R. and {Fudamoto}, Yoshinobu and {Oesch}, Pascal A. and
                 {Williams}, Christina C. and {Dayal}, Pratika and {Feldmann}, Robert and
                 {Greene}, Jenny E. and {Leja}, Joel and {Whitaker}, Katherine E. and
                 {Zitrin}, Adi and {Cutler}, Sam E. and {Furtak}, Lukas J. and
                 {Pan}, Richard and {Chemerynska}, Iryna and {Kokorev}, Vasily and
                 {Miller}, Tim B. and {Atek}, Hakim and {van Dokkum}, Pieter and
                 {Juneau}, St{\'e}phanie and {Kassin}, Susan and {Khullar}, Gourav and
                 {Marchesini}, Danilo and {Maseda}, Michael and {Nelson}, Erica J. and
                 {Setton}, David J. and {Smit}, Renske},
        title = "{{DUALZ}: Deep {UNCOVER}--{ALMA} Legacy High-$z$ Survey}",
      journal = {\apjs},
         year = 2025,
        month = jun,
       volume = {278},
       number = {2},
          eid = {45},
        pages = {45},
          doi = {10.3847/1538-4365/adc677},
archivePrefix = {arXiv},
       eprint = {2309.07834},
 primaryClass = {astro-ph.GA},
       adsurl = {https://ui.adsabs.harvard.edu/abs/2025ApJS..278...45F}
}
\bibliographystyle{aasjournalv7}
\end{document}